\documentclass[a4paper,11pt]{article}
\pdfoutput=1 

\usepackage{jheppub} 

\usepackage[T1]{fontenc} 
\usepackage{ytableau}
\usepackage{gensymb}
\usepackage{subcaption}
\usepackage{tikz}
\usetikzlibrary{arrows}

\tikzset{
  treenode/.style = {align=center, inner sep=0pt, text centered,
    font=\sffamily},
  arn_f/.style = {treenode, circle, black, draw=black,
    text width=1.5em, very thick},
  arn_x/.style = {treenode, circle, black, draw=red,
    text width=1.5em, very thick},
  arn_y/.style = {treenode, circle, black, draw=blue,
    text width=1.5em, very thick},
}

\DeclareMathOperator{\Tr}{Tr}

\preprint{KIAS-P18073}

\title{On 3d Seiberg-like Dualities with Two Adjoints}

\author[a]{Chiung Hwang,}
\author[b,c]{Hyungchul Kim}
\author[d]{and Jaemo Park}

\affiliation[a]{School of Physics, Korea Institute for Advanced Study,\\Seoul 02455, Korea}
\affiliation[b]{Center of Mathematical Sciences and Applications, Harvard University,\\Cambridge, MA 02138, USA}
\affiliation[c]{Jefferson Physical Laboratory, Harvard University,\\Cambridge, MA 02138, USA}
\affiliation[d]{Department of Physics, POSTECH,\\Pohang 37673, Korea}

\emailAdd{chwang@kias.re.kr}
\emailAdd{hyungchul\_kim@g.harvard.edu}
\emailAdd{jaemo@postech.ac.kr}

\abstract{We study $\mathcal N = 2$ 3-d theories with two adjoints and fundamental flavors along with D-type superpotential.  For superpotential
$W_{D_{n+2}} = \mathrm{Tr} \left(X^{n+1}+X Y^2\right)$ with $n$ odd, we propose the 3d dualities, which we motivate from
the dimensional reduction of the related 4-d theory. We consider the factorization of the superconformal index and
match precisely the vortex partition function  of the dual pairs. In the language of the Higgs branch localization, the nonzero contribution of the vortex partition function comes from the discrete Higgs vacua of the massively deformed theory, which
precisely matches with that of the dual theory. We also clarify the monopole operators parametrizing the Coulomb branch of
such theories. Existence of independent monopole operators of charge 2 is crucial to describe the Coulomb branch.}

\begin{document}
\maketitle
\flushbottom

\section{Introduction}
\label{sec:intro}

4d $\mathcal N = 1$ asymptotically free $SU(N_c)$ gauge theories with matters in the adjoint representation are classified by their superpotentials \cite{Intriligator:2003mi}
\begin{align}
\label{eq:sup}
\begin{aligned}
& W_{\widehat O} = 0, \qquad W_{\widehat A} = \mathrm{Tr} Y^2, \qquad W_{\widehat D} = \mathrm{Tr} X Y^2, \qquad W_{\widehat E} = \mathrm{Tr} Y^3, \\
& W_{A_n} = \mathrm{Tr} \left(X^{n+1}+Y^2\right), \qquad W_{D_{n+2}} = \mathrm{Tr} \left(X^{n+1}+X Y^2\right), \\
& W_{E_6} = \mathrm{Tr} \left(Y^3+X^4\right), \quad W_{E_7} = \mathrm{Tr} \left(Y^3+Y X^3\right), \quad W_{E_8} = \mathrm{Tr} \left(Y^3+X^5\right).
\end{aligned}
\end{align}
Those theories contain extra (anti-)fundamental matters and flow to interacting fixed points if the number of fundamentals is within a certain range. One might note that the above classification only involves two adjoints while the asymptotically free condition also allows three adjoint matters. In that case, however, there is no extra fundamentals allowed such that the theory is nothing but the $\mathcal N = 4$ super Yang-Mills theory. Also note that in $A$-type theories, the superpotential includes a quadratic term such that one of the adjoints can be integrated out. The $A$-type theories are thus effectively one adjoint theories.

Some of those classes are conjectured to have Seiberg-like dualities \cite{Kutasov:1995ve,Kutasov:1995np,Kutasov:1995ss,Brodie:1996vx,Kutasov:2014yqa}. For the $SU(N_c)$ gauge theory with $N_f$ fundamental/anti-fundamental pairs, the dual gauge group is given by
\begin{align}
SU(\alpha_G N_f-N_c)
\end{align}
where the coefficient $\alpha_G$ is determined by the superpotential type $G$:
\begin{align}
\begin{aligned}
\alpha_{A_n} &= n, \\
\alpha_{D_{n+2}} &= 3 n, \\
\alpha_{E_7} &= 30.
\end{aligned}
\end{align}
The dualities have passed various consistency checks such as 't Hooft anomaly matchings, mass deformations for flavors and $ a $-theorem. It is important to note that the $ A_n $-type theory and the $ D_{n+2} $-type theory with odd $ n $ have classical chiral ring truncations which come from the superpotential and are independent of $ N_c $. Thus the chiral rings between dual theories match simply for $ N_c $ large enough. However, the $ D_{n+2} $-type theories with even $ n $ and $ E_7 $ theories do not have the classical chiral ring truncation so they are conjectured to have quantum chiral ring truncations in order for the duality to hold. A related issue was studied in \cite{Intriligator:2016sgx} where subtleties for the latter dualities were pointed out by studying flat directions associated with the adjoint fields.\footnote{This is explained at the section 2.}

Along the development of supersymmetric localization technique, the exact computation of a supersymmetric partition function has been a powerful tool for testing a conjectural duality \cite{Kapustin:2011gh,Willett:2011gp,Bashkirov:2011vy,Hwang:2011qt,Benini:2011mf,Hwang:2011ht,Agarwal:2012wd,Cheon:2012be,Cheon:2011th,Amariti:2011uw,Gang:2011jj,Benini:2012ui,Gadde:2013ftv,Benini:2013nda,Benini:2013xpa,Benini:2014mia,Gomis:2014eya,Gadde:2015wta,Cho:2017bhd,Closset:2017vvl,Kim:2017zis}. The superconformal index \cite{Kinney:2005ej,Romelsberger:2005eg,Romelsberger:2007ec,Bhattacharya:2008zy,Bhattacharya:2008bja}, for example, is defined as the supersymmetric partition function on $S^d \times S^1$, which can be exactly computed using the localization technique \cite{Kim:2009wb,Imamura:2011su,Kim:2012gu,Closset:2013sxa,Assel:2014paa}. Usually such localization leads to a finite dimensional matrix integral with integrations as much as the rank of the gauge group. The 4d dualities we discuss above, however, involve very high gauge ranks. As a result, only the large $N$ limit of a partition function has been considered in the literature \cite{Kutasov:2014wwa} while there is no finite $N$ computation so far.
\\

We instead turn our attention to the 3d versions of those theories, and their dualities. Since it was shown that the 4d Seiberg duality \cite{Seiberg:1994pq} and the 3d Aharony duality \cite{Aharony:1997gp} are directly related, many 4d dualities have been discussed with emphasis on their relations to 3d dualities \cite{Aharony:2013dha,Aharony:2013kma,Nii:2014jsa,Amariti:2014iza,Amariti:2015yea,Amariti:2015mva,Amariti:2015xna} (see also \cite{Amariti:2016kat}). In particular, the 3d compactification of the $A_n$-type duality have been discussed with extra supporting evidence \cite{Niarchos:2008jb,Niarchos:2009aa,Kapustin:2011vz,Kim:2013cma,Hwang:2015wna}. One significant feature of such compactifications is the nontrivial role of monopole operators, which appear as local chiral operators in 3d \cite{Affleck:1982as,Aharony:1997bx,Kapustin:1999ha,Borokhov:2002ib,Borokhov:2002cg,Intriligator:2013lca}. For example, due to the lack of anomaly, a naive 3d compactified theory seems to have a larger global symmetry than the original 4d one. However, such extra components of global symmetry, which are anomalous in 4d, are forbidden in 3d as well by monopole superpotentials induced along the compactification \cite{Aharony:2013dha}. One can deform the theory to remove such monopole superpotentials, which leads to the following type of superpotentials in the dual theory,
\begin{align}
W \sim V \hat V
\end{align}
where $V$ is an elementary field and $\hat V$ is a monopole operator of the dual theory. The elementary field $V$ is mapped to a monopole operator of the original theory, which is a generic phenomenon in 3d Aharony-type dualities. Clarifying monopole operators and their superpotentials are thus crucial to obtain a correct 3d duality. In this note, we find monopole operators in two-adjoint theories in 3d and the correct form of their superpotentials required for the dualities. Furthermore, as we consider the compactification on a circle, we are able to turn on nontrivial holonomy along the circle, which leads to different theories depending on this holonomy value \cite{Aharony:2013dha,Hwang:2017nop,Aharony:2017adm,Hwang:2018riu}. In this note, however, we focus on the trivial holonomy while the effect of a nontrivial holonomy will be an interesting topic to study.

One of the technical advantages of studying 3d dualities is that in 3d the factorization property of partition functions, which could be a result of the Higgs branch localization \cite{Fujitsuka:2013fga,Benini:2013yva}, is studied quite extensively \cite{Pasquetti:2011fj,Beem:2012mb,Hwang:2012jh,Taki:2013opa}. For example, the factorization of the superconformal index in the presence of one adjoint is studied in \cite{Hwang:2015wna}. In most of the cases, the factorized expression is much more efficient than the usual contour integral form to compute exact partition functions. Furthermore, when there are decoupled operators and accidental symmetries in IR, the naive conformal dimensions of the decoupled operators read off from UV data could be negative, which yield a convergence issue of the trace formula:
\begin{align}
I = \mathrm{Tr} (-1)^F x^{\Delta+j},
\end{align}
which is basically the series expansion of the contour integral with respect to the conformal dimension(+angular momentum) fugacity $x$. On the other hand, the factorized expression doesn't require such series expansion and is still a useful quantity in spite of accidental symmetries hidden in UV. As we will see, a few of our examples are such cases having accidental symmetries in IR. In such cases, the factorized formula of the partition function is mandatory to study their dualities.
\\

In this note, therefore, we study the factorization of the superconformal index in the presence of two adjoints. While the contour integral form of the superconformal index is completely determined by the symmetry group and the representations of matters, the factorized expression in general is not known. Thus, we derive the factorized form of the superconformal index with two adjoints by explicitly evaluating the contour integral expression. We first notice that the index of the $\widehat O$-type theory suffers from double poles, which makes the evaluation of the contour integral more difficult. On the other hand, such double poles disappear once we introduce the $D_{n+2}$-type superpotential with odd $n$. As a result, we obtain the factorized form of the superconformal index with two adjoints for $D_\text{odd}$-type theories. The factorization leads to the fact that the Higgs vacua of the theory are labeled by so called \emph{growing trees}, whose definition will be explained in section \ref{sec:fact}. This classification of the Higgs vacua is also proven by explicitly solving the D- and F-term vacuum equations of the theory gapped by real masses. Indeed the Higgs vacua are shown to match precisely for
the massively deformed dual pairs, which in turn implies the matching of the growing trees. Using this one can show
the matching of the vortex partition function of the dual pairs, which leads to the factorized form of the  superconformal index.

Given the factorized form of the superconformal index, we perform a test of the $D_{n+2}$-type duality, i.e., the Brodie duality \cite{Brodie:1996vx}, which provides the first evidence of the duality for finite $N_c, \, N_f$. Also we numerically compute the $S^3$ partition function \cite{Kapustin:2009kz,Hama:2010av}, which can be used to determine the IR superconformal $R$-charges \cite{Jafferis:2010un,Jafferis:2011zi,Closset:2012vg} (see also \cite{Casini:2012ei}). Along the chain of RG flows $\widehat O \rightarrow \widehat D \rightarrow D_{n+2}$, we determine the IR superconformal values of the $R$-charges and clarify the relevance of the superpotential.

While the $D_\text{odd}$-type duality successfully passes the test, the fate of $D_\text{even}$- and $E_7$-type dualities is still unclear because double poles do not disappear in such cases and the factorized index is missing at the moment. Nevertheless, since our finite-$N$ analysis clearly distinguishes $D_\text{odd}$ and $D_\text{even}$, it signals that the finite-$N$ analysis of the superconformal index can teach us the mysterious quantum corrections for $D_\text{even}$ and $E_7$ if exist, which weren't manifest in the large-$N$ analysis in 4d. Indeed, such quantum corrections should be reflected in the Hilbert series, which captures the algebraic structure of the moduli space. For a theory with fundamental matters, the relation between the superconformal index and the Hilbert series was examined in \cite{Razamat:2014pta,Hanany:2015via}. It will be interesting if our computation of the superconformal index provides a better understanding of the Hilbert series of a theory with adjoints. In addition, recently it is observed that the correspondence of vortex states and elementary particle states under a 3d Seiberg-like duality is closely related to the FI wall-crossing of vortex quantum mechanics \cite{Hwang:2017kmk}. This relation is used to prove the vortex partition function equality for the Aharony duality. It will be interesting if that approach is also applied to adjoint matter cases. Also one might go further beyond 3d. There is an intriguing work considering the 2d reductions of 3d dualities \cite{Aharony:2017adm}. We relegate those interesting subjects to future works.
\\

The paper is organized as follows. In section \ref{sec:review}, we review the Brodie-Kustasov-Lin duality and, in particular, discuss the 3d reduction of the Brodie duality. In section \ref{sec:fact}, we derive the factorized form of the superconformal index in the presence of two adjoints with the $D_\text{odd}$-type superpotential and show that the Higgs vacua of the theory are labeled by the growing trees. In section \ref{sec:test}, we perform a test of the conjectured duality and clarify the monopole operators of the theory. In section \ref{sec:F-max}, we carry out the $F$-maximization to determine the IR superconformal $R$-charges and the relevance of the superpotential. We also provide supplementary discussions on contour integral, Higgs vacua and monopole operators in appendix \ref{sec:integral}, \ref{sec:vacua}, \ref{sec:monopole} respectively,  and lists of growing trees in appendix \ref{sec:growing trees}.
\\

\section{Brodie Duality and 3d Reduction}
\label{sec:review}

\subsection{$ D_{n+2} $ Theories and the Brodie Duality}
Let's review the $ D_{n+2} $ theories and their Brodie Duality in 4-dimensions. We are interested in the theories with $ D $-type superpotentials
\begin{align}
\label{eq:supD}
\begin{aligned}
& W_{\widehat O} = 0, \qquad W_{\widehat D} = \mathrm{Tr} X Y^2,  \qquad W_{D_{n+2}} = \mathrm{Tr} \left(X^{n+1}+X Y^2\right).
\end{aligned}
\end{align}
The $ D_{n+2} $ theory can be obtained from RG flows
\begin{align}
\widehat O \rightarrow \widehat D \rightarrow D_{n+2}
\end{align}
where each of the SCFTs is the IR fixed point of the theory with the corresponding superpotential. The theories without superpotential $ W_{\widehat{O}}=0 $ flow to an interacting RG fixed point, $ \widehat O $ SCFT for all $ N_f $ in the asymptotic free range: $ 0 \leq N_f < N_c $. The superpotential $ W_{\widehat D} $ is a relevant deformation of $ \widehat O $ SCFT and it is expected to flow an interacting IR fixed point, $ \widehat D $ SCFT for all $ N_f < N_c $. The $ \widehat D $ SCFT has a relevant deformation $ \Delta W = \mathrm{Tr} X^{n+1} $ for any $ n $, provided that $x= \frac{N_c}{N_f}  $ is sufficiently large. In other words, $ D_{n+2} $ SCFT can be obtained from the deformation of $ \widehat D $ SCFT for $ x > x^{\min}_{D_{n+2}} $ where $ x^{\min}_{D_{n+2}} $ can be determined by $ a $-maximization \cite{Intriligator:2003mi}.

For $ x > x^{\min}_{D_{n+2}} $, the $ D_{n+2} $ theory, which we call original or electric, has a dual (magnetic) description proposed in \cite{Brodie:1996vx}. The magnetic theory has $ SU(3nN_f - N_c) $ gauge group, two adjoints $ \hat X $ and $ \hat Y $, $ N_f $ fundamentals $ q $, $ N_f $ anti-fundamentals $ \tilde q $, and $ 3nN_f^2 $ singlets $ M_{st} $ which corresponds to the mesons $ QX^{s} Y^{t}\tilde Q $, $ s=0,\ldots,n-1 $, $ t=0,1,2 $ of the electric theory, where flavor indices are suppressed. The magnetic theory has a superpotential,
\begin{align}
W_{\textrm{mag}}= \mathrm{Tr} (\hat X^{n+1} +\hat X \hat Y^{2}) + \sum_{s=0}^{n-1}\sum_{t=0}^{2} M_{st}\tilde q \hat X^{n-s-1} \hat Y^{2-t} q
\end{align}
Both electric and magnetic theories have the anomaly free global symmetries, $ SU(N_f) \times SU(N_f) \times U(1)_B \times U(1)_R $. The matter fields transform under the global symmetries as given in table \ref{tab:4d}:
\begin{table}[tbp]
	\centering
	\begin{tabular}{|c|cccc|}
		\hline
		& $SU(N_f)$ & $SU(N_f)$ & $U(1)_B$ &  $U(1)_R$ \\
		\hline
		$Q$ & $\mathbf{N_f}$ & $\mathbf 1$ & $1$  & $1-\frac{N_c}{(n+1)N_f}$ \\
		$\tilde Q$ & $\mathbf 1$ & $\mathbf{\overline{N_f}}$ & $-1$ &  $1-\frac{N_c}{(n+1)N_f}$ \\
		$X$ & $\mathbf 1$ & $\mathbf 1$ & $0$ & $\frac{2}{n+1}$ \\
		$Y$ & $\mathbf 1$ & $\mathbf 1$ & $0$ &  $\frac{n}{n+1}$ \\
		\hline
		$q$ & $\mathbf{\overline{N_f}}$ & $\mathbf 1$ & $\frac{N_c}{3nN_f-N_c}$ & $1-\frac{3nN_f-N_c}{(n+1)N_f}$ \\
		$\tilde q$ & $\mathbf 1$ & $\mathbf{N_f}$ & $-\frac{N_c}{3nN_f-N_c}$ & $1-\frac{3nN_f-N_c}{(n+1)N_f}$ \\
		$\hat X$ & $\mathbf 1$ & $\mathbf 1$ & $0$ & $\frac{2}{n+1}$ \\
		$\hat Y$ & $\mathbf 1$ & $\mathbf 1$ & $0$ & $\frac{n}{n+1}$ \\
		$M_{s,t}$ & $\mathbf{N_f}$ & $\mathbf{\overline{N_f}}$ & $0$ & $2-\frac{2N_c}{(n+1)N_f}+\frac{2s+nt}{n+1}$ \\
		\hline
	\end{tabular}
	\caption{\label{tab:4d} The representations of the chiral operators under the global symmetry groups.}
\end{table}
The magnetic theory is asymptotic free for $N_f < 3n N_f - N_c  $. The term $ \mathrm{Tr} \hat X^{n+1} $ in the superpotential $ W_{\textrm{mag}} $ is relevant for $ x < 3n - \tilde x^{\min}_{D_{n+2}} $ where $ \tilde x^{\min}_{D_{n+2}} $ can be determined similarly to $ x^{\min}_{D_{n+2}}  $ of the electric theory. If the term $ \mathrm{Tr} \hat X^{n+1} $ is irrelevant we call the RG fixed point $ \widehat D$ magnetic SCFT. The phases of the $ D_{n+2} $ electric and magnetic theories can be summarized as \cite{Intriligator:2003mi}
\begin{align}
\begin{array}{cl}
x \leq 1, \quad & \textrm{free electric}, \\
1 < x < x^{\min}_{D_{n+2}}, \quad & \widehat D \textrm{ electric}, \\
x^{\min}_{D_{n+2}} < x < 3n - \tilde x^{\min}_{D_{n+2}}, \quad & D_{n+2} \textrm{ conformal window}, \\
3n - \tilde x^{\min}_{D_{n+2}} < x < 3n -1, \quad & \widehat D \textrm{ magnetic}, \\
3n -1 \leq x , \quad & \textrm{free magnetic}
\end{array}
\end{align}

$ D_{n+2} $ theories have qualitatively different classical chiral rings between $ n $ odd and $ n $ even. For $ n $ odd, the chiral ring is truncated classically by a constraint $ Y^3=0 $ coming from the $ F $-term condition, as we will review shortly. However, for $ n $ even, there is no such classical constraint. It was conjectured that the chiral ring is truncated quantum mechanically so that the electric and magnetic theories have the same quantum chiral rings as in odd $ n $ cases. But it is not yet known how such a quantum truncation occurs. Furthermore, moduli space of vacua for the theories and generic deformations of superpotential were studied in detail in \cite{Intriligator:2016sgx} and additional hurdles for the conjectured duality were pointed out. The subtleties arise from $ d>1 $ dimensional adjoint vacua where $ [X,Y]\neq0 $ and the adjoints $ X $ and $ Y $ are represented as irreducible $ d\times d $ matrices. Such higher dimensional adjoint flat directions are generic in the $ D_{n+2} $ theories with even $ n $. The higher dimensional vacua violates $ a $-theorem, which mean a SCFT on the flat direction has a bigger value of $ a $ than the $ UV $ theory. Besides, higher dimensional adjoint vacua of the electric and magnetic theories are inconsistent with the conjectured duality with even $ n $.

We focus on $ D_{n+2} $ theories with odd $ n $, whose chiral rings and superpotential deformations are understood well and consistent with the duality. The chiral ring and deformations associated with the adjoints $ X $ and $ Y $ can also be applied to $ 3$ dimensional theories. Let us review chiral rings and flat directions of $ U(N_c) $ gauge theories. The electric theory has $ F $-term conditions
\begin{align}
X^n +Y^2=0,\qquad XY + YX =0~.
\end{align}
We also have $ Y^3 = -YX^n = (-1)^{n+1} X^n Y = (-1)^{n+2} Y^3 $ so for odd $ n $ we have classical chiral ring truncation,
\begin{align}
Y^3=0~.
\end{align}
For $ n \leq N_c $, we have the following gauge invariant chiral ring generators:
\begin{align}
&\mathrm{Tr} X^{s} Y^t \\
&Q X^{s} Y^t \tilde Q
\end{align}
for $ s=0,\ldots, n-1 $ and $ t=0,1,2 $. For $ N_c < n $ the chiral operators involving more than $ X^{N_c} $ or $ Y^{N_c} $ are not independent due to the characteristic equation of the adjoint fields. In the magnetic theory, we have chiral ring relations $ \tilde q \hat X^{s} \hat Y^{t} q =0$ from the equation of motion of $ M_{st} $. Thus the chiral rings of the electric and magnetic theories match as
\begin{align}
&\mathrm{Tr} X^{s} Y^t \longleftrightarrow \mathrm{Tr} \hat X^{s} \hat Y^t\\
&Q X^{s} Y^t \tilde Q \longleftrightarrow M_{st}
\end{align}
If $ N_c<n $ or $ 3n N_f - N_c <n$  the electric and magnetic theories with different gauge groups ($ N_c \neq 3nN_f -N_c $) have different number of classical chiral ring generators. In that case, the quantum truncation of the chiral ring is expected in order for the duality to hold. In 3d analogues of the duality, the quantum truncation was shown explicitly using the superconformal index \cite{Kapustin:2011vz, Kim:2013cma}.

Generic deformations of $ D_{n+2} $ theories were discussed in \cite{Cachazo:2001gh,Intriligator:2016sgx}. Let us consider a deformation which lead to
\begin{align} \label{AdjointDeformation}
D_{n+2} \rightarrow (n+2) A_1^{1d} +\frac{n-1}{2}A_1^{2d}
\end{align}
where $ A_1 $ is the SQCD with matters only in fundamental and anti-fundamental representations. $ 1d $ and $ 2d $ refer one and two dimensional adjoint vacua respectively. We consider a generic deformation for $ X $
\begin{align}
W=\mathrm{Tr} XY^2 +\sum_{i=0}^{n}\frac{t_i}{i+1}\mathrm{Tr} X^{i+1} \label{SupotDeform}
\end{align}
where $ X $ is massive. We include the linear term $ t_0 \mathrm{Tr} X $ to prevent a solution at the origin $ X=Y=0 $ so that $ Y $ is also massive due to the term $ \mathrm{Tr} XY^2 $. Thus the adjoint fields are integrated out and the IR theories are SQCDs, i.e. $ A_1 $ theories. The F-term equations are given by
\begin{align}
Y^2 +\sum_{i=0}^{n}t_i X^i=0\label{AdjointFlatDirection1}\\
XY+YX =0~.
\end{align}
 The F-term equations show that $ X^2 $ and $ Y^2 $ are the Casimir, which means $ [X^2,Y]=[Y^2,X]=0 $. Let us first consider $ 1d $ solutions that have the form $ X=x\mathbf 1 $ and $ Y=y\mathbf 1 $. Firstly, if $ y=0 $ then there are $ n $ non-zero solutions for $ x $ from $ \sum_{i=0}^{n}t_i x^i=0 $, which we denote them by $ x^{1d}_i $, $ i=1,\ldots,n $. Secondly, if $ x=0 $ then there are two solutions, $ y=\pm \sqrt{t_0} $ denoted by $ y^{1d}_{n+1} $ and $ y^{1d}_{n+2} $. Thus there are $ (n+2) $ $ 1d $ solutions. Next, let's consider $ 2d $ solutions. We can take $ X=x\sigma_1 $ and $ Y=y\sigma_2 $ where $ \sigma_i $ are Pauli matrices. \eqref{AdjointFlatDirection1} becomes
\begin{align}
(y^2+t_{n-1}x^{n-1}+t_{n-3}x^{n-3}+\cdots+t_0)\mathbf{1} + (t_{n}x^n +t_{n-2}x^{n-2} +\cdots +t_1x^1)\sigma_1=0
\end{align}
There are $ \frac{n-1}{2} $ non-zero solutions for $ x^2 $ from $ \sum_{i=0}^{\frac{n-1}{2}}t_{2i+1}x^{2i}=0 $ and $ y^2 $ is fixed by $ y^2= -\sum_{i=0}^{\frac{n-1}{2}}t_{2i}x^{2i}$. This moduli space can be labeled by $ x^2 $ and $ y^2 $ modded out by gauge transformations $ x\rightarrow -x $ and $ y \rightarrow -y $. Thus there are $ \frac{n-1}{2} $ 2d solutions denoted by $ x^{2d}_{j} $ and $ y^{2d}_{j} $, $ j=1,\ldots,\frac{n-1}{2} $. In sum, the deformation lead to vacua \eqref{AdjointDeformation}. The $ N_c\times N_c $ matrices $ X $ and $ Y $ can be written as block diagonal matrices
\begin{align}
\begin{aligned}
\langle X \rangle \sim \mathrm{diag}\left(x^{1d}_1{\bf 1}_{m^{1d}_1}, \cdots, x^{1d}_n{\bf 1}_{m^{1d}_n}, 0 {\bf 1}_{m^{1d}_{n+1}}, 0 {\bf 1}_{m^{1d}_{n+2}}, x^{2d}_1 {\bf 1}_{m^{2d}_{1}} \otimes \sigma_1, \cdots, x^{2d}_{(n-1)/2} {\bf 1}_{m^{2d}_{(n-1)/2}}\otimes \sigma_1\right)\\
\langle Y \rangle \sim \mathrm{diag}\left(0{\bf 1}_{m^{1d}_1}, \cdots, 0{\bf 1}_{m^{1d}_n}, y^{1d}_{n+1} {\bf 1}_{m^{1d}_{n+1}}, y^{1d}_{n+2} {\bf 1}_{m^{1d}_{n+2}}, y^{2d}_1 {\bf 1}_{m^{2d}_{1}}\otimes \sigma_2, \cdots, y^{2d}_{(n-1)/2} {\bf 1}_{m^{2d}_{(n-1)/2}}\otimes \sigma_2\right)
\end{aligned}
\end{align}
where $ {\bf 1}_i $ are $ i\times i $ identity matrices, $ m^{jd}_i $ are the number of eigenvalues corresponding to $ x^{jd}_i $ or $ y^{jd}_i $, and $ N_c = \sum_{i=1}^{n+2} m^{1d}_i+\sum_{j=1}^{\frac{n-1}{2}}2m^{2d}_j $. Let's consider Higgsing of the gauge group for the expectation values of the adjoints. The part of the gauge group corresponding to one dimensional vacua breaks $ U(\sum_{i=1}^{n+2} m^{1d}_i) \rightarrow  \prod_{i=1}^{n+2} U(m^{1d}_i)$. A two dimensional $ X $ and $ Y $ expectation values, say $ m $ of $ x $ and $ y $ can be written as
\begin{align}
\begin{aligned}
x {\bf 1}_{m}\otimes \sigma_1 \rightarrow x \sigma_1\otimes {\bf 1}_{m}\\
y {\bf 1}_{m}\otimes \sigma_2 \rightarrow y \sigma_2\otimes {\bf 1}_{m}
\end{aligned}
\end{align}
by a similarity transformation. The matrix on the right hand side is $ 2\times 2 $ block matrix so it breaks $ U(2m)\rightarrow U(m)_1\times U(m)_2 $. $ x $ and $ y $ are in a representation $ ({\bf m}_1, \overline{{\bf m}}_2) $ so their expectation values break $ U(m)_1\times U(m)_2 \rightarrow U(m)_D $, which is the diagonal subgroup. Therefore, the $ U(N_c) $ gauge group is broken to
\begin{align}
U(N_c) \rightarrow \prod_{i=1}^{n+2} U(m^{1d}_i) \prod_{j=1}^{\frac{n-1}{2}} U(m^{2d}_j)
\end{align}
Note that as gauge groups corresponding to 2d vacua are broken $ U(2m)\rightarrow U(m)_1\times U(m)_2 \rightarrow U(m)_D$ the fundamental flavors decompose as $ {\bf 2m} \rightarrow ({\bf m,1}) + ({\bf 1,m}) \rightarrow 2 \cdot {\bf m} $. Thus there are $ 2N_f $ flavors which transform under each of $  U(m)_D $ gauge groups.

The deformation of superpotential can be applied to the magnetic theory. The expectation values for the adjoints and the breaking patterns of gauge groups are of  the same form. Thus we have the following map of RG flows and the duality.
\begin{align}
\begin{array}{ccc}
U(N_c) & \longrightarrow\ & {\displaystyle \prod_{i=1}^{n+2} U(m^{1d}_i) \prod_{j=1}^{\frac{n-1}{2}} U(m^{2d}_j)} \\
\big\updownarrow & & \big\updownarrow \\
U(3nN_f -N_c) & \longrightarrow & {\displaystyle \prod_{i=1}^{n+2} U(N_f-m^{1d}_i) \prod_{j=1}^{\frac{n-1}{2}} U(2N_f -m^{2d}_j)}
\end{array}
\end{align}
We have a consistent dual gauge group $ 3nN_f-N_c = \sum_{i=1}^{n+2}(N_f-m^{1d}_i) +\sum_{j=1}^{\frac{n-1}{2}}2(2N_f-m^{2d}_j)$.
\\

\subsection{Dimensional reduction of the deformed theories to 3-d}

One can  consider the circle compactification of the deformed theories and their dualities \cite{Aharony:2013dha, Nii:2014jsa}. Although the compactification of the full theory would be more complicated, for the effective description with the broken gauge group, one can adopt the procedure described in \cite{Aharony:2013dha} because the effective description doesn't involve adjoints. Along the compactification, as argued in \cite{Affleck:1982as}, the 3d theory acquires the Affleck-Harvey-Witten (AHW) superpotential $W= \eta V_+ V_-$ where $ V_\pm $ are monopole operators. The AHW superpotential can be discarded by giving large vector real mass to one flavor \cite{Csaki:2017cqm}.

On the original side, we start with the $U(N_c)$ theory with $N_f+1$ non-chiral flavors $ Q$, $\tilde Q $ and AHW superpotential $W_A=\eta V_+^{\text{high}}V_-^{\text{high}}$ where $V_\pm^{\text{high}}$ are monopole operators defined at high energy. We give a vector real mass to $ (N_f+1) $-th flavor, $ \mathfrak m(Q^{N_f+1})=m $, $ \mathfrak m(\tilde Q_{N_f+1})=-m $ and focus on a vacuum at the origin of the Coulomb branch $ \sigma=0 $. In that vacuum, the effect of the large real mass is turning off the effective superpotential as well as integrating out the $N_f+1$-th massive flavor. Thus we are left with the $ U(N_c) $ theory with $ N_f $ flavors without any superpotential, which is the 3d theory of interest. On the other hand, Seiberg dual theory is the $U(N_f+1-N_c)$ theory with $N_f+1$ non-chiral flavors $ q$, $\tilde q $ and $(N_f+1)^2$ singlets $ M $, and superpotentials $W_B=Mq\tilde q + \tilde \eta \hat V_+^{\text{high}}\hat V_-^{\text{high}}$. The chiral fields also get real masses $ \mathfrak m(q_{N_f+1}) =-m$, $ \mathfrak m(\tilde q^{N_f+1}) =m$, $ \mathfrak m(M^{N_f+1}_i) =m$ and $ \mathfrak m(M^i_{N_f+1}) =-m$ for $ i=1,\ldots,N_f $. A vacuum which is dual to the one obtained in the original theory is at $ \hat \sigma=\text{diag}(0,\ldots,0,m) $ as shown in \cite{Aharony:2013dha}. In that vacuum, the gauge group $ U(N_f+1-N_c) $ is broken to $U(N_f-N_c) \times U(1)$ where massless charged fields $ q $, $ \tilde q $, $ q_{N_f+1} $, $ \tilde q^{N_f+1} $ are $N_f \times \mathbf{(N_f-N_c)}_0 \oplus N_f \times \mathbf{\overline{(N_f-N_c)}}_0 \oplus \mathbf 1_1 \oplus \mathbf 1_{-1}$ and massless singlets are $M^{a = 1,\ldots,N_f}_{b = 1,\ldots,N_f}$ and $M_{N_f+1}^{N_f+1}$. The Coulomb branch of the theory also get correction through an AHW-type superpotential between the two gauge sectors \cite{Amariti:2017gsm}, $ W=V_+\hat V_- +V_-\hat V_+ $ where $ V_\pm $ from $ U(1) $ sector and $ \hat V_\pm $ from $ U(N_f-N_c) $ sector. Thus the effective superpotential of the $U(N_f-N_c) \times U(1)$ theory is
\begin{align}
W_B=Mq\tilde q + M^{N_f+1}_{N_f+1} q_{N_f+1} \tilde q^{N_f+1} + V_+\hat V_- +V_-\hat V_+
\end{align}
The $U(1)$ sector is then dualized into three chiral fields interacting through a superpotential $W=V_+ V_- \hat M$ where $ \hat M=q_{N_f+1}\tilde q^{N_f+1} $. In this description, $ \hat M $ and $M_{N_f+1}^{N_f+1}$ are massive due to the superpotential $W=M_{N_f+1}^{N_f+1} \hat M$ so they are integrated out and result in $ \hat M=0 $ and $M_{N_f+1}^{N_f+1}=V_+ V_-$. As a result, one obtains $ U(N_f-N_c) $ gauge theory with the chiral multiplets and superpotential
\begin{align}
W_B=Mq\tilde q + V_+\hat V_- +V_-\hat V_+~,
\end{align}
which is the original form of dual theory of the Aharony duality.

For our case, the procedure is almost the same, with a small modification for the $U(m^{2d}_j)$ sector. Since the number of flavors is doubled for this sector, we start from the $U(m^{2d}_j)$ theory with $2 N_f+2$ flavors. We give the same large mass to the last two flavors such that they are integrated out at the origin of the Coulomb branch. On the dual side, we start from the $U(2 N_f+2-m^{2d}_j)$ theory with $2 N_f+2$ flavors and $(2 N_f+2)^2$ singlets. Turning on the large mass, the gauge group is broken to $U(2 N_f-m^{2d}_j) \times U(2)$ at the vacuum discussed before, under which only $2 N_f \times (\mathbf{2 N_f-m^{2d}_j},\mathbf 1) \oplus 2 N_f \times (\mathbf{\overline{2 N_f-m^{2d}_j}},\mathbf 1) \oplus 2 \times (\mathbf 1,\mathbf 2) \oplus 2 \times (\mathbf 1,\mathbf{\overline{2}})$ and $4 N_f^2+4$ singlets are massless and the superpotential,
\begin{align}
W_B=M^a_b q_a\tilde q^b + M^{\alpha}_{\beta} q_{\alpha} \tilde q^{\beta} + W_+\hat W_- +W_-\hat W_+
\end{align}
where $ a,b=1,\ldots,2N_f $, $ \alpha, \beta = 2N_f+1, 2N_f+2$, and $ W_\pm $ and $ \hat W_\pm $ are monopole operators of $ U(2) $ and $ U(2N_f-m_j^{2d}) $ sectors respectively.
Again, the $U(2)$ sector is dualized into six chiral fields $W_+, \, W_-, \, \hat M_{\alpha}^{\beta}$ with the superpotential $W=W_+ W_- \det \hat M$ where $\hat M_\alpha^\beta = q_\alpha \tilde q^\beta$. Because of the superpotential $W=M^\alpha_\beta \hat M_\alpha^\beta$, the singlet fields  $M^\alpha_\beta$ and $\hat M_\alpha^\beta$ become massive and are integrated out. Their equations of motion fix them as $\partial_{M^\alpha_\beta}W=\hat M_\alpha^\beta=0$ and $\partial_{\hat M_\alpha^\beta}W= M^\alpha_\beta + W_+W_-\epsilon_{\alpha\alpha'}\epsilon_{\beta\beta'}\hat M_{\alpha'}^{\beta'}=M^\alpha_\beta=0$. Thus, the remaining massless singlets are two $W_\pm$ and $4 N_f^2$ $M$'s with a superpotential
\begin{align}
W_B=M^a_b q_a\tilde q^b + W_+\hat W_- +W_-\hat W_+
\end{align}

One should note that the topological charge of $U(2 N_f-m^{2d}_j)$ and that of the full gauge group $U(3 n N_f-N_c)$ are identified up to the multiplication by 2 because the former is the diagonal part of $U(2 N_f-m^{2d}_j) \times U(2 N_f-m^{2d}_j) \subset U(4 N_f-2 m^{2d}_j) \subset U(3 n N_f-N_c)$. Thus, $W_\pm$ carry topological charge 1 with respect to $U(2 N_f-m^{2d}_j)$ while they carry topological charge 2 with respect to the full gauge group $U(3 n N_f-N_c)$. Therefore, we expect that there are $\frac{n-1}{2}$ monopole operators of topological charge 2 from the 2d vacua and $ n+2 $ monopole operators of topological charge 1 from the 1d vacua as turning off the deformation parameters $ t_i $, $ i\neq n $ in the superpotential \eqref{SupotDeform}. Indeed, we confirm this by the explicit computation of the superconformal indices and their match under the duality.
\\

\subsection{Proposal of 3d Duality for $ D_{n+2} $ Theories}
Combining the results of the previous subsection, we propose the following duality between {\it theory A} and {\it theory B}, which is an analogue of the Brodie duality.
\paragraph{theory A} $U(N_c)$ gauge theory with chiral superfields, $N_f$ fundamentals $Q$ and $N_f$ anti-fundamentals $\tilde Q$, and two adjoints $X$ and $Y$ with superpotentials,
\begin{align}
W=\Tr X^{n+1} + \Tr X Y^2
\end{align}
where $n\geq 3$ is odd. The theory A has gauge invariant chiral operators, $3nN_f^2$ mesons $M_{s,t}=Q X^s Y^t \tilde Q$ for $s=0,\ldots,n-1$, $t=0,1,2$. It has monopole operators of charge 1, $V_{s,t}^+$, $V_{s,t}^-$ for $s=0,\ldots,n-1$, $t=0,1,2$ with $ st=0 $, and monopole operators of charge 2 $W_u^+$, $W_u^-$ for $u=0,\ldots, \frac{n-3}{2}$.

\paragraph{theory B} $U(3nN_f - N_c)$ gauge theory with chiral superfields, $N_f$ fundamentals $q$ and $N_f$ anti-fundamentals $\tilde q$, two adjoints $\hat X$ and $\hat Y$, $3nN_f^2$ singlets $ M_{s,t} $ and $2(n+2)$ singlets $ V^\pm_{s,t} $ with $ st=0 $, and $n-1$ singlets $W_u^\pm$ with superpotentials,
\begin{align}
W=&\Tr \hat X^{n+1} + \Tr \hat X \hat Y^2
+ \sum_{s=0}^{n-1}\sum_{t=0}^{2} M_{s,t}\tilde q \hat X^{n-1-s} \hat Y^{2-t} q
\\
&+ \sum_{s=1}^{n-1} V_{s,0}^\pm \hat V_{n-s,0}^\mp
+ \sum_{t=0}^{2} V_{0,t}^\pm \hat V_{0,2-t}^\mp
+\sum_{u=0}^{\frac{n-3}{2}} W_u^\pm \hat W_{\frac{n-3}{2}-u}^\mp \nonumber
\end{align}
where $\hat V_{s,t}^\pm$ and $\hat W_u^\pm$ are monopoles in the theory B. $M_{st}$ correspond to the mesons in the theory A, $V_{s,t}^\pm$ and $ W_u^\pm$ correspond to the monopoles in the theory A. Mesons $\hat M_{s,t}=\tilde q \hat X^{s} \hat Y^{t} q$ and monopoles $\hat V_{s,t}^\pm$, $\hat W_u^\pm$ are not chiral due to the superpotential.
A list of chiral operators appearing in a duality pair is given in table \ref{tab:3d}.
\begin{table}[tbp]
	\centering
	\begin{tabular}{|c|ccccc|}
		\hline
		& $SU(N_f)$ & $SU(N_f)$ & $U(1)_A$ & $U(1)_T$ & $U(1)_R$ \\
		\hline
		$Q$ & $\mathbf{\overline{N_f}}$ & $\mathbf 1$ & $1$ & $0$ & $r$ \\
		$\tilde Q$ & $\mathbf 1$ & $\mathbf{N_f}$ & $1$ & $0$ & $r$ \\
		$X$ & $\mathbf 1$ & $\mathbf 1$ & $0$ & $0$ & $\frac{2}{n+1}$ \\
		$Y$ & $\mathbf 1$ & $\mathbf 1$ & $0$ & $0$ & $\frac{n}{n+1}$ \\
		$M_{s,t}$ & $\mathbf{\overline{N_f}}$ & $\mathbf{N_f}$ & $2$ & $0$ & $2r+\frac{2s+nt}{n+1}$ \\
		$V_{s,t}^\pm$ & $\mathbf 1$ & $\mathbf 1$ & $-N_f$ & $\pm 1$ & $(1-r)N_f -\frac{1}{n+1} (N_c -1) +\frac{2 s+n t}{n+1}$ \\
		$W_u^\pm$ & $\mathbf 1$ & $\mathbf 1$ & $-2N_f$ & $\pm 2$ & $2(1-r)N_f -\frac{2}{n+1} (N_c -1) +\frac{2+4 u}{n+1}$ \\
		\hline
		$q$ & $\mathbf 1$ & $\mathbf{\overline{N_f}}$ & $-1$ & $0$ & $\frac{2-n}{n+1}-r$ \\
		$\tilde q$ & $\mathbf{N_f}$ & $\mathbf 1$ & $-1$ & $0$ & $\frac{2-n}{n+1}-r$ \\
		$\hat X$ & $\mathbf 1$ & $\mathbf 1$ & $0$ & $0$ & $\frac{2}{n+1}$ \\
		$\hat Y$ & $\mathbf 1$ & $\mathbf 1$ & $0$ & $0$ & $\frac{n}{n+1}$ \\
		$\hat M_{s,t}$ & $\mathbf{N_f}$ & $\mathbf{\overline{N_f}}$ & $-2$ & $0$ & $-2r+\frac{4-2n+2s+nt}{n+1}$ \\
		$\hat V_{s,t}^\pm$ & $\mathbf 1$ & $\mathbf 1$ & $N_f$ & $\pm 1$ & $(r-1)N_f +\frac{1}{n+1} (N_c +1) +\frac{2 s+n t}{n+1}$ \\
		$\hat W_u^\pm$ & $\mathbf 1$ & $\mathbf 1$ & $2N_f$ & $\pm 2$ & $2(r-1)N_f +\frac{2}{n+1} (N_c +1) +\frac{2+4 u}{n+1}$ \\
		\hline
	\end{tabular}
	\caption{\label{tab:3d} The representations of the chiral operators under the symmetry group. The upper lines are devoted to the original theory while the lower lines are devoted to the dual theory. In the original theory side, $M_{s,t}$, $V_{s,t}^\pm$ and $W_u^\pm$ are gauge invariant operators rather than elementary fields. For $V_{s,t}^\pm$, the indices satisfy $ st=0 $. Note that $W_u^\pm$ carries topological charge $\pm 2$.}
\end{table}

\section{The Superconformal Index with Two Adjoints}
\label{sec:fact}

\subsection{The Superconformal Index and Factorization}

In this section, we examine the superconformal indices of theories with multiple adjoint matters as well as (anti-)fundamental matters. Especially we will focus on their factorization, or equivalently their Higgs branch localization. The superconformal index is defined by the trace of $(-1)^F$ over the Hilbert space on $S^2$ with appropriate twists \cite{Bhattacharya:2008zy,Bhattacharya:2008bja}:
\begin{align}
I = \mathrm{Tr} (-1)^F e^{\beta' \{Q,S\}} x^{E+j} \prod_i t_i^{F_i}
\end{align}
where $F$ is the fermion number operator. One chooses a supercharge $Q$ carrying quantum numbers $E = \frac{1}{2}, \, j = -\frac{1}{2}, \, R = 1$ under the bosonic subgroup of the 3d $\mathcal N = 2$ superconformal group, $SO(3,2) \times SO(2)$, such that only the BPS states saturating
\begin{align}
\{Q,S\} = E-R-j \geq 0
\end{align}
contribute to the index. $S$ is another supercharge satisfying $S = Q^\dagger$. The index is therefore independent of $\beta'$. $F_i$'s are various global charges commuting with $Q$ and $S$.
\\

The superconformal index can be regarded as a supersymmetric partition function on $S^2 \times S^1$ with the periodic boundary condition of the fields along $S^1$. One can compute this partition function exactly, using the supersymmetric localization, whose final representation depends on which $Q$-exact deformation is used for the localization. Two of the most useful representations are the ones usually called the Coulomb branch localization \cite{Kim:2009wb,Imamura:2011su} and the Higgs branch localization \cite{Fujitsuka:2013fga,Benini:2013yva} respectively. Although their representations are completely different, they in fact are exactly the same.

The Coulomb branch localization of the index is given by
\begin{align}
\label{eq:Coulomb}
& I(x,t) = \nonumber \\
& \sum_{m \in \text{monopole background}} \frac{1}{|\mathcal W_m|} \int \frac{dz}{2 \pi i z} ~ Z_\text{classical}(x,t;z,m) ~ Z_\text{vector}(x;z,m) ~ Z_\text{chiral}(x,t;z,m)
\end{align}
with a classical action contribution and 1-loop determinants that can be found in appendix \ref{sec:integral}. \eqref{eq:Coulomb} is completely determined if one knows the symmetry group and the representations of matters. It allows us to obtain information of the strongly interacting IR theory from handful UV data. One caveat here, on the other hand, is that sometimes the IR symmetry group is quantum corrected such that it is different from the UV symmetry group. In that case, only the UV symmetry is visible to us, and we cannot obtain the correct superconformal index unless we know the exact relation between the UV symmetry and the IR symmetry.

In particular, the superconformal index is supposed to have a well-defined series expansion with respect to $x$, which is associated with quantum number $R+2 j$. It can be formally written as follows:
\begin{align}
\label{eq:series}
I(x,t) = \sum_{R+2 j \geq 0} a_{R+2 j}(t) x^{R+2 j}.
\end{align}
Note that $a_0$ should be $1$, which corresponds to the unique vacuum on $S^2$. Here $R$ should be the IR superconformal value $R_\text{IR}$ while the manifest one is the UV $R$-symmetry charge $R_\text{UV}$. Those two are related by nontrivial shifts involving global symmetry charges
\begin{align}
\label{eq:R}
R_\text{IR} = R_\text{UV}+\sum_i \alpha_i F_i.
\end{align}
The coefficients $\alpha_i$'s can be determined by the $F$-maximization \cite{Jafferis:2010un}. Therefore, in principle, it is enough to know $R_\text{UV}$ and $F_i$'s. However, as we mentioned, sometimes the full IR symmetry is not visible in UV. In such cases, we cannot determine the relation \eqref{eq:R} because we cannot introduce every allowed $\alpha_i$.

Nevertheless, assuming the well-defined IR superconformal index \eqref{eq:series}, the integral formula \eqref{eq:Coulomb} with UV $R$-charge can be regarded as a unrefined index with a ugly choice of fugacities:
\begin{align}
I^\text{unrefined}(x,t_i) = \left.I(x,t_i x^{-\alpha_i},t_\text{acc} x^{-\alpha_\text{acc}})\right|_{t_\text{acc} = 1}
\end{align}
where $t_\text{acc}$ is the fugacity for an accidental IR symmetry whose contribution to the IR $R$-charge is as follows:
\begin{align}
R_\text{IR} = R_\text{UV}+\sum_i \alpha_i F_i+\alpha_\text{acc} F_\text{acc}.
\end{align}
Thus, the integral formula \eqref{eq:Coulomb} should be persistent across the duality once we impose the correct UV symmetry map. One should note that the series expansion of \eqref{eq:Coulomb} might not be well-defined at $x = 0$ if we use the UV $R$-charge $R_\text{UV}$. In that case, one cannot compare the dual pair of the indices by the perturbative series expansion, and different approaches are required.\footnote{Some of such examples are worked out in\cite{Hwang:2015wna}.}

One of the alternatives, especially for $U(N)$ theories, is the Higgs branch localization. The Higgs branch localization of the superconformal index, which we will also call the factorized index in this note, is given by
\begin{align}
\label{eq:Higgs}
I(x,t) = \sum_{\sigma \in \text{Higgs vacua}} Z_\text{pert}(x,\sigma(t)) ~ Z_\text{vort}(x,\sigma(t)) ~ Z_\text{anti-vortex}(x,\sigma(t))
\end{align}
where $Z_\text{pert}$ is the 1-loop determinant computed at a given Higgs vacuum, and $Z_\text{vort}$ is the vortex partition function defined on omega deformed $\mathbb R_\Omega^2 \times S^1$. For the superconformal index, $Z_\text{anti-vortex}$ is given by $Z_\text{vortex}$ with arguments all inverted. Unlike the Coulomb branch localization, the general expression of $Z_\text{vort}$ for matters of arbitrary representations is not known. Nevertheless, $Z_\text{vort}$ is known for many interesting theories containing matters of fundamental-like representations such as (bi-)fundamental and adjoint \cite{Beem:2012mb,Hwang:2012jh,Hwang:2015wna}.

Once we impose the correct UV symmetry map, a dual pair of each component agree regardless of the analyticity at $x = 0$. This is why \eqref{eq:Higgs} is often more useful than \eqref{eq:Coulomb} if we are not able to know exact IR symmetries, especially the superconformal $R$-symmetry. Indeed, the Brodie-Kutasov-Lin duality we reviewed in the previous section is such an example. The factorization will turn out to be crucial to use the superconformal index as a testbed for the duality.
\\

The superconformal index and its factorization with a single adjoint matter is examined in \cite{Hwang:2015wna}. By evaluating the Coulomb expression, it is shown that the index is written in terms of vortex partition functions as \eqref{eq:Higgs} with
\begin{align}
&\quad Z_\text{pert}^{{}p}(x,t = e^{i M},\tilde t,\tau,\upsilon = e^{i \nu}) \nonumber \\
&= \left(\prod_{a,b = 1}^{N_f} \prod_{q = 1}^{p_a} \prod_{\substack{r = 1 \\ (\neq q \text{ if } a = b)}}^{p_b} 2 \sinh \frac{i M_{a}-i M_{b}+i \nu (q-r)}{2}\right) \left(\prod_{a,b = 1}^{N_f} \prod_{q = 1}^{p_a} \prod_{r = 1}^{p_b} \frac{\left(t_{a} t_{b}^{-1} \upsilon^{q-r-1} x^2;x^2\right)_\infty}{\left(t_{a}^{-1} t_{b} \upsilon^{-q+r+1};x^2\right)'_\infty}\right) \nonumber \\
&\quad \times \left(\prod_{a = 1}^{N_f} \prod_{q = 1}^{p_a} \frac{\prod_{b = 1}^{N_f} \left(t_a t_b^{-1} \upsilon^{q-1} x^2;x^2\right)_\infty}{\prod_{b = 1}^{N_f} \left(t_a \tilde t_b \tau^2 \upsilon^{q-1};x^2\right)_\infty} \frac{\prod_{b = 1}^{N_f} \left(t_a^{-1} \tilde t_b^{-1} \tau^{-2} \upsilon^{-q+1} x^2;x^2\right)_\infty}{\prod_{b = 1}^{N_f} \left(t_a^{-1} t_b \upsilon^{-q+1};x^2\right)'_\infty}\right),
\end{align}
\begin{align}
Z_\text{vortex}^{{}p}(x,t,\tilde t,\tau,\upsilon,\mathfrak w) &= \sum_{\mathfrak n_j \geq 0} \mathfrak w^{\sum_{j = 1}^{N_c} \sum_{n = 0}^{l_j-1} \mathfrak n_j^n} \mathfrak Z^{{}p}_{(\mathfrak n_j)}(x,t,\tilde t,\tau,\upsilon), \\
Z_\text{antivortex}^{{}p}(x,t,\tilde t,\tau,\upsilon,\mathfrak w) &= \sum_{\bar{\mathfrak n}_j \geq 0} \mathfrak w^{-\sum_{j = 1}^{N_c} \sum_{n = 0}^{l_j-1} \bar{\mathfrak n}_j^n} \mathfrak Z^{{}p}_{(\bar{\mathfrak n}_j)}(x^{-1},t^{-1},\tilde t^{-1},\tau^{-1},\upsilon^{-1}),
\end{align}
\begin{align}
\begin{aligned}
&\quad \mathfrak Z^{{}p}_{(\mathfrak n_j)}(x = e^{-\gamma},t = e^{i M},\tilde t = e^{i \tilde M},\tau = e^{i \mu},\upsilon = e^{i \nu}) \\
&= \left(\prod_{a,b = 1}^{N_f} \prod_{q = 1}^{p_a} \prod_{\substack{r = 1 \\ (\neq q \text{ if } a = b)}}^{p_b} \prod_{k = 1}^{\sum_{n = 1}^r \mathfrak n_{(b,n)}} \frac{\sinh \frac{i M_a-i M_b+i \nu (q-r)+2 \gamma k}{2}}{\sinh \frac{i M_a-i M_b+i \nu (q-r)+ 2 \gamma (k-1-\sum_{n = 1}^q \mathfrak n_{(a,n)})}{2}}\right) \\
&\quad \times \left(\prod_{a,b = 1}^{N_f} \prod_{q = 1}^{p_a} \prod_{\substack{r = 1 \\ (\neq q \text{ if } a = b)}}^{p_b} \prod_{k = 1}^{\sum_{n = 1}^r \mathfrak n_{(b,n)}} \frac{\sinh \frac{i M_a-i M_b+i \nu (q-r-1)+2 \gamma (k-1-\sum_{n = 1}^q \mathfrak n_{(a,n)})}{2}}{\sinh \frac{i M_a-i M_b+i \nu (q-r+1)+2 \gamma k}{2}}\right) \\
&\quad \times \left(\prod_{b = 1}^{N_f} \prod_{r = 1}^{p_b} \prod_{k = 1}^{\sum_{n = 1}^r \mathfrak n_{(b,n)}} \frac{\prod_{a = 1}^{N_a} \sinh \frac{-i \tilde M_a-i M_b-2 i \mu-i \nu (r-1)+2 \gamma (k-1)}{2}}{\prod_{a = 1}^{N_f} \sinh \frac{i M_a-i M_b-i \nu (r-1)+2 \gamma k}{2}}\right)
\end{aligned}
\end{align}
where $(\ldots)'$ indicates that the zero factors in a q-Pochhammer symbol are omitted\footnote{q-Pochhammer symbol $(a;q)_\infty$ is  defined by
\begin{align}
(a;q)_\infty = \prod_{k = 0}^\infty \left(1-a q^k\right).
\end{align}.} The Higgs vacua are labeled by the forest graphs consisting of $N_f$ one-branch trees where $N_f$ is the number of the fundamental matters. Given a forest graph, $p_a$ and $\mathfrak n_j^n$ (or equivalently $\mathfrak n_{(b,n)}$) are determined. For detailed explanations, see the original paper \cite{Hwang:2015wna}. If one introduces superpotential $W = X^{n+1}$ for the adjoint matter $X$, the height of each tree cannot exceed $n$. Using this factorized index, the duality of the theories with the A-type superpotential is tested and shows the exact match of the superconformal indices.
\\

\begin{figure}[tbp]
\centering 
\begin{subfigure}{.3\textwidth}
\centering
\begin{tikzpicture}[grow'=up,<-,>=stealth',level/.style={sibling distance = 2cm/#1,level distance = 1cm}]
\node [arn_f] {$i_1$}
    child{ node [arn_x] {$i_2$}
    }
    child[missing]
;
\end{tikzpicture}
\caption{}
\end{subfigure}
\begin{subfigure}{.3\textwidth}
\centering
\begin{tikzpicture}[grow'=up,<-,>=stealth',level/.style={sibling distance = 2cm/#1,level distance = 1cm}]
\node [arn_f] {$i_1$}
    child[missing]
    child{ node [arn_y] {$i_2$}
    }
;
\end{tikzpicture}
\caption{}
\end{subfigure}
\caption{\label{fig:U(2)} All possible tree graphs of $U(2)$. The root node is denoted by a black circle. A node connected to its parent node by a type-X branch is denoted by a red circle while a node connected by a type-Y branch is denoted by a blue circle. The root node corresponds to a pole factor taken from the 1-loop determinant of fundamental matters while nodes connected by a type-X branch and a type-Y branch correspond to pole factors taken from the 1-loop determinants of adjoints $X$ and $Y$ respectively. An arrow indicates the parent node.}
\end{figure}
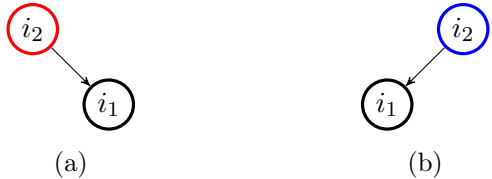

For a two-adjoint case, one can also perform a similar computation and obtain the following result:
\begin{align}
\label{eq:fact}
&\quad I(x,t,\tilde t,\tau,\upsilon_X,\upsilon_Y, w) \nonumber \\
&= \sum_{\mathsf f \in \mathsf G} Z^{\mathsf f}_{\text{pert}}(x,t,\tilde t,\tau,\upsilon_X,\upsilon_Y) Z^{\mathsf f}_{\text{vortex}}(x,t,\tilde t,\tau,\upsilon_X,\upsilon_Y, w) Z^{\mathsf f}_{\text{anti-vortex}}(x,t,\tilde t,\tau,\upsilon_X,\upsilon_Y, w),
\end{align}
\begin{align}
\label{eq:pert}
&\quad Z_{\text{pert}}^{\mathsf f} (x,t,\tilde t,\tau,\upsilon_X,\upsilon_Y) \nonumber \\
&= \left(\prod_{i \neq j}^{N_c} \left(1-t_{c_i} t_{c_j}^{-1} \upsilon_i^{-1} \upsilon_j\right)\right) \left(\prod_{A = X,Y} \prod_{i,j = 1}^{N_c} \frac{\left(t_{c_{i}}^{-1} t_{c_j} \upsilon_A^{-1} \upsilon_i \upsilon_j^{-1} x^2;x^2\right)_\infty}{\left(t_{c_i} t_{c_j}^{-1} \upsilon_A \upsilon_i^{-1} \upsilon_j;x^2\right)'_\infty}\right) \nonumber \\
&\quad \times \left(\prod_{a = 1}^{N_f} \prod_{j = 1}^{N_c} \frac{\left(t_{c_j}^{-1} t_{a} \upsilon_j x^2;x^2\right)_\infty}{\left(t_{c_j} t_a^{-1} \upsilon_j^{-1};x^2\right)'_\infty}\right) \left(\prod_{b = 1}^{N_f} \prod_{j = 1}^{N_c} \frac{\left(t_{c_j} \tilde t_b^{-1} \tau^{-2} \upsilon_j^{-1} x^2;x^2\right)_\infty}{\left(t_{c_j}^{-1} \tilde t_b \tau^2 \upsilon_j;x^2\right)_\infty}\right),
\end{align}
\begin{align}
Z^{\mathsf f}_{\text{vortex}} (x,t,\tilde t,\tau,\upsilon_X,\upsilon_Y, w)
&= \sum_{\mathsf n_j \geq 0} w^{\sum_{j = 1}^{N_c} \sum_{k = 0}^{l(j)-1} \mathsf n_{p^k(j)}} \mathsf Z^{\mathsf f}_{\left(\sum_{k = 0}^{l(j)-1} \mathsf n_{p^k(j)}\right)} (x,t,\tilde t,\tau,\upsilon_X,\upsilon_Y),
\end{align}
\begin{align}
Z^{\mathsf f}_{\text{anti-vortex}} (x,t,\tilde t,\tau,\upsilon_X,\upsilon_Y, w)
&= \sum_{\bar{\mathsf n}_j \geq 0} w^{-\sum_{j = 1}^{N_c} \sum_{k = 0}^{l(j)-1} \bar{\mathsf n}_{p^k(j)}} \mathsf Z^{\mathsf f}_{\left(\sum_{k = 0}^{l(j)-1} \bar{\mathsf n}_{p^k(j)}\right)} (x^{-1},t^{-1},\tilde t^{-1},\tau^{-1},\upsilon_X^{-1},\upsilon_Y^{-1}),
\end{align}
\begin{align}
\label{eq:vortex}
&\quad \mathsf Z^{\mathsf f}_{(n_j)} (x,t,\tilde t,\tau,\upsilon_X,\upsilon_Y) \nonumber \\
&= \left(\prod_{\substack{i,j = 1 \\ (i \neq j)}}^{N_c} \frac{\left[t_{c_i}^{-1} t_{c_j} \upsilon_i^{-1} \upsilon_j x^2; x^2\right]_{n_j}}{\left[t_{c_i}^{-1} t_{c_j} \upsilon_i^{-1} \upsilon_j x^{-2 n_i};x^2\right]_{n_j}} \right) \left(\prod_{A = X,Y} \prod_{\substack{i,j = 1 \\ (i \neq j)}}^{N_c} \frac{\left[t_{c_i}^{-1} t_{c_j} \upsilon_A \upsilon_i^{-1} \upsilon_j x^{-2 n_i};x^2\right]_{n_j}}{\left[t_{c_i}^{-1} t_{c_j} \upsilon_A^{-1} \upsilon_i^{-1} \upsilon_j x^2;x^2\right]_{n_j}}\right) \nonumber \\
&\quad \times \left(\prod_{j = 1}^{N_c} \frac{\prod_{b = 1}^{N_f} \left[\tilde t_b t_{c_j} \tau^2 \upsilon_j x^2; x^2\right]_{n_j}}{\prod_{a = 1}^{N_f} \left[t_{a}^{-1} t_{c_j} \upsilon_j;x^2\right]_{n_j}}\right)
\end{align}
where we use a shifted q-Pochhammer symbol
\begin{align}
\label{eq:shiftedq}
[a;q]_n = \prod_{k = 0}^{n-1} \left(a^{-\frac{1}{2}} q^{-\frac{k}{2}}-a^\frac{1}{2} q^\frac{k}{2}\right) = a^{-\frac{n}{2}} q^{-\frac{n(n-1)}{4}} (a;q)_n.
\end{align}
$t, \, \tilde t, \, \tau, \, \upsilon_X, \, \upsilon_Y, \, w$ are fugacities for the global symmetries $SU(N_f)_1 \times SU(N_f)_2 \times U(1)_A \times U(1)_X \times U(1)_Y \times U(1)_T$ respectively. Again $(\ldots)'$ indicates that the zero factors in a q-Pochhammer symbol are omitted. $\mathsf G$ is a set of forest graphs where each forest graph is defined as an ordered set of $N_f$ growing-tree graphs: $\mathsf f = (\mathsf t_1, \mathsf t_2, \ldots , \mathsf t_{N_f})$ whose total number of nodes is $N_c$.\footnote{A tree graph $\mathsf t_i$ is defined as a graph of nodes connected by branches without cycles. More precisely, our tree is a rooted tree with two types of branches, in which a certain node is designated as the root node and each branch has a specific type between the two. Since we can regard a tree as branched out from the root node, there is a natural notion of a parent node and a child node. Considering two adjacent nodes, one closer to the root node is a parent node of the other and the other one is a child node of the closer one. Examples can be found in figure \ref{fig:U(2)} and a few more subsequent ones.} Every node is labeled by $i = 1, \ldots, N_c$ while a permutation of the labeling doesn't affect the final answer. Given a labeling of the nodes, $c_j$ is defined such that the $j$-th node belongs to the $c_j$-th tree. Note that the tree graphs allowed here are those called growing-trees, whose definition will be explained shortly in the subsequent section.

For a given forest graph, $\upsilon_j$ and $n_j$ are defined as follows. The significant difference compared to the one adjoint case is that each tree now contains two types of branches, each of which corresponds to adjoint $X$ or $Y$. We define $p(i)$ such that $p(i) = j$ if the parent node of the $i$-th node is the $j$-th node. $B(i) = A$ if the $i$-th node is attached to the parent node by the branch of type-$A$. For a root node, since it doesn't have the parent node, we conventionally define $B(i) = 0$ if the $i$-th node is a root node. $\upsilon_j$ is then defined by
\begin{align}
\upsilon_j = \prod_{k = 0}^{l(j)-1} \upsilon_{B(p^k(j))}
\end{align}
with $\upsilon_0 = 1$.\footnote{Thus $\upsilon_j$ is the product of $\upsilon_A$ or $\upsilon_B$ for a given node of a tree
and the product is done over the all nodes between the root node and the given node.} $l(j)$, the level of the $j$-th node, is defined as the smallest positive integer such that $p^{l(j)-1}(i)$ is the root node of the tree. Also we assign a non-negative integer $\mathsf n_j$ to every node. $n_j$ is then defined by
\begin{align}
\label{eq:vorticity}
n_j = \sum_{k = 0}^{l(j)-1} \mathsf n_{p^k(j)}.
\end{align}
\\

The simplest example would be the $U(2)$ theory with one fundamental/anti-fundamental pair and two adjoints. Recall that the Coulomb branch localization of the index is given by \eqref{eq:Coulomb}. Each component of the integrand is given in appendix \ref{sec:integral}. In particular, the 1-loop determinant of chiral multiplets is given by
\begin{align}
Z_\text{chiral} (x,t;z,m) = \prod_{\rho \in R_G} \prod_{\sigma \in R_F} \left((-z)^\rho t^\sigma x^{r-1}\right)^{-|\rho(m)|/2} \frac{\left(z^{-\rho} t^{-\sigma} x^{2-r+|\rho(m)|};x^2\right)_\infty}{\left(z^\rho t^\sigma x^{r+|\rho(m)|};x^2\right)_\infty}.
\end{align}
For $U(2)$, this simplifies to
\begin{align}
Z_\text{chiral} (x,t;z,m) = \prod_{i = 1,2} \frac{\left(z_i^{-1} t^{-1} x^{2+|m_i|};x^2\right)_\infty}{\left(z_i t x^{|m_i|};x^2\right)_\infty} \prod_{A = X,Y} \prod_{i,j = 1,2} \frac{\left(z_i^{-1} z_j \upsilon_A^{-1} x^{2+|m_i-m_j|};x^2\right)_\infty}{\left(z_i z_j^{-1} \upsilon_A x^{|m_i-m_j|};x^2\right)_\infty}.
\end{align}
In this case two types of poles contribute to the integral. The relevant forest graphs are shown in figure \ref{fig:U(2)}.
Since we are considering one flavor, each forest graph contains only one tree graph. Tree graph (a) corresponds to the pole where the following factors in the denominator vanish:
\begin{align}
\label{eq:U(2)pole1}
z_{i_1}-t^{-1} x^{-|m_{i_1}|-2 k_{i_1}} = 0, \qquad z_{i_2}-z_{i_1} \upsilon_X^{-1} x^{-|m_{i_2}-m_{i_1}|-2 k_{i_2}} = 0
\end{align}
while tree graph (b) corresponds to the pole where the following factors vanish:
\begin{align}
\label{eq:U(2)pole2}
z_{i_1}-t^{-1} x^{-|m_{i_1}|-2 k_{i_1}} = 0, \qquad z_{i_2}-z_{i_1} \upsilon_Y^{-1} x^{-|m_{i_2}-m_{i_1}|-2 k_{i_2}} = 0.
\end{align}
 One can consider the other cases where $z_1$ and $z_2$ would take vanishing factors only from the fundamental part or only from the adjoint part, but the residue vanishes as the single adjoint case does \cite{Hwang:2015wna}. In \eqref{eq:U(2)pole1} or \eqref{eq:U(2)pole2}, a non-zero $m_i$ and $k_i$ case contributes to the vortex/anti-vortex parts while a vanishing $m_i$ and $k_i$ case contributes to the perturbative part. If $z_i$ is taken from the fundamental part, it is represented by the root node of a tree. If $z_i$ is taken from the adjoint $X$/$Y$, it is represented by a red/blue node respectively. Evaluating $Z_\text{pert}, \, Z_\text{(anti-)vortex}$ for each tree graph, one can obtain the superconformal index for this example. Moreover, in this case, \eqref{eq:Coulomb} can be computed by series expanding the integrand at $x = 0$ without any subtlety. Thus, we can explicitly compare \eqref{eq:Coulomb} and \eqref{eq:fact}, and see the exact agreement up to $x^4$. This can be done up to $U(3)$.

\begin{figure}[tbp]
\centering 
\begin{subfigure}{.3\textwidth}
\centering
\begin{tikzpicture}[grow'=up,<-,>=stealth',level/.style={sibling distance = 2cm/#1,level distance = 1cm}]
\node [arn_f] {$i_1$}
    child{ node [arn_x] {$i_2$}
        child{ node [arn_x] {$i_4$}
        }
        child[missing]
    }
    child{ node [arn_y] {$i_3$}
    }
;
\end{tikzpicture}
\caption{}
\end{subfigure}
\begin{subfigure}{.3\textwidth}
\centering
\begin{tikzpicture}[grow'=up,<-,>=stealth',level/.style={sibling distance = 2cm/#1,level distance = 1cm}]
\node [arn_f] {$i_1$}
    child{ node [arn_x] {$i_2$}
    }
    child{ node [arn_y] {$i_3$}
        child{ node [arn_x] {$i_4$}
        }
        child[missing]
    }
;
\end{tikzpicture}
\caption{}
\end{subfigure}
\begin{subfigure}{.3\textwidth}
\centering
\begin{tikzpicture}[grow'=up,<-,>=stealth',level/.style={sibling distance = 2cm/#1,level distance = 1cm}]
\node [arn_f] {$i_1$}
    child{ node [arn_x] {$i_2$}
        child[missing]
        child{ node [arn_y] {$i_4$}
        }
    }
    child{ node [arn_y] {$i_3$}
    }
;
\end{tikzpicture}
\caption{}
\end{subfigure}
\caption{\label{fig:U(4)} The tree graphs of $U(4)$ with $n = 3$ contributing to the index. (b) and (c) correspond to the same singularity and make a double pole there if we don't turn on the superpotential.}
\end{figure}
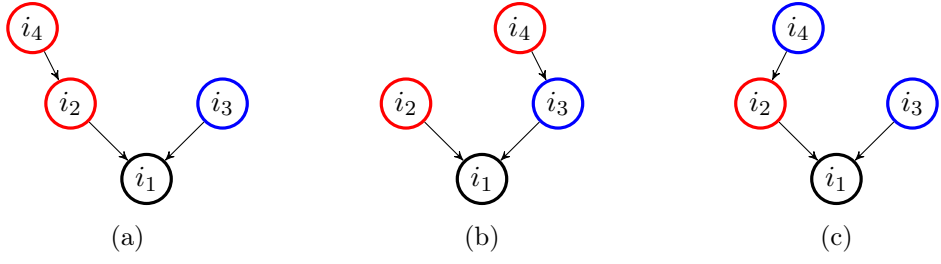

However, considering $U(4)$, one immediately sees that the perturbative part \eqref{eq:pert} diverges.\footnote{One can consider the other possibilities, i.e.,  the perturbative part vanishes but the vortex part would diverge such that the combined contribution remains finite or even diverges. However, we will see by examples in the next section that such a contribution is canceled by other contributions so that it doesn't affect the final result. We assume this is always the case and only focus on the perturbative part that doesn't vanish.} In particular, among the tree graphs of $U(4)$, there are two divergent trees, which are the last two graphs shown in figure \ref{fig:U(4)}.
Although those two trees naively seem to correspond to two different poles:
\begin{align}
\label{eq:dp1}
z_{i_1} = t^{-1}, \quad z_{i_2} = z_{i_1} \upsilon_X^{-1}, \quad z_{i_3} = z_{i_1} \upsilon_Y^{-1}, \quad z_{i_4} = z_{i_3} \upsilon_X^{-1}
\end{align}
and
\begin{align}
\label{eq:dp2}
z_{i_1} = t^{-1}, \quad z_{i_2} = z_{i_1} \upsilon_X^{-1}, \quad z_{i_3} = z_{i_1} \upsilon_Y^{-1}, \quad z_{i_4} = z_{i_2} \upsilon_Y^{-1},
\end{align}
both trees indeed correspond to the same singularity
\begin{align}
\label{eq:double pole}
z_{i_1} = t^{-1}, \quad z_{i_2} = t^{-1} \upsilon_X^{-1}, \quad z_{i_3} = t^{-1} \upsilon_Y^{-1}, \quad z_{i_4} = t^{-1} \upsilon_X^{-1} \upsilon_Y^{-1}
\end{align}
and make it a double pole.\footnote{We mean by a double pole here that the degree of a pole is $N_c+1$.  \eqref{eq:double pole} is a double pole because five different factors
\begin{align}
z_{i_1} = t^{-1}, \quad z_{i_2} = z_{i_1} \upsilon_X^{-1}, \quad z_{i_3} = z_{i_1} \upsilon_Y^{-1}, \quad z_{i_4} = z_{i_3} \upsilon_X^{-1}, \quad z_{i_4} = z_{i_2} \upsilon_Y^{-1},
\end{align}
which are the union of \eqref{eq:dp1} and \eqref{eq:dp2}, become zero simultaneously at the pole \eqref{eq:double pole}.} For this reason, from $U(4)$ with generic fugacities for adjoint matters,\footnote{In other words, there is no superpotential such that independent $U(1)$ symmetries rotating each adjoint are allowed.} the derivation of \eqref{eq:fact}, which only assumes simple poles, only works for up to $U(3)$ and doesn't work for higher ranks.
\\

In this note, however, we are interested in the theories with nontrivial superpotentials \eqref{eq:sup}, in particular, the D-type superpotential:
\begin{align}
\label{eq:sup_D}
W = \mathrm{Tr} X^{n+1}+\mathrm{Tr} X Y^2
\end{align}
with odd $n$. We will see that those superpotential tems reduce the double pole at \eqref{eq:double pole} to a simple pole. Let us recall the double pole tree graphs in figure \ref{fig:U(4)}. For both trees, the perturbative part could be written as follows:
\begin{align}
\label{eq:pert_U(4)}
&Z_\text{pert} = \nonumber \\
&\frac{\prod_{i \neq j} (1-{\upsilon_i}^{-1} \upsilon_j) \prod_{A = X,Y} \prod_{i,j = 1}^{N_c} ({\upsilon_A}^{-1} \upsilon_i {\upsilon_j}^{-1} x^2;x^2) \prod_{j = 1}^{N_c} (\upsilon_j x^2;x^2) \prod_{j = 1}^{N_c} (\tau^{-2} {\upsilon_j}^{-1} x^2;x^2)}{\prod_{A = X,Y} \prod_{i,j = 1}^{N_c} (\upsilon_A {\upsilon_i}^{-1} \upsilon_j;x^2) \prod_{j = 1}^{N_c} ({\upsilon_j}^{-1};x^2) \prod_{j = 1}^{N_c} (\tau^2 \upsilon_j;x^2)}
\end{align}
where $\upsilon_1 = 1, \, \upsilon_2 = \upsilon_X, \, \upsilon_3 = \upsilon_Y, \, \upsilon_4 = \upsilon_X \upsilon_Y$ and $(a;q) = (a;q)_\infty$. Note that $Z_\text{pert}$ contains five zeros in the denominator. If the pole were simple, the denominator should contain only four zeros, and we would just drop them from the expression. In the current case, however, we have one more zero in the denominator, which indicates that we have a double pole. In the integral form, the relevant part is the following:
\begin{align}
\int_{|z_j = 1|} \left(\prod_{j = 1}^{4} \frac{d z_j}{z_j}\right) \frac{(\ldots)}{(1-z_{i_1} t) (1-z_{i_2} z_{i_1}^{-1} \upsilon_X) (1-z_{i_3} z_{i_1}^{-1} \upsilon_Y) (1-z_{i_4} z_{i_2}^{-1} \upsilon_Y) (1-z_{i_4} z_{i_3}^{-1} \upsilon_X)}.
\end{align}
For simplicity, let us fix $i_j = j$ and take the residue at $z_1 = t^{-1}, \, z_2 = z_1 \upsilon_X^{-1} = t^{-1} \upsilon_X^{-1}, \, z_3 = z_1 \upsilon_Y^{-1} = t^{-1} \upsilon_Y^{-1}$ successively. We then have
\begin{align}
\int_{|z_4| = 1} \frac{d z_4}{z_4} \frac{(\ldots)}{(1-z_4 t \upsilon_X \upsilon_Y)^2}
\end{align}
where $(\ldots)$ doesn't vanish at $z_4 = t^{-1} \upsilon_X^{-1} \upsilon_Y^{-1}$. Now we have a double pole for the $z_4$ integration, and our result \eqref{eq:fact} fails to hold for high gauge rank theories with two adjoint and without superpotential.

However, once we introduce the superpotential \eqref{eq:sup_D}, which forces us to set $\upsilon_X = x^\frac{2}{n+1}, \, \upsilon_Y = x^\frac{n}{n+1}$, we have an additional zero of the integrand:
\begin{align}
\int_{|z_j| = 1} \left(\prod_{j = 1}^{4} \frac{d z_j}{z_j}\right) \frac{(1-z_1^{-1} z_4 x^\frac{n+2}{n+1}) (\ldots)}{(1-z_1 t) (1-z_2 z_1^{-1} x^\frac{2}{n+1}) (1-z_3 z_1^{-1} x^\frac{n}{n+1}) (1-z_4 z_2^{-1} x^\frac{n}{n+1}) (1-z_4 z_3^{-1} x^\frac{2}{n+1})}
\end{align}
where the zero comes from $(1-\upsilon_Y^{-1} z_1^{-1} z_4 x^2)$, which belongs to the 1-loop determinant of adjoint $Y$. Note that every factor of the integrand is organized so that the residue at
\begin{align}
z_1 = t^{-1}, \quad z_2 = t^{-1} x^{-\frac{2}{n+1}}, \quad z_3 = t^{-1} x^{-\frac{n}{n+1}}, \quad z_4 = t^{-1} x^{-\frac{n+2}{n+1}}
\end{align}
is simply 1. Thus, in \eqref{eq:pert_U(4)} with $\upsilon_X = x^\frac{2}{n+1}, \, \upsilon_Y = x^\frac{n}{n+1}$, although $\frac{0}{0}$'s in principle are ambiguous, in our context, if the numbers of extra zeros in the numerator are equal to those in the denominator, we can simply drop them. If the number of extra zeros is larger than that of poles, simply the corresponding forest graph doesn't contribute to the index. On the other hand, if the number of extra poles is larger, there exists a higher degree pole in spite of the superpotential \eqref{eq:sup_D}. However, this is not the case of examples we are considering in this note.
\\

Following this rule, one can obtain the correct answer from \eqref{eq:fact}, which is confirmed comparing the factorized index \eqref{eq:fact} and the series expansion of \eqref{eq:Coulomb} for a few examples. For the $U(4)$ theory with a single flavor and the $n = 3$ superpotential, the series expansion of \eqref{eq:Coulomb} gives the following result:
\begin{align}
\label{eq:U(4)_Coulomb}
\begin{array}{ll}
m = 0, \quad & 1+x^\frac{1}{6} \tau^2+x^\frac{1}{3} \left(\tau^4+\tau^{-2}\right)+x^\frac{1}{2} \left(\tau^6+2\right)+\mathcal O(x^\frac{2}{3}), \\
m = \pm 1, \quad & x^\frac{1}{6} \tau^{-1}+x^\frac{1}{3} \tau+x^\frac{1}{2} \left(\tau^3+\tau^{-3}\right)+\mathcal O(x^\frac{2}{3}), \\
m = \pm 2, \quad & x^\frac{1}{3} \tau^{-2}+x^\frac{1}{2}+\mathcal O(x^\frac{2}{3}), \\
m = \pm 3, \quad & x^\frac{1}{2} \tau^{-3}+\mathcal O(x^\frac{2}{3})
\end{array}
\end{align}
where $m$ is the total monopole flux. We have chosen the $R$-charge $r = 1/12$ for the fundamental matter such that the lowest monopole operator and the lowest mesonic operator have the same trial UV $R$-charge for convenience. The superconformal $R$-charge $R_\text{IR}$ can be restored by shifting the $U(1)_A$ fuagacity $\tau \rightarrow \tau x^\alpha$ where $\alpha$ satisfies $R_\text{IR} = R_\text{UV}+\alpha F_A$, where $F_A$ is the $U(1)_A$ charge.

On the Higgs branch localization side, we have two contributing tree graphs as shown in figure \ref{fig:U(4)}. As mentioned, (b) and (c) are equivalent graphs associated with the same singularity and give the same index contribution as expected. In section \ref{sec:test}, we will argue that such equivalent trees always give the same contribution in general. For tree (a), we have the following index contribution:
\begin{align}
\label{eq:U(4)_a}
\begin{array}{ll}
m = 0, \quad & 1+x^\frac{1}{6} \tau^2+x^\frac{1}{4}+x^\frac{1}{3} \left(\tau^4+\tau^{-2}\right)+x^\frac{5}{12} \tau^2+x^\frac{1}{2} \left(\tau^6+4\right)+x^\frac{7}{12} \left(\tau^4+\tau^{-2}\right)+\mathcal O(x^\frac{2}{3}), \\
m = \pm 1, \quad & x^\frac{1}{6} \tau^{-1}+x^\frac{1}{3} \tau+x^\frac{5}{12} \tau^{-1}+x^\frac{1}{2} \left(\tau^3+\tau^{-3}\right)+x^\frac{7}{12} \tau+\mathcal O(x^\frac{2}{3}), \\
m = \pm 2, \quad & x^\frac{1}{3} \tau^{-2}+x^\frac{1}{2}+x^\frac{7}{12} \tau^{-2}+\mathcal O(x^\frac{2}{3}), \\
m = \pm 3, \quad & x^\frac{1}{2} \tau^{-3}+\mathcal O(x^\frac{2}{3}).
\end{array}
\end{align}
For tree (b),
\begin{align}
\label{eq:U(4)_b}
\begin{array}{ll}
m = 0, \quad & -x^\frac{1}{4}-x^\frac{5}{12} \tau^2-2 x^\frac{1}{2}-x^\frac{7}{12} \left(\tau^4+\tau^{-2}\right)+\mathcal O(x^\frac{2}{3}), \\
m = \pm 1, \quad & -x^\frac{5}{12} \tau^{-1}-x^\frac{7}{12} \tau+\mathcal O(x^\frac{2}{3}), \\
m = \pm 2, \quad & -x^\frac{7}{12} \tau^{-2}+\mathcal O(x^\frac{2}{3}), \\
m = \pm 3, \quad & \mathcal O(x^\frac{2}{3}).
\end{array}
\end{align}
One obtains the same result as \eqref{eq:U(4)_Coulomb} by adding \eqref{eq:U(4)_a} and \eqref{eq:U(4)_b}, which confirms the simple pole computation \eqref{eq:fact} gives the correct answer once we introduce the superpotential \eqref{eq:sup_D}.
\\

We would like to comment on a higher rank example. Consider the $U(5)$ theory with a single flavor. One might want to compare \eqref{eq:Coulomb} and \eqref{eq:fact} in this case as well, but, unfortunately, it is not feasible because for $U(5)$ the naive conformal dimension of a monopole operator becomes negative.\footnote{One can avoid this problem by increasing the number of flavors but the computation will be heavier.} For this reason, \eqref{eq:Coulomb} is not analytic at $x = 0$, and one cannot evaluate the series expansion of \eqref{eq:Coulomb} with respect to $x$. Physically this is a signal of decoupling of the monopole operator because the conformal dimension of a gauge invariant operator at the IR fixed point shouldn't be less than 1/2. To be consistent, an operator of naive conformal dimension less than 1/2 should decouple from the interacting sector. Then there is an accidental $U(1)$ symmetry that freely rotates the decoupled operator. This accidental $U(1)$ is mixed with the $R$-symmetry and corrects the conformal dimension of the operator to be 1/2, which is the conformal dimension of a free operator.

Since \eqref{eq:Coulomb} is only sensitive to UV data and cannot capture those accidental symmetries appearing in IR, \eqref{eq:Coulomb} as a series expansion is not powerful at all in this case. However, as we mentioned, one should note that this singularity at $x = 0$ is not intrinsic and originates from the fact that we assign wrong conformal dimensions to decoupled operators. One can still have a meaningful answer from \eqref{eq:Coulomb} by exactly evaluating this contour integral rather than expanding it as a series in $x$, which is indeed the factorized index \eqref{eq:fact}. One can use \eqref{eq:fact} to compute the (analytically continued) superconformal index of higher gauge rank theories containing decoupled operators. Indeed, this allows us to test the duality for higher rank gauge groups in section \ref{sec:test}. Moreover, if we look at table \ref{tab:3d}, the dual chiral fields, $q, \, \tilde q$ carry negative $R$-charges. Due to the superpotential, their gauge invariant mesonic combination becomes non-BPS and doesn't suffer from the unitarity bound because its conformal dimension $\Delta$ doesn't saturate $\Delta \geq R+j$. Nevertheless, this negative $R$-charge yields a practical difficulty in computing the series expansion of \eqref{eq:Coulomb}. For this reason, the superconformal index of any dual theory cannot be computed using the series expansion method. This is another critical reason that we require the factorized index to test the duality. See section \ref{sec:test} for details of the duality test.
\\

\subsection{Growing Trees}

In the previous subsection, we have seen that the integral formula of the Coulomb branch localization, \eqref{eq:Coulomb}, suffers from double poles, for a high rank gauge group in particular, with multiple adjoint matters. However, we also argued that such double poles become simple ones if we introduce the superpotential \eqref{eq:sup_D}. Due to such cancelations between zeros and poles, the expressions in \eqref{eq:pert}, strictly speaking, are not well defined; we have to explain how to deal with zeros and poles in those expressions, and their cancelations. As we argued in the previous section, we can simply drop extra zeros and poles if the numbers of them are equal. If the number of extra zeros is larger than that of poles, simply the corresponding forest graph doesn't contribute to the index. On the other hand, if the number of extra poles is larger, there exists a higher degree pole in spite of the superpotential \eqref{eq:sup_D}, which doesn't appear in our examples.
\\

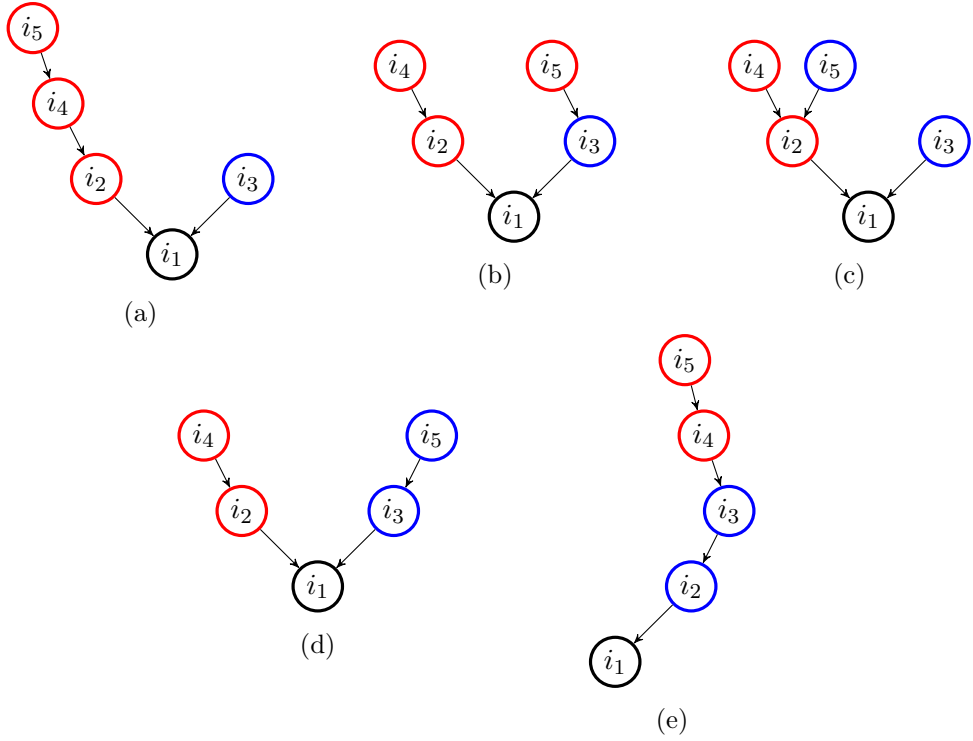
\begin{figure}[tbp]
\centering 
\begin{subfigure}{.3\textwidth}
\centering
\begin{tikzpicture}[grow'=up,<-,>=stealth',level/.style={sibling distance = 2cm/#1,level distance = 1cm}]
\node [arn_f] {$i_1$}
    child{ node [arn_x] {$i_2$}
        child{ node [arn_x] {$i_4$}
            child{ node [arn_x] {$i_5$}
            }
            child[missing]
        }
        child[missing]
    }
    child{ node [arn_y] {$i_3$}
    }
;
\end{tikzpicture}
\caption{}
\end{subfigure}
\begin{subfigure}{.3\textwidth}
\centering
\begin{tikzpicture}[grow'=up,<-,>=stealth',level/.style={sibling distance = 2cm/#1,level distance = 1cm}]
\node [arn_f] {$i_1$}
    child{ node [arn_x] {$i_2$}
        child{ node [arn_x] {$i_4$}
        }
        child[missing]
    }
    child{ node [arn_y] {$i_3$}
        child{ node [arn_x] {$i_5$}
        }
        child[missing]
    }
;
\end{tikzpicture}
\caption{}
\end{subfigure}
\begin{subfigure}{.3\textwidth}
\centering
\begin{tikzpicture}[grow'=up,<-,>=stealth',level/.style={sibling distance = 2cm/#1,level distance = 1cm}]
\node [arn_f] {$i_1$}
    child{ node [arn_x] {$i_2$}
        child{ node [arn_x] {$i_4$}
        }
        child{ node [arn_y] {$i_5$}
        }
    }
    child{ node [arn_y] {$i_3$}
    }
;
\end{tikzpicture}
\caption{}
\end{subfigure}
\begin{subfigure}{.3\textwidth}
\centering
\begin{tikzpicture}[grow'=up,<-,>=stealth',level/.style={sibling distance = 2cm/#1,level distance = 1cm}]
\node [arn_f] {$i_1$}
    child{ node [arn_x] {$i_2$}
        child{ node [arn_x] {$i_4$}
        }
        child[missing]
    }
    child{ node [arn_y] {$i_3$}
        child[missing]
        child{ node [arn_y] {$i_5$}
        }
    }
;
\end{tikzpicture}
\caption{}
\end{subfigure}
\begin{subfigure}{.3\textwidth}
\centering
\begin{tikzpicture}[grow'=up,<-,>=stealth',level/.style={sibling distance = 2cm/#1,level distance = 1cm}]
\node [arn_f] {$i_1$}
    child[missing]
    child{ node [arn_y] {$i_2$}
        child[missing]
        child{ node [arn_y] {$i_3$}
            child{ node [arn_x] {$i_4$}
                child{ node [arn_x] {$i_5$}
                }
                child[missing]
            }
            child[missing]
        }
    }
;
\end{tikzpicture}
\caption{}
\end{subfigure}
\caption{\label{fig:U(5)} Nontrivial trees of $U(5)$ with $n = 3$. The perturbative part corresponding to each tree has the same number of extra zeros in the numerator and in the denominator.}
\end{figure}

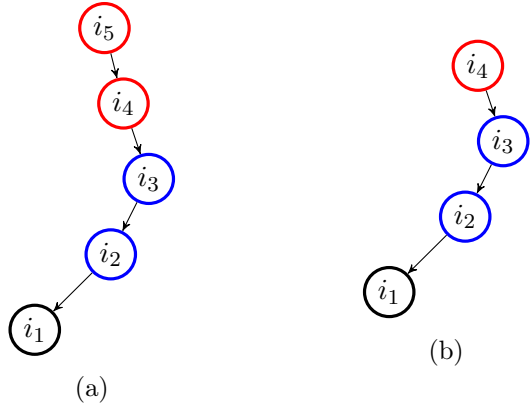
\begin{figure}[tbp]
\centering 
\begin{subfigure}{.3\textwidth}
\centering
\begin{tikzpicture}[grow'=up,<-,>=stealth',level/.style={sibling distance = 2cm/#1,level distance = 1cm}]
\node [arn_f] {$i_1$}
    child[missing]
    child{ node [arn_y] {$i_2$}
        child[missing]
        child{ node [arn_y] {$i_3$}
            child{ node [arn_x] {$i_4$}
                child{ node [arn_x] {$i_5$}
                }
                child[missing]
            }
            child[missing]
        }
    }
;
\end{tikzpicture}
\caption{}
\end{subfigure}
\begin{subfigure}{.3\textwidth}
\centering
\begin{tikzpicture}[grow'=up,<-,>=stealth',level/.style={sibling distance = 2cm/#1,level distance = 1cm}]
\node [arn_f] {$i_1$}
    child[missing]
    child{ node [arn_y] {$i_2$}
        child[missing]
        child{ node [arn_y] {$i_3$}
            child{ node [arn_x] {$i_4$}
            }
            child[missing]
        }
    }
;
\end{tikzpicture}
\caption{}
\end{subfigure}
\caption{\label{fig:non-growing tree} (a) is a non-growing tree of $U(5)$, which is obtained by attaching one more node to (b), a trivial tree of $U(4)$.}
\end{figure}

For the equal number of extra poles and zeros, however, this is not the end of the story. To provide a concrete example, we consider $U(5)$ with a single flavor and the $n = 3$ superpotential. In this case, the trees having the same number of extra zeros and poles in the perturbative part are listed in figure \ref{fig:U(5)}.
First note that (b), (c) are equivalent as similar to $U(4)$ and (a), (d) are equivalent once we introduce the superpotential \eqref{eq:sup_D}. One can check that each equivalent pair correspond to the same singularity and indeed give the same index contribution. This is because we have the F-term conditions
\begin{align}
X^n \sim Y^2, \qquad X Y \sim Y X
\end{align}
from the superpoential \eqref{eq:sup_D}. Also note that, except for the last one, those trees can be obtained from nontrivial trees of $U(4)$, which are shown in figure \ref{fig:U(4)}, by attaching one more node. On the other hand, the last tree is obtained by attaching a node to a trivial tree of $U(4)$. See figure \ref{fig:non-growing tree}.

Here we claim that among the trees having the same number of extra zeros and poles, only the \emph{growing} trees can contribute to the index. Up to $U(4)$ we define that every tree having the same number of extra zeros and poles is a growing tree. From $U(5)$ we define that a tree of the same number of extra zeros and poles is a growing tree if and only if it is obtained by attaching one more node to a growing tree of one node less. Non-growing trees can have nonzero contributions but their contributions are canceled among them.

We would like to explain the details more. Let us consider the non-growing tree shown in figure \ref{fig:non-growing tree}. Ignoring the superpotential for a moment, the tree is associated with the residue,
\begin{align}
\label{eq:res_non-growing 0}
\mathrm{Res}_{z_j = z_j^*} \frac{1}{\prod_{j = 1}^{5} z_j} \frac{(\ldots)}{(1-z_{i_1} t)(1-z_{i_2} z_{i_1}^{-1} \upsilon_Y)(1-z_{i_3} z_{i_2}^{-1} \upsilon_Y)(1-z_{i_4} z_{i_3}^{-1} \upsilon_X)(1-z_{i_5} z_{i_4}^{-1} \upsilon_X)}
\end{align}
where $z_j^*$ is the pole determined by denominators, and $(\ldots)$ is an expression regular at the pole. The sequence of indices $(i) \equiv (i_1,i_2,\ldots,i_5)$ is any permutation of $(1,2,\ldots,5)$. Indeed, the residue \eqref{eq:res_non-growing 0} should be understood as iterative residue evaluations with respect to the fixed order of $z_j$'s, e.g., $z_1 \rightarrow z_2 \rightarrow \ldots \rightarrow z_5$. Thus, the permutation of $(1,2,\ldots,5)$ is effectively changing the order of evaluations. Since each residue evaluation is done at a simple pole, it can be done without any difficulty.

Once we introduce the superpotential, however, we have more zeros and poles:
\begin{align}
\label{eq:res_non-growing}
& \mathrm{Res}_{z_j = z_j^*} \frac{1}{\prod_{j = 1}^{5} z_j} \frac{(1-z_{i_1}^{-1} z_{i_3} x^\frac{2 n}{n+1}) (1-z_{i_2}^{-1} z_{i_4} x^\frac{n+2}{n+1})}{(1-z_{i_4} t x^2) (1-z_{i_5} z_{i_1}^{-1} x^\frac{2 n+4}{n+1})} \nonumber \\
& \times \frac{f(z_j)}{(1-z_{i_1} t)(1-z_{i_2} z_{i_1}^{-1} x^\frac{n}{n+1})(1-z_{i_3} z_{i_2}^{-1} x^\frac{n}{n+1})(1-z_{i_4} z_{i_3}^{-1} x^\frac{2}{n+1})(1-z_{i_5} z_{i_4}^{-1} x^\frac{2}{n+1})}
\end{align}
where $f(z_j)$ is the remaining part of the integrand that is regular at the singularity. For $(i) = (1,2,3,4,5)$, one can see that \eqref{eq:res_non-growing} vanishes because after $z_1,z_2$ integrations, there is no pole at $z_3^* = t^{-1} x^{-\frac{2 n}{n+1}}$. It shows that the non-growing tree with $(i) = (1,2,3,4,5)$ doesn't contribute to the index. If we take $(i) = (1,2,5,3,4)$, on the other hand, the residue doesn't vanish and gives $f(z_j^*)$. However, we will shortly explain that this possible non-growing tree contribution is canceled by another non-growing tree contribution.

Let us look at \eqref{eq:res_non-growing} more closely. It is in fact a part of the evaluation of the integral,
\begin{align}
& \int_{|z_j| = 1} \left(\prod_{j = 1}^{5} \frac{d z_j}{z_j}\right) \frac{(1-z_{i_1}^{-1} z_{i_3} x^\frac{2 n}{n+1}) (1-z_{i_2}^{-1} z_{i_4} x^\frac{n+2}{n+1})}{(1-z_{i_4} t x^2) (1-z_{i_5} z_{i_1}^{-1} x^\frac{2 n+4}{n+1})} \nonumber \\
& \times \frac{f(z_j)}{(1-z_{i_1} t)(1-z_{i_2} z_{i_1}^{-1} x^\frac{n}{n+1})(1-z_{i_3} z_{i_2}^{-1} x^\frac{n}{n+1})(1-z_{i_4} z_{i_3}^{-1} x^\frac{2}{n+1})(1-z_{i_5} z_{i_4}^{-1} x^\frac{2}{n+1})}.
\end{align}
Since we want to isolate the residue contribution at the singularity,
\begin{align}
\label{eq:sing}
z_{i_1}^* = t^{-1}, \quad z_{i_2}^* = t^{-1} x^{-\frac{n}{n+1}}, \quad z_{i_3}^* = t^{-1} x^{-\frac{2 n}{n+1}}, \quad z_{i_4}^* = t^{-1} x^{-\frac{2 n+2}{n+1}}, \quad z_{i_5}^* = t^{-1} x^{-\frac{2 n+4}{n+1}},
\end{align}
let us forget about the poles from $f(z_j)$ and regard $f(z_j)$ as a constant for a moment to capture the essential logic of the argument:
\begin{align}
\label{eq:int_isol}
& \int_{|z_j| = 1} \left(\prod_{j = 1}^{5} \frac{d z_j}{z_j}\right) \frac{(1-z_{i_1}^{-1} z_{i_3} x^\frac{2 n}{n+1}) (1-z_{i_2}^{-1} z_{i_4} x^\frac{n+2}{n+1})}{(1-z_{i_4} t x^2) (1-z_{i_5} z_{i_1}^{-1} x^\frac{2 n+4}{n+1})} \nonumber \\
& \times \frac{C}{(1-z_{i_1} t)(1-z_{i_2} z_{i_1}^{-1} x^\frac{n}{n+1})(1-z_{i_3} z_{i_2}^{-1} x^\frac{n}{n+1})(1-z_{i_4} z_{i_3}^{-1} x^\frac{2}{n+1})(1-z_{i_5} z_{i_4}^{-1} x^\frac{2}{n+1})}
\end{align}
where $C = f(z_j^*)$. The integral \eqref{eq:int_isol} is easily computed after the series expansion with respect to $x$, and gives 1. Also it is independent of the order of the integrations unlike \eqref{eq:res_non-growing}.

However, one should note that this is not exactly the residue we originally compute. In order to see that, let us evaluate \eqref{eq:int_isol} without using the series expansion. For small $|t|,|x| < 1$, the contributing residues are located at the singularity \eqref{eq:sing} or at asymptotic infinity where some of $z_j$ take infinite values.\footnote{In this note, when we evaluate the unit circle contour integral, we take the residues of the poles located outside the unit circle.} However, one should note that originally we don't have poles at asymptotic infinity. See appendix \ref{sec:integral}. It means $f(z_j)$ should have zeros at asymptotic infinity, which cancel the corresponding poles in \eqref{eq:int_isol}. Therefore, the residue associated with the singularity \eqref{eq:sing} is the only relevant contribution and can be singled out by subtracting the asymptotic infinity contribution from \eqref{eq:int_isol}. We will demonstrate that the contributions from asymptotic infinity are summed up to 1 while those from the singularity \eqref{eq:sing} are summed up to 0. Thus, the non-growing tree doesn't contribute.
\\

\subsubsection*{Case I}

First let us consider $(i_1,i_2,i_3,i_4,i_5) = (1,2,3,4,5)$. We perform the integrations with respect to the order $z_1 \rightarrow \ldots \rightarrow z_5$. After $z_1, \, z_2$ integrations, the only relevant $z_3$ pole is $z_3 = \infty$. Taking the residue at $z_3 = \infty$, and subsequently performing the integrations over $z_4$ and $z_5$, the final answer is 1. Thus, 1, the result of \eqref{eq:int_isol}, comes from asymptotic infinity. There is no contribution from the singularity \eqref{eq:sing}.

\subsubsection*{Case II}

\begin{figure}[tbp]
\centering 
\begin{subfigure}{.3\textwidth}
\centering
\begin{tikzpicture}[grow'=up,<-,>=stealth',level/.style={sibling distance = 2cm/#1,level distance = 1cm}]
\node [arn_f] {$i_1$}
    child{ node [arn_x] {$i_2$}
    }
    child{ node [arn_y] {$i_3$}
        child[missing]
        child{ node [arn_y] {$i_4$}
            child{ node [arn_x] {$i_5$}
            }
            child[missing]
        }
    }
;
\end{tikzpicture}
\caption{}
\end{subfigure}
\caption{\label{fig:exceptional tree} The non-growing tree corresponds to the second chain of poles in \eqref{eq:pole chains}.}
\end{figure}
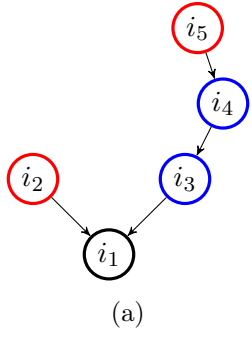

For $(i_1,i_2,i_3,i_4,i_5) = (1,2,5,3,4)$, the integral \eqref{eq:int_isol} can be computed by collecting the residues at the following chains of poles:
\begin{align}
\label{eq:pole chains}
\begin{array}{lllllllll}
z_1 = t^{-1} & \rightarrow & z_2 = t^{-1} x^{-\frac{n}{n+1}} & \rightarrow & z_3 = z_5 x^{-\frac{2}{n+1}} & \rightarrow & z_4 = z_5 x^{-\frac{4}{n+1}} & \rightarrow & z_5 = t^{-1} x^{-\frac{2 n}{n+1}}, \\
z_1 = t^{-1} & \rightarrow & z_2 = t^{-1} x^{-\frac{n}{n+1}} & \rightarrow & z_3 = z_5 x^{-\frac{2}{n+1}} & \rightarrow & z_4 = t^{-1} x^{-\frac{2 n+4}{n+1}} & \rightarrow & z_5 = t^{-1} x^{-\frac{2 n}{n+1}}, \\
z_1 = t^{-1} & \rightarrow & z_2 = t^{-1} x^{-\frac{n}{n+1}} & \rightarrow & z_3 = z_5 x^{-\frac{2}{n+1}} & \rightarrow & z_4 = t^{-1} x^{-\frac{2 n+4}{n+1}} & \rightarrow & z_5 = \infty.
\end{array}
\end{align}
The residue from each chain of poles is 1, -1, 1 respectively, whose total sum reproduces the result of \eqref{eq:int_isol}. Substituting the explicit values of $z_j$ at the end, one can immediately note that the first two are from the singularity \eqref{eq:sing} while the last one is from asymptotic infinity and is an irrelevant contribution. Therefore, again, we show that the residues associated with the singularity \eqref{eq:sing} are summed up to zero. We would like to comment on the second chain of poles. In our tree representation language, it corresponds to the non-growing tree shown in figure \ref{fig:exceptional tree} with non-zero vortex numbers while the first chain corresponds to the non-grown tree (a) in figure \ref{fig:non-growing tree}.
Roughly speaking, for this tree, the perturbative part have zeros while the vortex/anti-vortex contributions have poles such that their product gives a finite answer. We expect such non-physical contributions, i.e.,where the perturbative part and the vortex part are not well-defined independently so that they spoil the factorization, should be canceled out by the mechanism we just explained.

\subsubsection*{Case III}

For $(i_1,i_2,i_3,i_4,i_5) = (1,5,2,3,4)$, after $z_1$ integration, we have
\begin{align}
\label{eq:int_dp 1}
& \int_{|z_j| = 1} \left(\prod_{j = 2}^{5} \frac{d z_j}{z_j}\right) \frac{(1-t z_2 x^\frac{2 n}{n+1}) (1-z_5^{-1} z_3 x^\frac{n+2}{n+1})}{(1-z_3 t x^2) (1-z_4 t x^\frac{2 n+4}{n+1})} \nonumber \\
& \times \frac{C}{(1-z_5 t x^\frac{n}{n+1}) (1-z_2 z_5^{-1} x^\frac{n}{n+1}) (1-z_3 z_2^{-1} x^\frac{2}{n+1}) (1-z_4 z_3^{-1} x^\frac{2}{n+1})}
\end{align}
For $z_2$ integration, one can check that the relevant poles are at $z_2 = z_5 x^{-\frac{n}{n+1}}$ and at $z_2 = \infty$. Taking the first one, and subsequently taking the pole at $z_3 = t^{-1} x^{-2}$, we have
\begin{align}
\label{eq:int_dp 2}
& \int_{|z_j| = 1} \left(\prod_{j = 4}^{5} \frac{d z_j}{z_j}\right) \frac{C}{(1-z_4 t x^\frac{2 n+4}{n+1})^2},
\end{align}
which includes a double pole at $z_4 = t^{-1} x^{-\frac{2 n+4}{n+1}}$. Now one should remind that $C$ is not really a constant but is a function of $z_j$, $f(z_j)$. Up to $z_3$ integration, $f(z_j)$ is regular; we just substitute the pole values of $z_1, z_2, z_3$. Let us call it $g(z_4,z_5)$. Since $g(z_4,z_5)$ is regular at $z_4 = t^{-1} x^{-\frac{2 n+4}{n+1}}$, the residue at this point is given by
\begin{align}
\label{eq:int_dp 3}
&-\int_{|z_5| = 1} \frac{d z_5}{z_5} t^{-2} x^{-\frac{4 n+8}{n+1}} \left.\frac{\partial}{\partial z_4} z_4^{-1} g(z_4,z_5)\right|_{z_4 = t^{-1} x^{-\frac{2 n+4}{n+1}}} \nonumber \\
&= \int_{|z_5| = 1} \frac{d z_5}{z_5} \left(g(t^{-1} x^{-\frac{2 n+4}{n+1}},z_5)-t^{-1} x^{-\frac{2 n+4}{n+1}}  g'(t^{-1} x^{-\frac{2 n+4}{n+1}},z_5)\right)
\end{align}
where $g'(z_4,z_5) = \frac{\partial}{\partial z_4} g(z_4,z_5)$. Our interest is whether there is a pole at $z_5 = t^{-1} x^{-\frac{n}{n+1}}$ or not. Indeed, since $g(z_4,z_5)$ is regular at $z_4 = t^{-1} x^{-\frac{2 n+4}{n+1}}, \, z_5 = t^{-1} x^{-\frac{n}{n+1}}$, so is $g'(z_4,z_5)$ at the same point. Thus, \eqref{eq:int_dp 3} doesn't receive any contribution from the singularity \eqref{eq:sing}.

As a consistency check, we evaluate the residue of \eqref{eq:int_dp 2} at $z_4 = t^{-1} x^{-\frac{2 n+4}{n+1}}, \, z_5 = \infty$ and obtain 1. Also we evaluate the residue of \eqref{eq:int_dp 1} at $z_2 = \infty$ and obtain 0. Their sum is 1, which is the same as the result of \eqref{eq:int_isol}.
\\

\subsubsection*{Numerical Comparison}

We have demonstrated three different choices of $(i_1,i_2,i_3,i_4,i_5)$, and the other choices are similar. In those cases, the residues associated with the singularity \eqref{eq:sing} are summed up to zero. Thus, the non-growing tree associated with this singularity doesn't contribute to the index.

We also confirm this claim: the non-growing tree doesn't contribute, by an explicit comparison of \eqref{eq:Coulomb} and \eqref{eq:fact}. For $U(5)$ with a single flavor, the $R$-charges of monopole operators are positive only when $n > 3$. Thus, let us consider $n = 5$. First we evaluate the \eqref{eq:Coulomb} as a series exapnsion, which gives rise to
\begin{align}
\label{eq:U(5)_Coulomb}
&1+2 \sqrt[12]{x}+3 \sqrt[6]{x}+4 \sqrt[4]{x}+6 \sqrt[3]{x}+10 x^{5/12}+17 \sqrt{x}+24 x^{7/12}+33 x^{2/3}+46 x^{3/4}+67 x^{5/6} \nonumber \\
&+96 x^{11/12}+134 x+180 x^{13/12}+244 x^{7/6}+330 x^{5/4}+447 x^{4/3}+596 x^{17/12}+788 x^{3/2} \nonumber \\
&+1030 x^{19/12}+1349 x^{5/3}+1754 x^{7/4}+2273 x^{11/6}+2920 x^{23/12}+3733 x^2+4746 x^{25/12} \nonumber \\
&+6024 x^{13/6}+7600 x^{9/4}+9550 x^{7/3}+11932 x^{29/12}+14861 x^{5/2}+18428 x^{31/12}+22781 x^{8/3} \nonumber \\
&+28022 x^{11/4}+34351 x^{17/6}+41942 x^{35/12}+51053 x^3+O\left(x^{37/12}\right).
\end{align}
We have chosen the $R$-charge $r = \frac{1}{4}$ for the fundamental matter. For simplicity we only keep the fugacity $x$ and turn off the others.

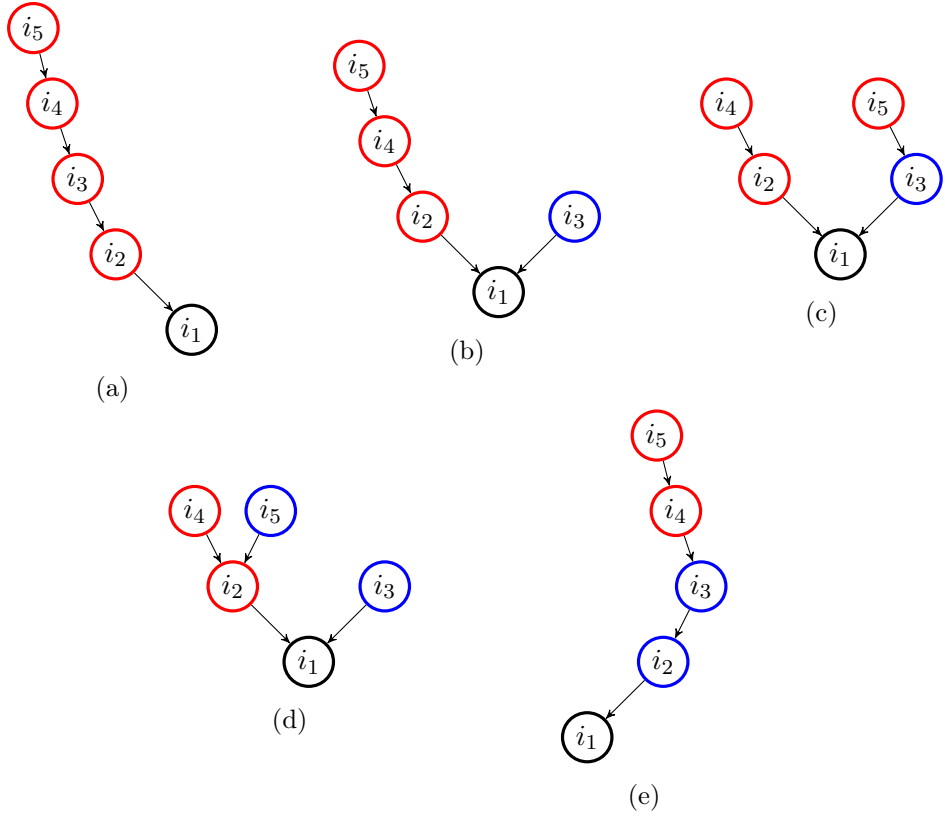
\begin{figure}[tbp]
\centering 
\begin{subfigure}{.3\textwidth}
\centering
\begin{tikzpicture}[grow'=up,<-,>=stealth',level/.style={sibling distance = 2cm/#1,level distance = 1cm}]
\node [arn_f] {$i_1$}
    child{ node [arn_x] {$i_2$}
        child{ node [arn_x] {$i_3$}
            child{ node [arn_x] {$i_4$}
                child{ node [arn_x] {$i_5$}
                }
                child[missing]
            }
            child[missing]
        }
        child[missing]
    }
    child[missing]
;
\end{tikzpicture}
\caption{}
\end{subfigure}
\begin{subfigure}{.3\textwidth}
\centering
\begin{tikzpicture}[grow'=up,<-,>=stealth',level/.style={sibling distance = 2cm/#1,level distance = 1cm}]
\node [arn_f] {$i_1$}
    child{ node [arn_x] {$i_2$}
        child{ node [arn_x] {$i_4$}
            child{ node [arn_x] {$i_5$}
            }
            child[missing]
        }
        child[missing]
    }
    child{ node [arn_y] {$i_3$}
    }
;
\end{tikzpicture}
\caption{}
\end{subfigure}
\begin{subfigure}{.3\textwidth}
\centering
\begin{tikzpicture}[grow'=up,<-,>=stealth',level/.style={sibling distance = 2cm/#1,level distance = 1cm}]
\node [arn_f] {$i_1$}
    child{ node [arn_x] {$i_2$}
        child{ node [arn_x] {$i_4$}
        }
        child[missing]
    }
    child{ node [arn_y] {$i_3$}
        child{ node [arn_x] {$i_5$}
        }
        child[missing]
    }
;
\end{tikzpicture}
\caption{}
\end{subfigure}
\begin{subfigure}{.3\textwidth}
\centering
\begin{tikzpicture}[grow'=up,<-,>=stealth',level/.style={sibling distance = 2cm/#1,level distance = 1cm}]
\node [arn_f] {$i_1$}
    child{ node [arn_x] {$i_2$}
        child{ node [arn_x] {$i_4$}
        }
        child{ node [arn_y] {$i_5$}
        }
    }
    child{ node [arn_y] {$i_3$}
    }
;
\end{tikzpicture}
\caption{}
\end{subfigure}
\begin{subfigure}{.3\textwidth}
\centering
\begin{tikzpicture}[grow'=up,<-,>=stealth',level/.style={sibling distance = 2cm/#1,level distance = 1cm}]
\node [arn_f] {$i_1$}
    child[missing]
    child{ node [arn_y] {$i_2$}
        child[missing]
        child{ node [arn_y] {$i_3$}
            child{ node [arn_x] {$i_4$}
                child{ node [arn_x] {$i_5$}
                }
                child[missing]
            }
            child[missing]
        }
    }
;
\end{tikzpicture}
\caption{}
\end{subfigure}
\caption{\label{fig:U(5)+n=5} Nontrivial trees of $U(5)$ with $n = 5$. The perturbative part of each tree has the same number of extra zeros in the numerator and in the denominator.}
\end{figure}

Next we evaluate \eqref{eq:fact} for each tree allowed for $U(5)$. The list of trees having the same number of extra zeros and poles in the perturbative part is given in figure \ref{fig:U(5)+n=5}.
In figure \ref{fig:U(5)+n=5}, (c) and (d) are equivalent; i.e., they correspond to the same singularity. (e) is a non-growing tree and shouldn't contribute to the index. For (a), (b) and (c)=(d), we have the following index contributions respectively:
\begin{align}
&x^{2/3}+2 x^{3/4}+4 x^{5/6}+6 x^{11/12}+9 x+14 x^{13/12}+24 x^{7/6}+36 x^{5/4}+53 x^{4/3}+74 x^{17/12} \nonumber \\
&+104 x^{3/2}+146 x^{19/12}+207 x^{5/3}+282 x^{7/4}+381 x^{11/6}+506 x^{23/12}+671 x^2+884 x^{25/12} \nonumber\\
&+1164 x^{13/6}+1508 x^{9/4}+1944 x^{7/3}+2480 x^{29/12}+3157 x^{5/2}+3994 x^{31/12}+5041 x^{8/3} \nonumber \\
&+6306 x^{11/4}+7855 x^{17/6}+9722 x^{35/12}+12000 x^3+O\left(x^{37/12}\right),
\end{align}
\begin{align}
&1+2 \sqrt[12]{x}+4 \sqrt[6]{x}+6 \sqrt[4]{x}+11 \sqrt[3]{x}+18 x^{5/12}+32 \sqrt{x}+48 x^{7/12}+74 x^{2/3}+108 x^{3/4} \nonumber \\
&+164 x^{5/6}+238 x^{11/12}+348 x+488 x^{13/12}+689 x^{7/6}+952 x^{5/4}+1321 x^{4/3}+1798 x^{17/12} \nonumber \\
&+2444 x^{3/2}+3268 x^{19/12}+4364 x^{5/3}+5760 x^{7/4}+7589 x^{11/6}+9894 x^{23/12}+12858 x^2 \nonumber \\
&+16558 x^{25/12}+21260 x^{13/6}+27090 x^{9/4}+34412 x^{7/3}+43412 x^{29/12}+54586 x^{5/2} \nonumber \\
&+68212 x^{31/12}+84959 x^{8/3}+105232 x^{11/4}+129931 x^{17/6}+159626 x^{35/12}+195479 x^3 \nonumber \\
&+O\left(x^{37/12}\right),
\end{align}
\begin{align}
&-\sqrt[6]{x}-2 \sqrt[4]{x}-5 \sqrt[3]{x}-8 x^{5/12}-15 \sqrt{x}-24 x^{7/12}-42 x^{2/3}-64 x^{3/4}-101 x^{5/6}-148 x^{11/12} \nonumber \\
&-223 x-322 x^{13/12}-469 x^{7/6}-658 x^{5/4}-927 x^{4/3}-1276 x^{17/12}-1760 x^{3/2}-2384 x^{19/12} \nonumber \\
&-3222 x^{5/3}-4288 x^{7/4}-5697 x^{11/6}-7480 x^{23/12}-9796 x^2-12696 x^{25/12}-16400 x^{13/6} \nonumber \\
&-20998 x^{9/4}-26806 x^{7/3}-33960 x^{29/12}-42882 x^{5/2}-53778 x^{31/12}-67219 x^{8/3} \nonumber \\
&-83516 x^{11/4}-103435 x^{17/6}-127406 x^{35/12}-156426 x^3+O\left(x^{37/12}\right)
\end{align}
up to $x^3$. We immediately note that their sum is exactly the same as \eqref{eq:U(5)_Coulomb}. On the other hand, the naive contribution from (e) is given by
\begin{align}
-x^3+O\left(x^4\right),
\end{align}
which will be canceled by the contributions of other non-growing trees so that we obtain the correct $x^3$ term. Hence, we only need to keep the growing trees.
\\

For higher rank gauge groups, it is not easy to perform the same comparision due to the difficulty of evaluating the integral formula of the Coulomb branch localization. Although we don't give a general proof for higher rank gauge groups that non-growing trees do not contribute to the index, we will see in the next section that this growing tree prescription is perfectly consistent with the proposed duality. A quick check is that we need to have the same number of contributing forest graphs for a duality pair in order to match the superconformal index. Adopting the growing tree prescription, we indeed find a one-to-one map between the contributing forest graphs of a duality pair. This can be easily seen using the box representation of a forest graph, which will be explained in the subsequent section shortly.

We should comment about the (anti-)vortex part as well. Once we set $\upsilon_X = x^\frac{2}{n+1}$ and $\upsilon_Y = x^\frac{n}{n+1}$, the vortex part also contains zeros and poles while it doesn't for generic $\upsilon_X, \, \upsilon_Y$. Since we have a regular expression for generic $\upsilon_X, \, \upsilon_Y$, we can take the limit $\upsilon_X \rightarrow x^\frac{2}{n+1}, \, \upsilon_Y \rightarrow x^\frac{n}{n+1}$ so that all extra zeros and poles cancel each other. The expression for the vortex part with $\upsilon_X = x^\frac{2}{n+1}, \,\upsilon_Y = x^\frac{n}{n+1}$ should be understood in this way.
\\

\subsection{Higgs Vacua} \label{Sec:Higgs_Vacua}
To provide extra evidence of our growing tree prescription, let us examine the structure of the vacuum solutions of the theory, especially the solutions for the vector multiplet real scalar $\sigma$. The full solutions including chiral fields will be discussed in appendix \ref{sec:vacua}. Considering the $U(N_c)$ gauge theory with (anti-)fundamentals and two adjoints, the vacuum solutions are determined by two sets of equations. As the FI parameter $\zeta > 0$ is turned on, the D-term equation is given by
\begin{align}
\label{eq:D}
\zeta \mathbf 1_N-\sum_{a = 1}^{N_f} Q^a Q^a{}^\dagger +\sum_{b = 1}^{N_a} \tilde Q^b{}^\dagger \tilde Q^b-[X,X^\dagger]-[Y,Y^\dagger] = 0
\end{align}
with mass term conditions
\begin{gather}
\label{eq:mass}
\begin{gathered}
(\sigma_i-m_a) Q^a_i = 0,  \qquad (-\sigma_i+\tilde m_b) \tilde Q^b_i = 0, \\
(\sigma_i-\sigma_j-m_X) X_{ij} = 0, \qquad (\sigma_i-\sigma_j-m_Y) Y_{ij} = 0.
\end{gathered}
\end{gather}
In addtion, the F-term equation is given by
\begin{align}
\label{eq:F}
XY+YX=0, \qquad X^n + Y^2= 0.
\end{align}
It will be shown that the second condition requires $n m_X = 2 m_Y$.\footnote{More precisely, the real masses of adjoints, $m_X, \, m_Y$, are restricted as follows once we turn on the superpotential $W = \mathrm{Tr} \left(X^{n+1}+X Y^2\right)$:
\begin{align}
\label{eq:discrete mass}
(n+1) m_X = 0, \qquad m_X+2 m_Y = 0,
\end{align}
which demand that $m_X = m_Y = 0$ as long as we consider a 3d theory on a flat spacetime $\mathbb R^3$. Nevertheless, if we consider a spacetime that requires periodic masses $m \sim m+\Lambda$, the condition \eqref{eq:discrete mass} allows discrete values of the masses $m_X = \frac{k}{n+1} \Lambda, \, m_Y = \frac{1}{2} (l-\frac{k}{n+1}) \Lambda$ where $k = 0, \ldots, n, \, l = 0, 1$. For example, if we consider $\mathbb R^2 \times S^1$, one can turn on constant holonomies along the circle, which lift real masses to be complexified and periodic, so that our analysis of the vacuum equations works. Note that the vortex
partition function is defined on  $\mathbb R^2 \times S^1$. When there are no superpotential terms for $X, Y$, the analysis
of the vacua is identical for $\mathbb R^3$ and $\mathbb R^2 \times S^1$.
}
\\

For simplicity let us consider $N_f = N_a = 1$; the generalization to general $N_f$ is straightforward. In order to satisfy \eqref{eq:D}, some of chiral fields should have nonzero vevs. Especially, if one takes the trace of \eqref{eq:D}, one obtains
\begin{align}
N \zeta-\sum_{i = 1}^N |Q_i|^2+\sum_{i = 1}^N |\tilde Q_i|^2 = 0
\end{align}
and sees that at least one component of $Q$ should have the nonvanishing vev. The real scalars $\sigma$'s then should be pinned to masses of those chiral fields having nonzero vevs. If there are coincident $\sigma$'s, however, the effective low energy theory contains a nonabelian factor with few massless charged fields, which leads to a runaway superpotential \cite{Affleck:1982as}. Therefore, $\sigma_i \neq \sigma_j$ for different $i, \, j$.

Since now all $\sigma$'s are distinct, from real mass terms \eqref{eq:mass}, we know that only one component of $Q$, say $i = 1$, is nonzero:
\begin{align}
Q_i = \delta_{i1} Q.
\end{align}
Furthermore, if $X_{ij}$ is nonzero, $X_{ik}$ and $X_{kj}$ should vanish for $k \neq i, \, j$. The commutator $[X,X^\dagger]$ is thus a diagonal matrix:
\begin{align}
[X,X^\dagger]_{ij} = \sum_k X_{ik} X_{jk}^*-X_{ki}^* X_{kj} = \delta_{ij} \left(|X_{is}|^2-|X_{ti}|^2\right)
\end{align}
where $X_{is}$ ($X_{ti}$) is the unique nonzero element in the $i$-th row (column) if it exists. Also note that $X_{ii} = 0$. Those conditions are satisfied by $Y$ as well. As a result, only the diagonal components of the D-term equation \eqref{eq:D} are nontrivial:
\begin{align}
\label{eq:diagD}
\zeta-\delta_{i1} |Q|^2-|X_{is}|^2+|X_{ti}|^2-|Y_{is'}|^2+|Y_{t'i}|^2 = 0.
\end{align}
For each $i \neq 1$, $\zeta$ must be compensated either by $X_{is}$ or by $Y_{is'}$. Such nonzero element $X_{is}$ (or $Y_{is'}$) then determines each and every $\sigma_i$ by \eqref{eq:mass}.

Indeed, those conditions ensure that the solutions for $\sigma$'s are written in the following form: $\sigma_i = m+a_i m_X+b_i m_Y$ where $m$ is the mass of $Q$. The solutions are labeled by distinct non-negative integer pairs $(a_i,b_i)$ that satisfy the following conditions. First, $\sigma_1$ should be $m$ because we choose the first component of $Q$ to be nonzero; i.e., $(a_1,b_1) = (0,0)$. Second, for any $i \neq 1$ with $(a_i,b_i)$, there must exist $j$ with $(a_j,b_j)$ such that $a_i-a_j = 1$ or that $b_i-b_j = 1$. Since $\sigma$'s are all distinct, it is convenient to represent those integer pairs by box diagrams; e.g., for $N_c = 3$,
\begin{align}
\ytableausetup{boxsize=0.9cm}
\begin{ytableau}
$(2,0)$ \\
$(1,0)$ \\
$(0,0)$
\end{ytableau} \qquad
\begin{ytableau}
\none \\
$(1,0)$ \\
$(0,0)$ & $(0,1)$
\end{ytableau} \qquad
\begin{ytableau}
\none \\
$(1,0)$ & $(1,1)$ \\
$(0,0)$
\end{ytableau} \qquad
\begin{ytableau}
\none \\
\none & $(1,1)$ \\
$(0,0)$ & $(0,1)$
\end{ytableau} \qquad
\begin{ytableau}
\none \\
\none \\
$(0,0)$ & $(0,1)$ & $(0,2)$
\end{ytableau}
\end{align}
where corresponding $(a_i,b_i)$ is written in each box. Note that each box would have carried another label $i$, which is irrelevant due to the Weyl symmetry permuting $\sigma$'s among themselves.
\\

We have considered the D-term equation so far and now move on to the F-term equation \eqref{eq:F}. Let us examine the first equation: $X Y+Y X = 0$. Since only one element is nontrivial in each row and column of $X$ and $Y$, the equation reduces to
\begin{align}
\label{eq:F1}
\sum_k X_{ik} Y_{kj}+\sum_l Y_{il} X_{lj} = X_{is} Y_{sj}+Y_{is'} X_{s'j} = 0.
\end{align}
If $X_{is} Y_{sj} \neq 0$, then $\sigma_i = \sigma_s+m_X$ and $\sigma_s = \sigma_j+m_Y$, which are represented by boxes stacked in the following way:
\begin{align}
\label{eq:illegal boxes}
\ytableausetup{boxsize=1.2cm}
\begin{ytableau}
\none[\cdots] & \none & (a_i,b_i) & \none \\
\none & (a_j,b_j) & (a_s,b_s) & \none \\
\none & \none & \none[\vdots] & \none
\end{ytableau}
\end{align}
where
\begin{gather}
\begin{gathered}
(a_s,b_s) = (a_j,b_j+1), \\
(a_i,b_i) = (a_s+1,b_s) = (a_j+1,b_j+1).
\end{gathered}
\end{gather}
In this case, $Y_{is'} X_{s'j} = -X_{is} Y_{sj}$ is nonzero as well. Thus, there must be $\sigma_{s'}$ that satisfies $\sigma_i = \sigma_{s'}+m_Y$ and $\sigma_{s'} = \sigma_j+m_X$. Thus, the boxes stacked as in \eqref{eq:illegal boxes} cannot be complete; there should be $(a_{s'},b_{s'})$ such that the allowed box diagram contains
\begin{align}
\ytableausetup{boxsize=1.2cm}
\begin{ytableau}
\none[\cdots] & (a_i,b_i) & (a_i,b_i) & \none \\
\none & (a_j,b_j) & (a_s,b_s) & \none \\
\none & \none & \none[\vdots] & \none
\end{ytableau}
\end{align}
where
\begin{gather}
\begin{gathered}
(a_s,b_s) = (a_j,b_j+1), \\
(a_{s'},b_{s'}) = (a_j+1,b_j), \\
(a_i,b_i) = (a_s+1,b_s) = (a_{s'},b_{s'}+1) = (a_j+1,b_j+1).
\end{gathered}
\end{gather}
For the same reason, the boxes of shape \ytableausetup{smalltableaux}\ydiagram{2,1} \, are not allowed. Therefore, the first equation of \eqref{eq:F} tells us that the solutions for $\sigma$'s are labeled by Young tableaux. One should note that whenever $\sigma_i$ satisfies \eqref{eq:mass} such that $X_{ij}$ or $Y_{ij}$ can be nonzero, then such $X_{ij}$ or $Y_{ij}$ should be nonzero; otherwise, equation \eqref{eq:diagD} and \eqref{eq:F1} cannot be satisfied simultaneously.

The second equation then imposes an extra condition that the allowed Young tableaux should satisfy. We mentioned that the adjoint masses satisfy $n m_X = 2 m_Y$. This can be shown as follows. Since only a single element can be nontrivial in each row and column, again, the second equation reduces to
\begin{align}
\sum_{k_1,\ldots,k_{n-1}} X_{ik_{n-1}} X_{k_{n-1} k_{n-2}} \cdots X_{k_1j}+\sum_l Y_{il} Y_{l j} = X_{is_{n-1}} \cdots X_{s_2 s_1} X_{s_1j}+Y_{is'} Y_{s'j} = 0.
\end{align}
This equation is satisfied in two ways:  $X^n = -Y^2 \neq 0$ or $X^n = Y^2 = 0$. In the former case, the condition $\left[X^n\right]_{ij} = X_{is_{n-1}} \cdots X_{s_2 s_1} X_{s_1j} \neq 0$ requires that $\sigma$'s should satisfy
\begin{align}
\begin{aligned}
\sigma_i &= \sigma_{s_{n-1}}+m_X, \\
\sigma_{s_p} &= \sigma_{s_{p-1}}+m_X, \qquad 2 \leq p \leq n-1, \\
\sigma_{s_1} &= \sigma_j+m_X.
\end{aligned}
\end{align}
In the same manner, the condition $\left[Y^2\right]_{ij} = Y_{is'} Y_{s'j} \neq 0$ requires that
\begin{align}
\begin{aligned}
\sigma_i &= \sigma_{s'}+m_Y, \\
\sigma_{s'} &= \sigma_j+m_Y.
\end{aligned}
\end{align}
From those relations, we observe that $\sigma_i-\sigma_j = n m_X = 2 mY$. In other words, $\sigma_i = m+n m_X$ and $\sigma_j = m+2 m_Y$ are indistinguishable; i.e., $(a_i,b_i) = (n,0) \sim (0,2)$. When we represent such a solution by a box diagram, we are thus able to use both of the diagrams:
\begin{align}
\ytableausetup{boxsize=1.1cm}
\begin{ytableau}
$(n,0)$ & \none & \none \\
$(n-1,0)$ & \none & \none \\
\none[\vdots] & \none & \none \\
$(0,0)$ & $(0,1)$ & \none
\end{ytableau} \qquad
\begin{ytableau}
\none & \none & \none \\
$(n-1,0)$ & \none & \none \\
\none[\vdots] & \none & \none \\
$(0,0)$ & $(0,1)$ & $(0,2)$
\end{ytableau}
\end{align}
among which we are going to use the right one for the subsequent discussion. Accordingly the heights of the Young tableaux we are using are restricted up to $n$.

Furthermore, for odd $n$, combining the two equations in \eqref{eq:F}, one can see that
\begin{align}
Y^3 = Y (-X^n) = X^n Y = -Y^3,
\end{align}
which implies that $\left[Y^3\right]_{ij} = Y_{is} Y_{st} Y_{tj} = 0$. This condition demands that the solution should not contain more than three $\sigma$'s that are connected by $m_Y$. In other words, $b_i$ must be less than 3. Hence, the maximum size of the allowed Young tableaux is $n \times 3$. Lastly if the solution contains $(a_i,b_i) = (0,2)$, it should contain both $(0,1)$ and $(n-1,0)$ because $(0,2) \sim (n-1,0)$.

In the latter case, i.e., $X_{is_{n-1}} \cdots X_{s_2 s_1} X_{s_1j} = Y_{is'} Y_{s'j} = 0$, a solution for $\sigma$ satisfies that $a_i < n$ and that $b_i < 2$. Hence, the Young tableaux contain at most two columns in this case. Combining the two cases, the general rule finding the allowed box diagrams is as follows: the allowed box diagrams are Young tableaux of the maximum size $n \times 3$ which satisfy that if the third column is non-empty, the first column is always fully occupied.
\\

From this rule, we can count the number of allowed Young tableaux, or equivalently that of Higgs vacua, which is given by
\begin{align}
\label{eq:nov0}
p(n,2;N_c)+p(n-1,2;N_c-n-2)
\end{align}
where $p(n,m;N)$ denotes the number of partitions of $N$ into at most $m$ integers which are not greater than $n$. The first term counts the solutions satisfying $X^n = Y^2 = 0$ while the second term counts those satisfying $X^n = -Y^2 \neq 0$.

Now let us evaluate $p(n,m;N)$ for $m = 2$. When $0 \leq N \leq n$, every partition that divides $N$ into at most two parts is allowed. The number of such partitions is as follows:
\begin{align}
p(n,2;N) = \left\lfloor\frac{N}{2}\right\rfloor+1.
\end{align}
When $n+1 \leq N \leq 2 n$, one simply finds that
\begin{align}
p(n,2;N) = p(n,2;2 n-N) = \left\lfloor\frac{2 n-N}{2}\right\rfloor+1,
\end{align}
which is counting the empty positions among $n \times 2$ available slots in a given Young tableau.
$p(n,2;N)$ vanishes otherwise. Each term in \eqref{eq:nov0} is then given by
\begin{align}
p(n,2;N_c) &= \left\{\begin{array}{cc}
\left\lfloor\frac{N_c}{2}\right\rfloor+1, & 0 \leq N_c \leq n \\
\left\lfloor\frac{2 n-N_c}{2}\right\rfloor+1, & n+1 \leq N_c \leq 2 n \\
0, & 2 n+1 \leq N_c
\end{array}\right., \\
p(n-1,2;N_c-n-2) &= \left\{\begin{array}{cc}
0, & 0 \leq N_c \leq n+1, \\
\left\lfloor\frac{N_c-n-2}{2}\right\rfloor+1, & n+2 \leq N_c \leq 2 n+1 \\
\left\lfloor\frac{3 n-N_c}{2}\right\rfloor+1, & 2 n+2 \leq N_c \leq 3 n \\
0, & 3 n+1 \leq N_c
\end{array}\right..
\end{align}
Their sum is therefore
\begin{align}
p(n,2;N_c)+p(n-1,2;N_c-n-2) &= \left\{\begin{array}{cc}
\left\lfloor\frac{N_c}{2}\right\rfloor+1, & 0 \leq N_c \leq n \\
\left\lfloor\frac{2 n-N_c}{2}\right\rfloor+1, & N_c = n+1 \\
\left\lfloor\frac{2 n-N_c}{2}\right\rfloor+\left\lfloor\frac{N_c-n-2}{2}\right\rfloor+2, & n+2 \leq N_c \leq 2 n \\
\left\lfloor\frac{N_c-n-2}{2}\right\rfloor+1, & N_c = 2 n+1 \\
\left\lfloor\frac{3 n-N_c}{2}\right\rfloor+1, & 2 n+2 \leq N_c \leq 3 n \\
0, & 3 n+1 \leq N_c
\end{array}\right..
\end{align}
While the right hand side looks complicated, indeed it is simplified to $\frac{n+1}{2}$ for $n+1 \leq N_c \leq 2 n+1$ because $n$ is odd. Thus, the final result is that
\begin{align}
\label{eq:nov1}
p(n,2;N_c)+p(n-1,2;N_c-n-2) &= \left\{\begin{array}{cc}
\left\lfloor\frac{N_c}{2}\right\rfloor+1, & 0 \leq N_c \leq n \\
\frac{n+1}{2}, & n+1 \leq N_c \leq 2 n+1 \\
\left\lfloor\frac{3 n-N_c}{2}\right\rfloor+1, & 2 n+2 \leq N_c \leq 3 n \\
0, & 3 n+1 \leq N_c
\end{array}\right..
\end{align}

So far we have considered the solutions for $\sigma$. Indeed, this Young tableau classification of $\sigma$'s vev is expected to have a one-to-one map to the growing trees we found in the previous section. This one-to-one map is confirmed for a few low $n$ cases, which we will illustrate in the next section shortly.
\\

\section{Test of Dualities}
\label{sec:test}

\subsection{$N_f = 1$}

So far we have explained how to obtain the factorized form of the superconformal index for a theory with two adjoints. In this section, using this factorized index, we perform a test of the 3d version of the Brodie duality, especially for the D(odd)-type superpotential. Comparing the vortex partition functions of a dual pair, we also find a list of monopole operators showing up in IR physics.

Let us first consider the single flavor case with $n = 3$. Since there is only one flavor, each forest graph includes a single tree, which should be a growing tree as we explained. The list of growing trees is given in appendix \ref{sec:growing trees}. We also list the numbers of growing trees for a few low $n$ in table \ref{tab:growing trees}.
\begin{table}[tbp]
\centering
\begin{tabular}{|c|c|c|c|}
\hline
gauge group & \multicolumn{3}{|c|}{\# of growing trees} \\ \cline{2-4}
 & $n = 1$ & $n = 3$ & $n = 5$ \\
\hline
$U(1)$ & 1 & 1 & 1 \\
$U(2)$ & 1 & 2 & 2 \\
$U(3)$ & 1 & 2 & 2 \\
$U(4)$ & - & 2 & 3 \\
$U(5)$ & - & 2 & 3 \\
$U(6)$ & - & 2 & 3 \\
$U(7)$ & - & 2 & 3 \\
$U(8)$ & - & 1 & 3 \\
$U(9)$ & - & 1 & 3 \\
$U(10)$ & - & - & 3 \\
$U(11)$ & - & - & 3 \\
$U(12)$ & - & - & 2 \\
$U(13)$ & - & - & 2 \\
$U(14)$ & - & - & 1 \\
$U(15)$ & - & - & 1 \\
\hline
\end{tabular}
\caption{\label{tab:growing trees} The number of growing trees. One can see exact matches for duality pairs between $U(N_c)$ and $U(3 n-N_c)$ with $N_f = 1$.}
\end{table}
One can see that the number of growing trees is exactly the same as that of the Higgs vacua we found in the previous section. As we expect the one-to-one correspondence between the growing trees and the Higgs vacua, indeed the number of growing trees for general $n$, from that of Higgs vacua \eqref{eq:nov1}, is also given by
\begin{align}
\label{eq:nov2}
\begin{array}{rl}
\left[\frac{N_c}{2}\right]+1, \qquad & 1 \leq N_c \leq n-2, \\
\frac{n+1}{2}, \qquad & n-1 \leq N_c \leq 2 n+1, \\
\left[\frac{3 n-N_c}{2}\right]+1, \qquad & 2 n+2 \leq N_c \leq 3 n.
\end{array}
\end{align}
Note that this is consistent with the proposed duality for $N_f = 1$:
\begin{align}
U(N_c) \leftrightarrow U(3 n-N_c).
\end{align}
From table \ref{tab:growing trees}, or from equation \eqref{eq:nov2} for general $n$, one can immediately see that a dual pair have exactly the same number of growing trees, and equivalently the same number of Higgs vacua.

\begin{figure}[tbp]
\centering 
\begin{subfigure}{.3\textwidth}
\centering
\begin{tikzpicture}[grow'=up,<-,>=stealth',level/.style={sibling distance = 2cm/#1,level distance = 1cm}]
\node [arn_f] {$i_1$}
    child{ node [arn_x] {$i_2$}
    }
    child{ node [arn_y] {$i_3$}
        child{ node [arn_x] {$i_4$}
        }
        child[missing]
    }
;
\end{tikzpicture}
\caption{}
\end{subfigure}
\begin{subfigure}{.3\textwidth}
\centering
\begin{tikzpicture}[grow'=up,<-,>=stealth',level/.style={sibling distance = 2cm/#1,level distance = 1cm}]
\node [arn_f] {$i_1$}
    child{ node [arn_x] {$i_2$}
        child[missing]
        child{ node [arn_y] {$i_4$}
        }
    }
    child{ node [arn_y] {$i_3$}
    }
;
\end{tikzpicture}
\caption{}
\end{subfigure}
\caption{\label{fig:U(4)tree} The forest graphs corresponding to the box diagram \eqref{eq:U(4)box}}
\end{figure}
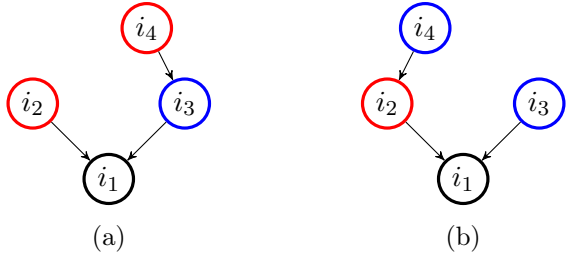

In order to compare the indices of a dual pair, we have to identify the mapping among growing trees. For $N_c = 0, 1, 8, 9$, the map is manifest because there is only a single growing tree. On the other hand, for general $N_c$, we first notice that the growing trees can be represented by stacked boxes of the maximum height $n$ and the maximum width 3 once we set $\upsilon_X = x^\frac{2}{n+1}, \, \upsilon_Y = x^\frac{n}{n+1}$, which are exactly the Young tableaux that we used to classify the Higgs vacua of the theory. For example, the growing trees of $U(2)$ shown in figure \ref{fig:U(2)} can be represented by the following stacked boxes:
\begin{align}
\ytableausetup{boxsize=normal}
\ydiagram{3+0,1,1} \qquad \ydiagram{3+0,0,2}
\end{align}
where boxes are stacked from the left-bottom conner. The box at the left-bottom corner corresponds to the black colored root node. A red circle of a tree is then translated into a box stacked on the top while a blue circle is translated into a box stacked on the right. Note that here we don't specify the labeling $i_j$ of boxes because the final result \eqref{eq:fact} is invariant under permutations of $i_j$ such that it doesn't depend on the labeling. Thus, $N_c!$ possible permutations all give the same contributions, which cancel the Weyl factor $N_c!$. In addition, given a box diagram, there would be multiple tree graphs depending on how one connects tree nodes by arrows. For example, $U(4)$ has a box diagram:
\begin{align}
\label{eq:U(4)box}
\ydiagram{3+0,2,2}
\end{align}
whose corresponding tree graphs are shown in figure \ref{fig:U(4)tree}.
Those multiple graphs are indeed all equivalent and correspond to a single singularity as we discussed in the previous section. Recall the perturbative part and the vortex part given in \eqref{eq:pert} and \eqref{eq:vortex}. While the independence of the perturbative part to the choice of arrow connections is manifest, that of the vortex part is less obvious because the ranges of vortex charges $n_j$ look different. For figure \ref{fig:U(4)tree} as an example, recall that $n_4$ is defined by \eqref{eq:vorticity} such that
\begin{align}
\begin{aligned}
n_{i_4} &= \mathsf n_{i_1}+\mathsf n_{i_3}+\mathsf n_{i_4} > n_{i_3} = \mathsf n_{i_1}+\mathsf n_{i_3} \qquad \text{for (a)}, \\
n_{i_4} &= \mathsf n_{i_1}+\mathsf n_{i_2}+\mathsf n_{i_4} > n_{i_2} = \mathsf n_{i_1}+\mathsf n_{i_2} \qquad \text{for (b)}.
\end{aligned}
\end{align}
Depending on which one is larger between $n_{i_2}$ and $n_{i_3}$, the allowed ranges of $n_{i_4}$ look different for (a) and (b). However, from the vortex part \eqref{eq:vortex}, especially the factor
\begin{align}
\label{eq:vanishing factor}
\left[t_{c_i}^{-1} t_{c_j} \upsilon_A \upsilon_i^{-1} \upsilon_j x^{-2 n_i};x^2\right]_{n_j},
\end{align}
one can see that the nontrivial contributions only come from the range
\begin{align}
n_{i_4} \geq \mathrm{Max}(n_{i_2},n_{i_3})
\end{align}
in both cases.\footnote{$[a;q]_n$ is defined in \eqref{eq:shiftedq}.} For (a), taking $i = i_4$, $j = i_2$ and $A = Y$, \eqref{eq:vanishing factor} becomes
\begin{align}
\left[x^{-2 n_{i_4}};x^2\right]_{n_{i_2}},
\end{align}
which doesn't vanish only when $n_{i_4} \geq n_{i_2}$. Similarly, taking $i = i_4$, $j = i_3$ and $A = X$ for (b), we have
\begin{align}
\left[x^{-2 n_{i_4}};x^2\right]_{n_{i_3}},
\end{align}
which again doesn't vanish only when $n_{i_4} \geq n_{i_3}$. Thus, $n_{i_4}$ must be larger than or equal to both of $n_{i_2}$ and $n_{i_3}$. Now we have a one-to-one correspondence between $n_{i_4}$ in (a) and that in (b), with which the vortex parts for (a) and (b) agree. Indeed, this argument doesn't depend on how the $i_2$-node and the $i_3$-node are connected to their parent nodes. Therefore, in general, given a box diagram, one can reconstruct a tree graph with any convenient choice of arrow connections between adjacent boxes as long as the parent node of any node is unique. For this reason, in this section, we use the box representation without specifying the arrow connections.
\\

Let us go back to the $U(2)$ theory and its dual. The growing trees of the dual gauge group $U(7)$ is listed in appendix \ref{sec:growing trees}. For a dual gauge group, we define the box representation of a growing tree in a similar way but also rotate the stacked boxes 180\degree. In other words, boxes are stacked from the right-top conner, a red circle is translated into a box stacked on the bottom, and a blue circle is translated into a box stacked on the left. Defining the box representation in this way, we have found that a growing tree and its dual growing tree together always form an $n \times 3$ stack of a rectangular shape. For example, the dual partner of the first $U(2)$ growing tree is given by
\begin{align}
\ydiagram{3,3,3} \quad - \quad \ydiagram{3+0,1,1} \quad = \quad \ydiagram{3,1+2,1+2}
\end{align}
which corresponds to one of the growing trees of $U(7)$. The other dual pairs satisfy the same relation regardless of the gauge group rank.

Upon this duality map of growing trees, now one can compare the perturbative parts and the vortex parts of a dual pair. The perturbative part will be discussed in the following subsection for general $N_f$. In this subsection, let us focus on the vortex part. For $N_f = 1, \, n = 3$, we have found that the vortex partition functions satisfy the following relation
\begin{align}
& Z_\text{vortex}^{\mathsf t}/\hat Z_\text{vortex}^{\mathsf t^c} = \nonumber \\
& \mathrm{PE} \left[\frac{\left(1+x^\frac{1}{2}+x^\frac{3}{4}+x+x^\frac{3}{2}\right) \left(x^{-\frac{1}{4} N_c+\frac{5}{4}} \tau^{-1}-x^{\frac{1}{4} N_c-\frac{3}{4}} \tau\right) w}{1-x^2}+\frac{\left(x^{-\frac{1}{2} N_c+3} \tau^{-2}-x^{\frac{1}{2} N_c-1} \tau^2\right) w^2}{1-x^2}\right]
\end{align}
for any growing tree $\mathsf t$ and its dual tree $\mathsf t^c$.\footnote{We expect that this gives rise to the
half index of
the singlet states (or vortex partition function) in the dual theory, corresponding to monopole operators of the original theory. Thus from this expression
 one can read off the states of the monopole operators, which can be compared with the full superconformal index of the
 monopole opertors. The superconformal index can be obtained as the product of the two half-indices as usual.} The first term in the plethystic exponential corresponds to the contribution of monopole operators of monopole charge 1, which are $V^\pm_{s,t}$ for $s = 0,\ldots,n-1, \, t = 0,1,2$ and $s t = 0$. The second term, on the other hand, corresponds to the contribution of monopole operators of monopole charge 2, which we denote by $W_0^\pm$.
\\

To understand the appearance of those charge-2 monopole operators $W_0^\pm$, let us trace their counterparts in the Hilbert space on $S^2$. A monopole operator relates to a state on $S^2$ with nontrivial magnetic flux background by the radial quantization. By observing the global charges of $W^\pm_0$, it is immediately seen that possible candidates are
\begin{align}
\label{eq:W_candidates}
\mathrm{Tr}_1 X \left|\pm2,0,\ldots\right>, \qquad \mathrm{Tr}_2 X \left|\pm2,0,\ldots\right>, \qquad \left|\pm1,\pm1,\ldots\right>,
\end{align}
which carry exactly the same global charges as $W^\pm_0$. For the first two states, the gauge group is broken to $U(1) \times U(N_c-1)$ due to nonzero magnetic flux. Each trace above, $\mathrm{Tr}_1$ or $\mathrm{Tr}_2$, is taken over each unbroken factor, $U(1)$ or $U(N_c-1)$, respectively.

We suspect that the existence of charge-2 monopole operators is generic for theories with two adjoints regardless of the superpotential. Thus, we don't stick to the superpotential \eqref{eq:sup_D} and instead consider the superconformal indices of theories without the superpotential. For example, let us consider the $U(2)$ theory with $N_f = 1$. The superconformal index can be computed by either ways: the integral formula \eqref{eq:Coulomb} or the factorized index formula \eqref{eq:fact}. For convenience we assign the trial $R$-charges $\frac{1}{4}$, $\frac{2}{n+1} = \frac{1}{2}$ and $\frac{n}{n+1} = \frac{3}{4}$ for the (anti-)fundamentals and the two adjoints respectively. Since we turn off the superpotential, there are global symmetries $U(1)_X \times U(1)_Y$ rotating each adjoint field. We thus restore the fugacities $\upsilon_X$ and $\upsilon_Y$ for those global symmetries.

The chiral primaries can be read off from the superconformal index by taking the plethystic logarithm, which is the inverse operation of PE. For $U(2)$ with $N_f = 1$, PL of the index is given by
\begin{align}
\label{eq:U(2)_generic}
\mathrm{PL} [I] &= x^\frac{1}{2} \left(\tau^2+\upsilon_X\right)+x^\frac{3}{4} \upsilon_Y+x \left(\tau^2 \upsilon_X+\upsilon_X^2\right)+x^\frac{5}{4} \left(-\upsilon_Y^{-1}+\tau^2 \upsilon_Y+\upsilon_X \upsilon_Y\right) \nonumber \\
& +x^\frac{3}{2} \left(-\upsilon_X^{-1}-\upsilon_X^{-2} \upsilon_Y^{-2}-\tau^{-2} \upsilon_X^{-1} \upsilon_Y^{-2}+\upsilon_Y^2\right) \nonumber \\
& +w^\pm \left(x^\frac{1}{2} \tau^{-1} \upsilon_X^{-1} \upsilon_Y^{-1}+x \tau^{-1} \upsilon_Y^{-1}+x^\frac{5}{4} \tau^{-1} \upsilon_X^{-1}-x^\frac{3}{2} \tau \upsilon_Y^{-1}\right)
+w^{\pm 2} x^\frac{3}{2} \tau^{-2}
+\mathcal O(x^\frac{7}{4}).
\end{align}
Looking at bosonic contributions, one can recognize the following chiral operators:
\begin{align}
M_{0,0}, \quad \mathrm{Tr} X, \quad \mathrm{Tr} Y, \quad M_{1,0}, \quad \mathrm{Tr} X^2, \quad M_{0,1}, \quad \mathrm{Tr} XY, \quad \mathrm{Tr} Y^2, \quad V_{0,0}^\pm, \quad V_{1,0}^\pm, \quad V_{0,1}^\pm, \quad W^\pm_0
\end{align}
up to $x^\frac{3}{2}$. One should note that $W^\pm_0$ is not charged under $U(1)_X \times U(1)_Y$. An advantage of considering a theory without the superpotential is that the first two and last states in \eqref{eq:W_candidates} are distinguished by those $U(1)^2$  global charges. For a bare monopole state $\left|m_1,m_2\right>$, the $U(1)^2$ global charges are given by
\begin{align}
F_X = F_Y = -\frac{1}{2} \sum_{i,j = 1}^{2} |m_i-m_j|.
\end{align}
The first two states in \eqref{eq:W_candidates} thus carry the charges (-1,-2) and can be written in terms of $V_{0,1}^\pm$ and $\mathrm{Tr} X$:
\begin{align}
\begin{aligned}
\label{eq:V2}
\mathrm{Tr}_1 X \left|\pm2,0\right> &\sim \alpha_1 (V_{0,0}^\pm)^2 \mathrm{Tr} X+\beta_1 V_{0,0}^\pm V_{1,0}^\pm, \\
\mathrm{Tr}_2 X \left|\pm2,0\right> &\sim \alpha_2 (V_{0,0}^\pm)^2 \mathrm{Tr} X+\beta_2 V_{0,0}^\pm V_{1,0}^\pm.
\end{aligned}
\end{align}
On the other hand, the last state in \eqref{eq:W_candidates} carries charges (0,0) and can be identified with $W^\pm_0$:
\begin{align}
\label{eq:W}
\left|\pm1,\pm1\right> \sim W^\pm_0.
\end{align}
This suggests that the charge-2 monopole operator shouldn't appear in a $U(1)$ theory. Indeed, we have checked that there is no independent monopole operator carrying topological charge 2 in the $U(1)$ theory with $N_f = 1$.

For the nonzero superpotential \eqref{eq:sup_D}, the state/operator identifications \eqref{eq:V2} and \eqref{eq:W} are not distinguished by their global charges and can be mixed. There still exist independent charge-2 monopole operators, though.
\\

This comparison can be made for higher $n$ as well. We have found that a dual pair of vortex partition functions satisfy
\begin{align}
\label{eq:higher n}
&\quad Z_\text{vortex}^{\mathsf t}/\hat Z_\text{vortex}^{\mathsf t^c} \nonumber \\
&= \mathrm{PE} \left[\frac{\left(\sum_{s = 0}^{n-1} x^\frac{2 s}{n+1}+x^\frac{n}{n+1}+x^\frac{2 n}{n+1}\right) \left(x^{-\frac{1}{n+1} N_c+\frac{n+2}{n+1}} \tau^{-1}-x^{\frac{1}{n+1} N_c-\frac{n}{n+1}} \tau\right) w}{1-x^2}\right] \nonumber \\
&\quad \times \mathrm{PE} \left[\frac{\left(\sum_{u = 0}^\frac{n-3}{2} x^\frac{4 u}{n+1}\right) \left(x^{-\frac{2}{n+1} N_c+\frac{2 n+6}{n+1}} \tau^{-2}-x^{\frac{2}{n+1} N_c-\frac{2 n-2}{n+1}} \tau^2\right) w^2}{1-x^2}\right].
\end{align}
From this, we expect that the independent monopole operators are give by
\begin{align}
\begin{array}{ll}
V^\pm_{s,t}, \qquad & s = 0, \ldots, n-1, \quad t = 0, 1, 2, \quad s t = 0, \\
W^\pm_u, \qquad & u = 0, \ldots, \frac{n-3}{2}  \label{eq:monop}
\end{array}
\end{align}
for high enough gauge ranks. Those operators are mapped to the following monopole states on $S^2$:
\begin{align}
\begin{aligned}
\label{eq:monopoles1}
V^\pm_{s,t} &\sim \mathrm{Tr} X^s Y^t \left|1,0,0,\ldots,0\right>, \\
W^\pm_u &\sim \mathrm{Tr} X^{2 u}\left|1,1,0\ldots,0\right>
\end{aligned}
\end{align}
by radial quantization as we will explain at the appendix \ref{sec:monopole}. Note that there are more monopole operators of topological charge 2 as we
consider the higher $n$ case.
\\

\subsection{$N_f > 1$}

For $N_f > 1$, a forest graph now contains multiple trees, each of which should be one of the growing trees allowed in the $N_f = 1$ case. Any union of growing trees can appear as long as the total number of the nodes is kept to be the rank of the gauge group. For example, for the $U(3)$ theory with $N_f = 2$ and $n = 3$, we have the following list of the box representations of growing forests:
\begin{align}
\label{eq:forests}
\begin{array}{ll}
\text{Forest 1}: \qquad & $\ydiagram{1,1,1}$ \\
\\
\text{Forest 2}: \qquad & $\ydiagram{3+0,1,2}$ \\
\\
\text{Forest 3}: \qquad & $\ydiagram{3+0,1,1} \quad \ydiagram{3+0,3+0,1}$ \\
\\
\text{Forest 4}: \qquad & $\ydiagram{3+0,3+0,2} \quad \ydiagram{3+0,3+0,1}$
\end{array}
\end{align}
and their permutations among trees. In the previous subsection, we have identified the duality map for growing trees, which can be manifestly extended to general forest graphs for $N_f > 1$. In the current example, the dual theory is given by $U(3 n N_f-N_c) = U(15)$. Its growing forest graphs are then expressed in the box representation as follows:
\begin{align}
\begin{array}{ll}
\text{Forest 1}: \qquad & $\ydiagram{1+2,1+2,1+2} \quad \ydiagram{3,3,3}$ \\
\\
\text{Forest 2}: \qquad & $\ydiagram{3,1+2,2+1}$ \quad \ydiagram{3,3,3} \\
\\
\text{Forest 3}: \qquad & $\ydiagram{3,1+2,1+2} \quad \ydiagram{3,3,1+2}$ \\
\\
\text{Forest 4}: \qquad & $\ydiagram{3,3,2+1} \quad \ydiagram{3,3,1+2}$ \\
\end{array}
\end{align}
which are indeed the forest graphs dual to \eqref{eq:forests}. In general, a growing forest graph is defined as an ordered set of $N_f$ growing trees, $\mathsf f = (\mathsf t_1,\ldots,\mathsf t_{N_f})$, and its dual forest graph is given by $\mathsf f^c = (\mathsf t_1^c,\ldots,\mathsf t_{N_f}^c)$. Upon this duality map, we again compare the vortex partition functions for $N_f > 1$ and find
\begin{align}
&\quad Z_\text{vortex}^{\mathsf f}/\hat Z_\text{vortex}^{\mathsf f^c} \nonumber \\
&= \mathrm{PE} \left[\frac{\left(\sum_{s = 0}^{n-1} x^\frac{2 s}{n+1}+x^\frac{n}{n+1}+x^\frac{2 n}{n+1}\right) \left(x^{N_f-\frac{1}{n+1} N_c+\frac{1}{n+1}} \tau^{-N_f}-x^{-N_f+\frac{1}{n+1} N_c+\frac{1}{n+1}} \tau^{N_f}\right) w}{1-x^2}\right] \nonumber \\
&\quad \times \mathrm{PE} \left[\frac{\left(\sum_{u = 0}^\frac{n-3}{2} x^\frac{4 u}{n+1}\right) \left(x^{2 N_f-\frac{2}{n+1} N_c+\frac{4}{n+1}} \tau^{-2 N_f}-x^{-2 N_f+\frac{2}{n+1} N_c+\frac{4}{n+1}} \tau^{2 N_f}\right) w^2}{1-x^2}\right],
\end{align}
which reproduces the extra operators on the dual side expected from the analysis in section \ref{sec:review}. See table \ref{tab:3d}. This is numerically checked up to $w^4$ for several low values of $N_f$ and $n$. Note that $Z_\text{vortex}^{\mathsf f}$ and $\hat Z_\text{vortex}^{\mathsf f^c}$ now depend on the $SU(N_f) \times SU(N_f)$ flavor fugacities $t_a, \, \tilde t_a$ while the right hand side is independent of them as expected.
\\

So far we have discussed the vortex partition function part. One can also show that the perturbative parts agree under the duality. Since now we are using the box representation, the perturbative part can be expressed as follows:
\begin{align}
\label{eq:pert0}
& Z_\text{pert}^{\mathsf f}(x,t,\tilde t,\tau) = \nonumber \\
& \left(\prod_{\mathsf b,\mathsf c \in \mathsf f} \frac{\left(t_{\mathsf k(\mathsf b)} t_{\mathsf k(\mathsf c)}^{-1} x^{-\frac{n \mathsf j(\mathsf b)+2 \mathsf i(\mathsf b)}{n+1}+\frac{n \mathsf j(\mathsf c)+2 \mathsf i(\mathsf c)}{n+1}};x^2\right)'_\infty}{\left(t_{\mathsf k(\mathsf c)} t_{\mathsf k(\mathsf b)}^{-1} x^{-\frac{n \mathsf j(\mathsf c)+2 \mathsf i(\mathsf c)}{n+1}+\frac{n \mathsf j(\mathsf b)+2 \mathsf i(\mathsf b)}{n+1}+2};x^2\right)_\infty}\right)
\left(\prod_{A = X,Y} \prod_{\mathsf b,\mathsf c \in \mathsf f} \frac{\left(t_{\mathsf k(\mathsf b)}^{-1} t_{\mathsf k(\mathsf c)} \upsilon_A^{-1} x^{\frac{n \mathsf j(\mathsf b)+2 \mathsf i(\mathsf b)}{n+1}-\frac{n \mathsf j(\mathsf c)+2 \mathsf i(\mathsf c)}{n+1}+2};x^2\right)_\infty}{\left(t_{\mathsf k(\mathsf c)} t_{\mathsf k(\mathsf b)}^{-1} \upsilon_A x^{-\frac{n \mathsf j(\mathsf c)+2 \mathsf i(\mathsf c)}{n+1}+\frac{n \mathsf j(\mathsf b)+2 \mathsf i(\mathsf b)}{n+1}};x^2\right)'_\infty}\right) \nonumber \\
& \times \left(\prod_{a = 1}^{N_f} \prod_{\mathsf b \in \mathsf f} \frac{\left(t_{\mathsf k(\mathsf b)}^{-1} t_a x^{\frac{n \mathsf j(\mathsf b)+2 \mathsf i(\mathsf b)}{n+1}-\frac{n+2}{n+1}+2};x^2\right)_\infty}{\left(t_{\mathsf k(\mathsf b)} t_a^{-1} x^{-\frac{n \mathsf j(\mathsf b)+2 \mathsf i(\mathsf b)}{n+1}+\frac{n+2}{n+1}};x^2\right)'_\infty}\right)
\left(\prod_{b = 1}^{N_f} \prod_{\mathsf b \in \mathsf f} \frac{\left(t_{\mathsf k(\mathsf b)} \tilde t_b^{-1} \tau^{-2} x^{-\frac{n \mathsf j(\mathsf b)+2 \mathsf i(\mathsf b)}{n+1}+\frac{n+2}{n+1}+2};x^2\right)_\infty}{\left(t_{\mathsf k(\mathsf b)}^{-1} \tilde t_b \tau^2 x^{\frac{n \mathsf j(\mathsf b)+2 \mathsf i(\mathsf b)}{n+1}-\frac{n+2}{n+1}};x^2\right)_\infty}\right)
\end{align}
where $\mathsf b \in \mathsf f$ represents each box of forest graph $\mathsf f$. $\mathsf k(\mathsf b)$ indicates the corresponding flavor of the box $\mathsf b$. $(\mathsf i, \, \mathsf j)$ represents a location of $\mathsf b$ in a given tree with $\mathsf i = \mathsf j = 0$ for the root node. As before $(\ldots)'$ indicates that the zero factors in a q-Pochhammer symbol are omitted. Compared with \eqref{eq:pert}, here we substitute the explicit value of
$\upsilon_i$
\begin{equation}
\upsilon_i \rightarrow x^{\frac{n \mathsf j(\mathsf b)+2 \mathsf i(\mathsf b)}{n+1}}.  \label{vi}
\end{equation}
Previously we saw that $v_i$ is the product of $v_A$ and $v_B$ where the product is done over all the nodes between
the root node and the given node. In box diagram, $i(\mathsf b)$ is the number of the nodes with red circle connected by
X-branch and  $j(\mathsf b)$ is the number of the nodes with blue circle connected by
Y-branch. Thus $i(\mathsf b)+j(\mathsf b)$ is the number of nodes between
the root node and the given node including itself. Hence the eq. (\ref{vi}) is justified. We also rewrite
\begin{align}
\left(\prod_{i \neq j}^{N_c} \left(1-t_{c_i} t_{c_j}^{-1} \upsilon_i^{-1} \upsilon_j\right)\right) \rightarrow \left(\prod_{\mathsf b,\mathsf c \in \mathsf f} \frac{\left(t_{\mathsf k(\mathsf b)} t_{\mathsf k(\mathsf c)}^{-1} x^{-\frac{n \mathsf j(\mathsf b)+2 \mathsf i(\mathsf b)}{n+1}+\frac{n \mathsf j(\mathsf c)+2 \mathsf i(\mathsf c)}{n+1}};x^2\right)'_\infty}{\left(t_{\mathsf k(\mathsf c)} t_{\mathsf k(\mathsf b)}^{-1} x^{-\frac{n \mathsf j(\mathsf c)+2 \mathsf i(\mathsf c)}{n+1}+\frac{n \mathsf j(\mathsf b)+2 \mathsf i(\mathsf b)}{n+1}+2};x^2\right)_\infty}\right),
\end{align}
which follows from
$\frac{(a;x^2)'_{\infty}}{(ax^2;x^2)_{\infty}}=1-a$ if $a\neq 1$ and 1 if $a= 1$.
Thus the expression is entirely expressed in terms of the position of each box and is easier to manipulate.

We first rewrite the last factor which comes from the anti-fundamentals:
\begin{align}
\label{eq:pert1}
& \left(\prod_{b = 1}^{N_f} \prod_{\mathsf b \in \mathsf f} \frac{\left(t_{\mathsf k(\mathsf b)} \tilde t_b^{-1} \tau^{-2} x^{-\frac{n \mathsf j(\mathsf b)+2 \mathsf i(\mathsf b)}{n+1}+\frac{n+2}{n+1}+2};x^2\right)_\infty}{\left(t_{\mathsf k(\mathsf b)}^{-1} \tilde t_b \tau^2 x^{\frac{n \mathsf j(\mathsf b)+2 \mathsf i(\mathsf b)}{n+1}-\frac{n+2}{n+1}};x^2\right)_\infty}\right) = \nonumber \\
& \left(\prod_{b = 1}^{N_f} \prod_{\mathsf b \in \mathsf F} \frac{\left(t_{\mathsf k(\mathsf b)} \tilde t_b^{-1} \tau^{-2} x^{-\frac{n \mathsf j(\mathsf b)+2 \mathsf i(\mathsf b)}{n+1}+\frac{n+2}{n+1}+2};x^2\right)_\infty}{\left(t_{\mathsf k(\mathsf b)}^{-1} \tilde t_b \tau^2 x^{\frac{n \mathsf j(\mathsf b)+2 \mathsf i(\mathsf b)}{n+1}-\frac{n+2}{n+1}};x^2\right)_\infty}\right) \left(\prod_{b = 1}^{N_f} \prod_{\mathsf b \in \mathsf f^c} \frac{\left(t_{\mathsf k(\mathsf b)}^{-1} \tilde t_b \tau^2 x^{\frac{n \mathsf j(\mathsf b)+2 \mathsf i(\mathsf b)}{n+1}-\frac{n+2}{n+1}};x^2\right)_\infty}{\left(t_{\mathsf k(\mathsf b)} \tilde t_b^{-1} \tau^{-2} x^{-\frac{n \mathsf j(\mathsf b)+2 \mathsf i(\mathsf b)}{n+1}+\frac{n+2}{n+1}+2};x^2\right)_\infty}\right)
\end{align}
where $\mathsf F$ represents the full set of boxes, i.e., $\mathsf F  = \mathsf f \cup \mathsf f^c$ for any $\mathsf f$. In the first factor, we can decompose $\prod_{\mathsf b \in \mathsf F}$ as
\begin{align}
\prod_{\mathsf b \in \mathsf F} \rightarrow \prod_{a = 1}^{N_f} \prod_{s = 0}^{n-1} \prod_{t = 0}^2
\end{align}
such that $s = \mathsf i(b)-1, \, t = \mathsf j(b)-1$ and $a = \mathsf k(\mathsf b)$. In the second factor, we introduce new location function $(\mathsf i',\mathsf j')$ appropriate for a dual forest such that
\begin{align}
\mathsf i' = n+1-\mathsf i, \qquad \mathsf j' = 4-\mathsf j.
\end{align}
Then \eqref{eq:pert1} is simply
\begin{align}
\label{eq:pert1'}
\left(\prod_{a, b = 1}^{N_f} \prod_{s = 0}^{n-1} \prod_{t = 0}^2 \frac{\left(t_a \tilde t_b^{-1} \tau^{-2} x^{-\frac{2 s+n t}{n+1}+2};x^2\right)_\infty}{\left(t_a^{-1} \tilde t_b \tau^2 x^{\frac{2 s+n t}{n+1}};x^2\right)_\infty}\right) \left(\prod_{b = 1}^{N_f} \prod_{\mathsf b \in \mathsf f^c} \frac{\left(t_{\mathsf k(\mathsf b)}^{-1} \tilde t_b \tau^2 x^{-\frac{n \mathsf j'(\mathsf b)+2 \mathsf i'(\mathsf b)}{n+1}+\frac{n+2}{n+1}+2-\frac{4-2 n}{n+1}};x^2\right)_\infty}{\left(t_{\mathsf k(\mathsf b)} \tilde t_b^{-1} \tau^{-2} x^{\frac{n \mathsf j'(\mathsf b)+2 \mathsf i'(\mathsf b)}{n+1}-\frac{n+2}{n+1}+\frac{4-2 n}{n+1}};x^2\right)_\infty}\right)
\end{align}
where the first one is exactly the contribution of mesonic singlets $M_{s,t}$ while the second one is the contribution of dual fundamentals $q$.

Similarly, one can show that the remaining part of \eqref{eq:pert0} can be written in terms of dual field contributions:
\begin{align}
\label{eq:pert2}
& \left(\prod_{\mathsf b,\mathsf c \in \mathsf f^c} \frac{\left(t_{\mathsf k(\mathsf b)}^{-1} t_{\mathsf k(\mathsf c)} x^{-\frac{n \mathsf j'(\mathsf b)+2 \mathsf i'(\mathsf b)}{n+1}+\frac{n \mathsf j'(\mathsf c)+2 \mathsf i'(\mathsf c)}{n+1}};x^2\right)'_\infty}{\left(t_{\mathsf k(\mathsf c)}^{-1} t_{\mathsf k(\mathsf b)} x^{-\frac{n \mathsf j'(\mathsf c)+2 \mathsf i'(\mathsf c)}{n+1}+\frac{n \mathsf j'(\mathsf b)+2 \mathsf i'(\mathsf b)}{n+1}+2};x^2\right)_\infty}\right) \nonumber \\
& \times \left(\prod_{A = X,Y} \prod_{\mathsf b,\mathsf c \in \mathsf f^c} \frac{\left(t_{\mathsf k(\mathsf b)} t_{\mathsf k(\mathsf c)}^{-1} \upsilon_A^{-1} x^{\frac{n \mathsf j'(\mathsf b)+2 \mathsf i'(\mathsf b)}{n+1}-\frac{n \mathsf j'(\mathsf c)+2 \mathsf i'(\mathsf c)}{n+1}+2};x^2\right)_\infty}{\left(t_{\mathsf k(\mathsf c)}^{-1} t_{\mathsf k(\mathsf b)} \upsilon_A x^{-\frac{n \mathsf j'(\mathsf c)+2 \mathsf i'(\mathsf c)}{n+1}+\frac{n \mathsf j'(\mathsf b)+2 \mathsf i'(\mathsf b)}{n+1}};x^2\right)'_\infty}\right) \nonumber \\
& \times \left(\prod_{a = 1}^{N_f} \prod_{\mathsf b \in \mathsf f^c} \frac{\left(t_{\mathsf k(\mathsf b)} t_a^{-1} x^{\frac{n \mathsf j'(\mathsf b)+2 \mathsf i'(\mathsf b)}{n+1}-\frac{n+2}{n+1}+2};x^2\right)_\infty}{\left(t_{\mathsf k(\mathsf b)}^{-1} t_a x^{-\frac{n \mathsf j'(\mathsf b)+2 \mathsf i'(\mathsf b)}{n+1}+\frac{n+2}{n+1}};x^2\right)'_\infty}\right)
\end{align}
Combining \eqref{eq:pert1'} and \eqref{eq:pert2}, we show that
\begin{align}
Z_\text{pert}^\mathsf f/\hat Z_\text{pert}^{\mathsf f^c} = \mathrm{PE} \left[\sum_{a,b = 1}^{N_f} \frac{\left(\sum_{s = 0}^{n-1} \sum_{t = 0}^2 x^\frac{2 s+n t}{n+1}\right) \left(t_a \tilde t_b \tau^2-x^{-2+\frac{6}{n+1}} t_a^{-1} \tilde t_b^{-1} \tau^{-2}\right)}{1-x^2}\right]
\end{align}
where the right hand side is a rewriting of the first factor in \eqref{eq:pert1'}.
\\

In conclusion, we have found
\begin{align}
\begin{aligned}
Z_\text{pert}^\mathsf f &= \hat Z_\text{pert}^{\mathsf f^c} I_M, \\
Z_\text{vortex}^\mathsf f &= \hat Z_\text{vortex}^{\mathsf f^c} Z_V Z_W, \\
Z_\text{anti-vortex}^\mathsf f &= \hat Z_\text{anti-vortex}^{\mathsf f^c} \overline{Z_V} \, \overline{Z_W}
\end{aligned}
\end{align}
where
\begin{align}
I_M &= \mathrm{PE} \left[\sum_{a,b = 1}^{N_f} \frac{\left(\sum_{s = 0}^{n-1} \sum_{t = 0}^2 x^\frac{2 s+n t}{n+1}\right) \left(t_a \tilde t_b \tau^2-x^{-2+\frac{6}{n+1}} t_a^{-1} \tilde t_b^{-1} \tau^{-2}\right)}{1-x^2}\right], \\
Z_V &= \mathrm{PE} \left[\frac{\left(\sum_{s = 0}^{n-1} x^\frac{2 s}{n+1}+x^\frac{n}{n+1}+x^\frac{2 n}{n+1}\right) \left(x^{N_f-\frac{1}{n+1} N_c+\frac{1}{n+1}} \tau^{-N_f}-x^{-N_f+\frac{1}{n+1} N_c+\frac{1}{n+1}} \tau^{N_f}\right) w}{1-x^2}\right], \\
Z_W &= \mathrm{PE} \left[\frac{\left(\sum_{u = 0}^\frac{n-3}{2} x^\frac{4 u}{n+1}\right) \left(x^{2 N_f-\frac{2}{n+1} N_c+\frac{4}{n+1}} \tau^{-2 N_f}-x^{-2 N_f+\frac{2}{n+1} N_c+\frac{4}{n+1}} \tau^{2 N_f}\right) w^2}{1-x^2}\right]
\end{align}
and
\begin{align}
\overline{Z_{V,W}} = \left.Z_{V,W}\right|_{x \rightarrow x^{-1},\tau \rightarrow \tau^{-1},w \rightarrow w^{-1}}.
\end{align}
Note that $I_M$ is the index contribution of $M_{s,t}$. The contributions of $V_{s,t}$ and $W_u$ are given by
\begin{align}
I_V = Z_V \overline{Z_V}, \qquad I_W = Z_W \overline{Z_W}.
\end{align}
Combining those results, we finally obtain the equality of the superconformal indices of a conjectured duality pair:
\begin{align}
I = \hat I
\end{align}
where
\begin{align}
\begin{aligned}
I &= \sum_{\mathsf f \in \mathsf G} Z_\text{pert}^\mathsf f Z_\text{vortex}^\mathsf f Z_\text{anti-vortex}^\mathsf f, \\
  &= I_M I_V I_W \sum_{\mathsf f^c \in \hat{\mathsf G}} \hat Z_\text{pert}^{\mathsf f^c} \hat Z_\text{vortex}^{\mathsf f^c} \hat Z_\text{anti-vortex}^{\mathsf f^c}
\end{aligned}
\end{align}
with $\mathsf G, \, \hat{\mathsf G}$, sets of growing forests for the original theory and the dual theory respectively. The lists of growing trees are given in appendix \ref{sec:growing trees}.
\\

\section{$F$-maximization}
\label{sec:F-max}

In this section, we investigate more about the IR dynamics of the theories with two adjoints by performing the $F$-maximization, especially focusing on the $n = 3$ case. When we computed the superconformal index and tested the proposed duality in the previous section, we assumed that the superpotential \eqref{eq:sup_D} leads to a new IR fixed point. However, depending on $N_c, \, N_f, \, n$, the superpotential would be irrelevant, and the theory just flows to the same fixed point as the theory without superpotential flows to. In 4d, it was argued that this new IR fixed point can be reached by a series of relevant deformations and associated RG-flows:
\begin{align}
\label{eq:RG}
\widehat O \rightarrow \widehat D \rightarrow D_{n+2}.
\end{align}
We first start from the theory with no superpotential. Without any superpotential, the theory with $N_f$ pairs of fundamental and anti-fundamental matters and two adjoints flows to a nontrivial IR fixed point, which we call $\widehat O$. Then we introduce the deformation by the superpotential
\begin{align}
W_{\widehat D} = \mathrm{Tr} X Y^2.
\end{align}
If this deformation is relevant, an RG-flow is initiated and reaches to a different IR fixed point $\widehat D$. Then we also introduce the superpotential \begin{align}
W_{D_{n+2}} = \mathrm{Tr} X^{n+1}.
\end{align}
Again if this is relevant, another RG-flow is initiated and reaches to the fixed point $D_{n+2}$, which we are interested in.

Therefore, we have to ask whether each deformation is relevant or not. In 4d, this is answered by $a$-maximization, which determines the superconformal $R$-charge in IR. In 3d, instead, the superconformal $R$-charge is determined by the $F$-maximization. The free energy $F$ is given by
\begin{align}
F = -\log |Z|
\end{align}
where $Z$ is the supersymmetric partition function on $S^3$. The $S^3$ partition function $Z$ is also exactly computed by the supersymmetric localization \cite{Kapustin:2009kz}. We have to perform the $F$-maximization to check if the RG-flow \eqref{eq:RG} is nontrivial; i.e. the superpotential \eqref{eq:sup_D} is relevant so that we have a new fixed point $D_{n+2}$ rather than $\widehat O$.
\\

Let us start from the fixed point $\widehat O$. Because we don't have any superpotential, the adjoints $X$ and $Y$ appear symmetrically. Thus, we assign the same trial $R$-charge $\delta$ to them. Assigning the trial $R$-charge $\Delta$ to the fundamentals, we have the $S^3$ partition function of the form:
\begin{align}
\label{eq:Z_O}
Z_{\widehat O} (\Delta,\delta) &= \frac{1}{N_c!} \int_{-\infty}^\infty \left(\prod_{j = 1}^{N_c} \frac{d x_j}{2 \pi} \right) \left(\prod_{i < j} 2 \sinh \frac{x_i-x_j}{2}\right)^2 \nonumber \\
&\qquad \times \left(\prod_{j = 1}^{N_c} e^{N_f l\left(1-\Delta+i \frac{x_j}{2 \pi}\right)+N_f l\left(1-\Delta-i \frac{x_j}{2 \pi}\right)}\right) \left(\prod_{i,j = 1}^{N_c} e^{2 l\left(1-\delta+i \frac{x_i-x_j}{2 \pi}\right)}\right)
\end{align}
where $l(z)$ is defined by
\begin{align}
l(z) = -z \log(1-e^{2 \pi i z})+\frac{i}{2} \left(\pi z^2+\frac{1}{\pi} \mathrm{Li}_2(e^{2 \pi i z})\right)-\frac{i \pi}{12}.
\end{align}
Since the superconformal values of the $R$-charges maximize the free energy $F = -\log |Z|$, one can determine the correct IR values of $\Delta$ and $\delta$ by evaluating \eqref{eq:Z_O}. For some low values of $N_c, \, N_f$ with $n = 3$, we have found the results shown in table \ref{tab:Delta_O}.
\begin{table}[tbp]
\centering
\begin{tabular}{|c|ccccc|}
\hline
$\Delta/\delta$ & $N_f = 1$ & $N_f = 2$ & $N_f = 3$ & $N_f = 4$ & $N_f = 5$ \\
\hline
$U(1)$ & (0.333)/0.500 & (0.409)/0.500 & (0.437)/0.500 & (0.452)/0.500 & (0.461)/0.500 \\
$U(2)$ & 0.328/0.367 & 0.388/0.395 & 0.415/0.412 & 0.431/0.424 & 0.442/0.433 \\
$U(3)$ & 0.35/0.32 & 0.39/0.35 & 0.41/0.37 & 0.42/0.39 & 0.43/0.40 \\
\hline
\end{tabular}
\caption{\label{tab:Delta_O} The superconformal values of $\Delta$ and $\delta$ are computed by the $F$-maximization for the IR fixed point $\widehat O$.
Note that, for $U(1)$, the theory is decomposed into SQED with (anti-)fundamentals and decoupled free $X,Y$; $\Delta$ (in parenthesis) is for SQED while $\delta = 1/2$ is for free $X,Y$.}
\end{table}
We should comment that a $U(1)$ theory is decomposed into SQED with (anti-)fundamentals and decoupled free $X,Y$. The $U(1)$ results shown in table \ref{tab:Delta_O} means that $\Delta$ in parenthesis is that of SQED while $\delta = 1/2$ is that of free $X,Y$.

Our question is whether the superpotential $W_{\widehat D} = \mathrm{Tr} X Y^2$, whose dimension is $3 \delta$, is relevant or not. One can see that $3 \delta < 2$ for all the examples in table \ref{tab:Delta_O}. The deformation by $W_{\widehat D}$ thus leads to a new IR fixed point denoted by $\widehat D$. For $U(1)$, $W_{\widehat D}$ is for decoupled $X,Y$ and leads to a nontrivial Landau-Ginzburg theory of them in IR.
\\

Since $W_{\widehat D}$ is a relevant deformation, we can initiate the RG-flow to another IR fixed point $\widehat D$ by turning on the superpotential $W_{\widehat D}$. We should determine the $R$-charges of $X$ and $Y$ there. Because of the superpotential $W_{\widehat D}$, we should assign $\delta$ to $X$ and $1-\delta/2$ to $Y$. The $S^3$ partition function is thus given by
\begin{align}
\label{eq:Z_D}
Z_{\widehat D} (\Delta,\delta) &= \frac{1}{N_c!} \int_{-\infty}^\infty \left(\prod_{j = 1}^{N_c} \frac{d x_j}{2 \pi} \right) \left(\prod_{i < j} 2 \sinh \frac{x_i-x_j}{2}\right)^2 \left(\prod_{j = 1}^{N_c} e^{N_f l\left(1-\Delta+i \frac{x_j}{2 \pi}\right)+N_f l\left(1-\Delta-i \frac{x_j}{2 \pi}\right)}\right) \nonumber \\
&\qquad \times \left(\prod_{i,j = 1}^{N_c} e^{l\left(1-\delta+i \frac{x_i-x_j}{2 \pi}\right)+l\left(1-(1-\delta/2)+i \frac{x_i-x_j}{2 \pi}\right)}\right).
\end{align}
Maximizing $F = -\log |Z|$, we have found the IR values of $\Delta$ and $\delta$ as shown in table \ref{tab:Delta_D}.
\begin{table}[tbp]
\centering
\begin{tabular}{|c|ccccc|}
\hline
$\Delta/\delta$ & $N_f = 1$ & $N_f = 2$ & $N_f = 3$ & $N_f = 4$ & $N_f = 5$ \\
\hline
$U(1)$ & (0.333)/0.584 & (0.409)/0.584 & (0.437)/0.584 & (0.452)/0.584 & (0.461)/0.584 \\
$U(2)$ & 0.242/0.437 & 0.349/0.489 & 0.392/0.512 & 0.416/0.526 & 0.431/0.535 \\
$U(3)$ & 0.19/0.34 & 0.30/0.42 & 0.36/0.46 & 0.39/0.48 & 0.40/0.500 \\
\hline
\end{tabular}
\caption{\label{tab:Delta_D} The superconformal values of $\Delta$ and $\delta$ are computed by the $F$-maximization for the IR fixed point $\widehat D$. Note that, for $U(1)$, the theory is decomposed into SQED with (anti-)fundamentals and the Landau-Ginzburg theory of $X,Y$; $\Delta$ (in parenthesis) is for SQED while $\delta$ is for LG.}
\end{table}

Now we should check if the superpotential $W_{D_{n+2}} = \mathrm{Tr} X^{n+1}$ is a relevant deformation. For $U(1)$, (more precisely, for the decoupled $\widehat D$ LG theory,) $W_{D_{n+2}}$ is irrelevant because the dimension $(n+1) \delta = 4 \delta$ is greater than 2. Thus, we don't have the $D_{n+2}$ IR fixed point for the LG theory. For $U(2)$, $W_{D_{n+2}}$ is relevant for $N_f = 1, 2$ while it is irrelevant for $N_f > 2$. Thus, we have the rank-2 $D_{n+2}$ fixed point only for $N_f = 1, 2$. For $U(3)$, $W_{D_{n+2}}$ is relevant for $N_f < 5$ and is marginally relevant for $N_f = 5$. Turning on relevant $W_{D_{n+2}}$, we reach the fixed point $D_{n+2}$.
\\

We have figured out that for which low values of $N_c$ and $N_f$ we have a new IR fixed point due to the superpotential \eqref{eq:sup_D}. In the previous section, while we already tested the duality for this fixed point, we didn't fix the correct superconformal $R$-charge there. Let us find the superconformal value of the $R$-charge by the $F$-maximization. Now we assign the fixed $R$-charges to the adjoints $X,Y$ due to the superpotential \eqref{eq:sup_D}: $\frac{2}{n+1}$ and $\frac{n}{n+1}$ respectively. Then we should maximize the free energy $F = -\log |Z|$ where $Z$ is now
\begin{align}
\label{eq:Z_Dn}
Z_{D_{n+2}} (\Delta) &= \frac{1}{N_c!} \int_{-\infty}^\infty \left(\prod_{j = 1}^{N_c} \frac{d x_j}{2 \pi} \right) \left(\prod_{i < j} 2 \sinh \frac{x_i-x_j}{2}\right)^2 \left(\prod_{j = 1}^{N_c} e^{N_f l\left(1-\Delta+i \frac{x_j}{2 \pi}\right)+N_f l\left(1-\Delta-i \frac{x_j}{2 \pi}\right)}\right) \nonumber \\
&\qquad \times \left(\prod_{i,j = 1}^{N_c} e^{l\left(1-\frac{2}{n+1}+i \frac{x_i-x_j}{2 \pi}\right)+l\left(1-\frac{n}{n+1}+i \frac{x_i-x_j}{2 \pi}\right)}\right).
\end{align}
We found the superconformal values as shown in table \ref{tab:Delta}.
\begin{table}[tbp]
\centering
\begin{tabular}{|c|ccccc|}
\hline
$\Delta$ & $N_f = 1$ & $N_f = 2$ & $N_f = 3$ & $N_f = 4$ & $N_f = 5$ \\
\hline
$U(1)$ & (0.3333) & (0.4085) & (0.4370) & (0.4519) & (0.4611) \\
$U(2)$ & $0.232 \rightarrow 0.212$ & 0.348 & - & - & - \\
$U(3)$ & $0.144 \rightarrow 0.080$ & 0.293 & 0.352 & 0.384 & 0.404 \\
\hline
\end{tabular}
\caption{\label{tab:Delta} The superconformal values of $\Delta$ are computed for the IR fixed point $D_{n+2}$. For $\Delta < 1/4$, the meson operator $M_{0,0} = \tilde Q Q$ decouples in IR. In such cases, the $F$-maximization is performed again after factoring out the contribution of $M_{0,0}$, and possibly other decoupled operators as well, whose result is written on the right hand side of $\rightarrow$. Note that, for $U(1)$, the results are the same as those of SQED without adjoint \cite{Safdi:2012re} because the adjoints are gauge singlets.}
\end{table}

One should note that for $U(2), \, U(3)$ with $N_f = 1$, the superconformal value of $\Delta$ is less than 1/4. Thus, the conformal dimension $2 \Delta$ of the meson operator $M_{0,0} = \tilde Q Q$ seems less than 1/2, which is a signal of decoupling of $M_{0,0}$. In general, due to the unitarity, the conformal dimension of any gauge invariant operator at the IR superconformal fixed point shouldn't be less than 1/2. If the conformal dimension is less than 1/2, one possible scenario is that the operator decouples and the conformal dimension is corrected to 1/2 by an accidental symmetry which freely rotates the decoupled operator. Therefore, when we perform the $F$-maximization, we have to check whether there are gauge invariant operators violating the unitarity bound.

In our case, the following minimal operators must be checked first: the meson operator $M_{0,0}$, the monopole operators $V_{0,0}^\pm$, and $\mathrm{Tr} X$. In particular, the conformal dimension of $\mathrm{Tr} X$ is already determined by the superpotential regardless of the $F$-maximization, which is $1/2$. Thus, $\mathrm{Tr} X$ is decoupled at the IR fixed point. The conformal dimension of the meson operator $M_{0,0}$ is given by $2 \Delta$. From table \ref{tab:Delta}, one can see that $M_{0,0}$ has the conformal dimension less than 1/2 for $U(2)$ and $U(3)$ with $N_f = 1$. In such cases, $M_{0,0}$ should decouple, and the $F$-maximization should be performed again after factoring out the contribution of $M_{0,0}$. More precisely, in fact, we should check the conformal dimension of the monopole operators $V_{0,0}^\pm$ as well. As shown in table \ref{tab:Delta_V}, only the $U(3)$ theory with $N_f = 1$ has the unitarity violating $V_{0,0,}^\pm$.
\begin{table}[tbp]
\centering
\begin{tabular}{|c|ccccc|}
\hline
$\Delta_V$ & $N_f = 1$ & $N_f = 2$ & $N_f = 3$ & $N_f = 4$ & $N_f = 5$ \\
\hline
$U(1)$ & (0.6667) & (1.183) & (1.689) & (2.192) & (2.694) \\
$U(2)$ & $0.518 \rightarrow 0.538$ & 1.05 & - & - & - \\
$U(3)$ & $0.356 \rightarrow 0.42$ & 0.914 & 1.44 & 1.96 & 2.48 \\
\hline
\end{tabular}
\caption{\label{tab:Delta_V} The conformal dimension of the minimal monopole operators $V_{0,0}^\pm$.}
\end{table}
Thus, for $U(2)$ with $N_f = 1$, we perform the $F$-maximization again after factoring out $M_{0,0}$ only. The new value of $\Delta$ slightly decreases and is written on the right hand side of $\rightarrow$ in table \ref{tab:Delta}. On the other hand, for $U(3)$ with $N_f = 1$, both $M_{0,0}$ and $V_{0,0}^\pm$ violate the unitarity bound. We thus have two scenarios: $M_{0,0}$ decouples first or $V_{0,0}^\pm$ decouples first. In the current example, however, it turns out that both operators indeed decouple in the end. Therefore, we perform the $F$-maximization factoring out both $M_{0,0}$ and $V_{0,0}^\pm$, whose result is written on the right hand side of $\rightarrow$. The other operators have the conformal dimensions greater than those of the minimal operators, which are indeed greater than 1/2.

In conclusion, for a few low rank examples having the relevant superpotential with $n = 3$, we have figured out whether there are operators that decouple in IR or not. First, the operator $\mathrm{Tr} X$ always decouples in IR. Second, the meson operator $M_{0,0}$ decouples for $U(2)$ and $U(3)$ with $N_f = 1$. Third, the monopole operators $V_{0,0}^\pm$ decouple  for $U(3)$ with $N_f = 1$.
\\

When there is a decoupled operator, one can explicitly integrate out such an operator by introducing an extra operator playing a role of a Lagrangian multiplier. In particular, for $U(2)$ theory with $N_f = 1$, the meson operator $M_{0,0}$ decouples in IR, and we factor out its contribution from the partition function by hand to obtain the correct IR $R$-charges. Indeed, the same procedure can be understood as introducing a Lagrangian multiplier field. As we want to integrate out such a decoupled operator, we introduce an extra field $N$ with the superpotential
\begin{align}
\label{eq:N}
\Delta W = N \tilde Q Q
\end{align}
so that $M_{0,0}$ has a quadratic mass term:
\begin{align}
\left|\frac{\partial \Delta W}{\partial N}\right|^2 = |\tilde Q Q|^2 = |M_{0,0}|^2.
\end{align}
Due to the superpotential term, the extra field $N$ carries the global charges as shown in table \ref{tab:N}.
\begin{table}[tbp]
\centering
\begin{tabular}{|c|ccccc|}
\hline
 & $SU(N_f)_1$ & $SU(N_f)_2$ & $U(1)_A$ & $U(1)_T$ & $U(1)_R$ \\
\hline
$N$ & $\mathbf{N_f}$ & $\overline{\mathbf{N_f}}$ & $-2$ & $0$ & $2-2 r$ \\
\hline
\end{tabular}
\caption{\label{tab:N} The representations of the chiral field $N$ under the symmetry groups.}
\end{table}
Thus, its partition function contribution is indeed the same as the inverse of the contribution of $M_{0,0}$. As a result, factoring out the contribution of $M_{0,0}$ by hand is equivalent to introducing the Lagrangian multiplier field $N$ with the superpotential \eqref{eq:N}.

Introducing a Lagrangian multiplier field leads to an interesting consequence on the dual side. Originally we have the superpotential terms involving $M_{0,0}$ as follows:
\begin{align}
W = M_{0,0} \tilde q X^{n-1} Y^2 q+\ldots
\end{align}
Now we introduce the extra field $N$ and the deformation $\Delta W = N M_{0,0}$, which leads to the following F-term relation:
\begin{gather}
\begin{gathered}
M_{0,0} = 0, \\
N = \tilde q X^{n-1} Y^2 q.
\end{gathered}
\end{gather}
In other words, $M_{0,0}$ is excluded from the chiral ring while $N = \tilde q X^{n-1} Y^2 q$, one of the dual theory mesons, is instead included in the chiral ring. The moduli space of the theory is thus parametrized by $N$ rather than $M_{0,0}$. Introducing the extra field $N$ with the superpotential $\Delta W$ is called ``flipping'' of the operator $M_{0,0}$. In particular, the role of the flipping of a decoupled operator is recently emphasized in \cite{Benvenuti:2017lle}.

Note that the extra field $N$ is safe from the unitarity violation because its $R$-charge is $2-2 r = 1.576$, which is determined by the F-maximization. It would be interesting if we can confirm this in the dual theory point of view. Unfortunately, it is beyond our computational capacity because the dual theory is a $U(7)$ theory, whose gauge rank is too high. Instead, we have already observed that the factorized superconformal indices agree for a duality pair. Since the (squashed) $S^3$ partition function share the same factorization form, we would expect that the $S^3$ partition functions, and accordingly the free energies, also agree for a duality pair and give the same result for superconformal $R$-charges as in table \ref{tab:Delta}, which is obtained from the original theory point of view.
\\

\section*{Acknowledgements}

JP  is supported in part
by the NRF Grant 2018R1A2B6007159.
The work of HK was supported by Basic Science Research Program through the National Research Foundation of Korea(NRF) funded by the Ministry of Education(2017R1A6A3A03009422) and by the Center for Mathematical Sciences and Applications at Harvard University.
\\

\appendix
\section{More on Contour Integral}
\label{sec:integral}

In this appendix, we would like to explain more details of the contour integral evaluation of the superconformal index. Recall that using the Coulomb branch localization, one obtains the following expression of the superconformal index:
\begin{align}
& I(x,t) = \nonumber \\
& \sum_{m \in \text{monopole background}} \frac{1}{|\mathcal W_m|} \int \frac{dz}{2 \pi i z} ~ Z_\text{classical}(x,t;z,m) ~ Z_\text{vector}(x;z,m) ~ Z_\text{chiral}(x,t;z,m)
\end{align}
The classical action contribution $Z_\text{classical}$ includes the various Chern-Simons terms. For example, if we turn on CS level $\kappa$ for gauge group $U(N_c)$,
\begin{align}
Z_\text{classical}(x,t;z,m) = \prod_{i = 1}^{N_c} (-z_i)^{\kappa m_i}.
\end{align}
One can consider a flavor CS term as well, whose contribution is nontrivial once we turn on background flux for a flavor symmetry. One can also consider mixed CS terms between abelian factors of the symmetry group, which include the Fayet-Iliopoulos term.

The vector multiplet contribution is given by
\begin{align}
Z_\text{vector} (x;z,m) = \prod_{\alpha \in \mathfrak g} x^{-|\alpha(m)|/2} \left(1-z^\alpha x^{|\alpha(m)|}\right)
\end{align}
where $\alpha$ is a root vector of Lie algebra $\mathfrak g$. $z^\alpha$ is a short-hand notation of $e^{i \alpha(a)}$ where $z_j = e^{i a_j}$. The chiral multiplet contribution is given by
\begin{align}
Z_\text{chiral} (x,t;z,m) = \prod_{\rho \in R_G} \prod_{\sigma \in R_F} \left((-z)^\rho t^\sigma x^{r-1}\right)^{-|\rho(m)|/2} \frac{\left(z^{-\rho} t^{-\sigma} x^{2-r+|\rho(m)|};x^2\right)_\infty}{\left(z^\rho t^\sigma x^{r+|\rho(m)|};x^2\right)_\infty}
\end{align}
where $\rho$ and $\sigma$ are weights of the representations under gauge group $G$ and flavor group $F$ respectively. $(a;q)_\infty$ is a q-Pochhammer symbol defined by
\begin{align}
(a;q)_\infty = \prod_{k = 0}^\infty \left(1-a q^k\right).
\end{align}

As long as each chiral multiplet carries the positive $R$-charge, one can make a series expansion of the integrand at $x = 0$. Then the integration $\int \frac{dz}{2 \pi i z}$ is nothing but the projection onto gauge singlets, thus what we need to do is just reading off the $z^0$ sector of the integrand. This method is referred as Coulomb branch computation in the previous sections. As we discussed, this method is only applicable to cases with positive $R$-charges. If there is accidental IR symmetries, we usually have negative UV $R$-charges, which is indeed the case of examples considered in this note.

On the other hand, one can evaluate the integral without the series expansion, by taking the residue of the poles enclosed by the integration contour. Such poles are from the chiral multiplet contribution, $Z_\text{chiral}$. For our two adjoint example, the relevant poles are as follows:
\begin{align}
z_j = \left\{\begin{array}{lc}
t_{a_j}^{-1} \tau^{-1} x^{-|m_j|-2 k_j}, & 1 \leq a_j \leq N_f, \quad k_j \geq 0, \\
z_i \upsilon_A^{-1} x^{-|m_j-m_i|-2 k_j}, \qquad & 1 \leq i (\neq j) \leq N_c, \quad A = X,Y, \quad k_j \geq 0
\end{array}\right.
\end{align}
where $t_a, \tau, \upsilon_X, \upsilon_Y$ are fugacities for $SU(N_f)_1 \times U(1)_A \times U(1)_X \times U(1)_Y$ respectively. The final expression of the integral is obtained by evaluating the residues at those poles. Detailed computations are similar to those of the single adjoint case, which are found in \cite{Hwang:2015wna}. The only difference is how to deal with double poles, which is explained in section \ref{sec:fact}.
\\

We should comment about asymptotic singularities at $z = 0$ or $z = \infty$. To examine those singularities, one should introduce the approximated index with cutoff, $I_n$, in which every infinite q-Pochhammer symbol is replaced by a finite q-Pochhammer symbol. Formally,
\begin{align}
I_n \equiv I|_{(a;q)_\infty \rightarrow (a;q)_n}.
\end{align}
The original index is obtained as a limit of the approximated indices:
\begin{align}
I = \lim_{n \rightarrow \infty} I_n.
\end{align}
For the case with (anti-)fundamental matters, the asymptotic poles of $I_n$ are discussed in \cite{Hwang:2012jh,Benini:2013yva}. For $|\kappa| < \frac{|N_f-N_a|}{2}$, the integrand of $I_n$ has an asymptotic pole only at one of $z = 0$ or $z = \infty$. Thus, one can avoid the asymptotic pole by deforming the integration contour. For $|\kappa| > \frac{|N_f-N_a|}{2}$, the integrand of $I_n$ has asymptotic poles both at $z = 0$ and $z = \infty$. In such cases, one cannot avoid the asymptotic poles, and how to compute their residues is still an open question. For $|\kappa| = \frac{|N_f-N_a|}{2}$, again the integrand has poles both at $z = 0$ and $z = \infty$, but the residue of one of them becomes zero as $n$ goes to $\infty$.

In this note we are interested in $\kappa = N_f-N_a = 0$ with two extra adjoints matters. We have asymptotic poles both at $z = 0$ and at $z = \infty$. For example, the residue at $z = 0$ can be written as follows:
\begin{align}
\mathrm{Res}_{z = 0} Z_n &\sim \left(\prod_{j = 1}^{N_c} \lim_{z_j \rightarrow 0} \frac{1}{(N_c-1)!} \frac{\partial^{N_c-1}}{\partial z_j^{N_c-1}}\right) \left(\prod_{\substack{i,j = 1 \\ (i \neq j)}}^{N_c} \left(z_j-z_i x^{|m_i-m_j|}\right)\right) \\
&\quad \times \left(\prod_{j = 1}^{N_c} \prod_{a = 1}^{N_f} \prod_{k = 0}^{n-1} \frac{z_j-t_a^{-1} \tau^{-1} x^{|m_j|+2 k+2}}{1-z_j t_a \tau x^{|m_j|+2 k}} \frac{1-z_j \tilde t_a^{-1} \tau^{-1} x^{|m_j|+2 k+2}}{z_j-\tilde t_a \tau x^{|m_j|+2 k}}\right) \\
&\quad \times \left(\prod_{\substack{i,j = 1 \\ (i \neq j)}}^{N_c} \prod_{A = X,Y} \prod_{k = 0}^{n-1} \frac{z_i-z_j \upsilon_A^{-1} x^{|m_i-m_j|+2 k+2}}{z_j-z_i \upsilon_A x^{|m_i-m_j|+2 k}}\right)
\end{align}
where $Z_n$ is the integrand of $I_n$. For $N_c = 1$, there is no derivative such that the computation is simple:
\begin{align}
\mathrm{Res}_{z = 0} \, Z_n &\sim \prod_{a = 1}^{N_f} \prod_{k = 0}^{n-1} t_a^{-1} \tilde t_a^{-1} \tau^{-2} x^2.
\end{align}
The residue doesn't vanish for finite $n$ while it converges to zero as $n$ goes to infinity given that $|t_a^{-1} \tilde t_a^{-1} \tau^{-2} x^2| < 1$. For $N_c > 1$, the residue vanishes due to the factor $\prod \left(z_j-z_i x^{|m_i-m_j|}\right)$. Derivatives can kill a part of them but not all. Note that this is still true when a part of $z_j$ take finite values. Thus, in our examples, we don't have asymptotic residue contributions.
\\

\section{Solutions for Higgs Vacua}
\label{sec:vacua}
In this section, we present explicit solutions of the Higgs vacua of the $U(N)$ theory with $N_f=1$ studied in the section \ref{Sec:Higgs_Vacua}. Let us summarize the form of solutions for the scalar field in the vector multiplet $\sigma$ obtained in the previous section. The solutions for $\sigma$ have the form of $\sigma = \textrm{diag}(\sigma_1,\ldots, \sigma_N)$ and
\begin{align}
\sigma_i = m+a_i m_X+b_i m_Y \label{Higgs_vacua_sigma}
\end{align}
where $m_X$ and $m_Y$ are real masses for $X$ and $Y$ respectively, and $a_i$, $b_i$ are non-negative integers.
$\sigma_i$ are labelled by the restricted Young tableaux. The maximum size of the restricted Young tableaux is $n\times 3$. Furthermore, if the third column of the restricted Young tableaux is non-empty the first column is always fully occupied. We fix the Weyl symmetry of the Higgs vacua as $\sigma_i < \sigma_j$ for all $i,j$.

If a solution for $\sigma$ is obtained then the corresponding solutions for the chiral fields $Q$, $\tilde Q$, $X$, $Y$ are uniquely determined by the real mass term, the $D$-term and the $F$-term equations. Because of the solution \eqref{Higgs_vacua_sigma} unique solution for chiral fields $Q$ and $\tilde Q$ are given by
\begin{align}
|Q|^2= (N\zeta, 0,\ldots, 0)\quad \textrm{and}\quad   |\tilde Q|^2 =(0,\ldots, 0)~.
\end{align}
The other chiral fields $X$ and $Y$ have non-zero elements only at $X_{ij}$ and $Y_{i'j'}$ such that
\begin{align}
\sigma_i-\sigma_j = m_X, \qquad \sigma_{i'}-\sigma_{j'} =m_Y,
\end{align}
which comes from the real mass term. $X$ and $Y$ can have the unique nonzero element in each row and column.
Then, the non-trivial elements of $X$ and $Y$ are fixed by the $F$-term equations
\begin{align}
&X_{is} Y_{sj} + Y_{is'} X_{s'j}~, \\
&X_{is_{n-1}} \cdots X_{s_2 s_1} X_{s_1j}+Y_{is'} Y_{s'j} = 0
\end{align}
and the $D$-term equations
\begin{align}
\zeta-\delta_{i1} |Q|^2-|X_{is}|^2+|X_{ti}|^2-|Y_{is'}|^2+|Y_{t'i}|^2 = 0~.
\end{align}
Therefore, all Higgs vacuum solutions for the chiral fields can be determined from the solutions for $\sigma$, which are labelled by the restricted Young tableaux. Let's give examples for $n=3$ cases.

\subsection{Higgs vacua for the $U(N)$ theories with $n=3$, $N_f=1$}
In this section, we give explicit Higgs vacuum solutions for $\sigma$, $X$ and $Y$ of the $U(N)$ theories with $n=3$, $N_f=1$.
The $U(1)$ theory has a unique solution, which is given by
\begin{align}
\sigma=0,\quad X=Y=0~.
\end{align}

\subsection*{$U(2)$ theory}
\begin{figure}[htbp]
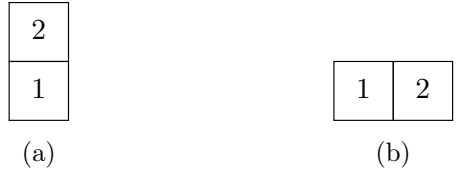

\centering
\begin{subfigure}{0.3\textwidth}
\centering
\ytableausetup{mathmode, boxsize=2em}
 \begin{ytableau}
 2 \\  1
  \end{ytableau}
\caption{ \label{Tree:n3_U(2)_1}}
\end{subfigure}
\begin{subfigure}{0.3\textwidth}
\centering
\ytableausetup{mathmode, boxsize=2em}
 \begin{ytableau}
\none \\ 1 & 2
  \end{ytableau}
\caption{ \label{Tree:n3_U(2)_2}}
\end{subfigure}
\caption{ \label{Tree:n3_U(2)} The box diagrams for the $U(2)$ theory with $n=3$, $N_f=1$. The numbers in the box represent gauge indices $i$ of $\sigma_i$.}
\end{figure}

The $U(2)$ theory has two solutions and corresponding box diagrams are given in figure \ref{Tree:n3_U(2)}.
Note that the numbers in the boxes are gauge indices and we fix the gauge indices by requiring $\sigma_i < \sigma_j$ for all $i,\,j$.

The first solution corresponding to figure \ref{Tree:n3_U(2)_1} is given by
\begin{align} \label{eq:vacuumU(2)Nf1n3-1}
\begin{aligned}
& \sigma = (m,  m+m_X)~,\\
&X=\left(\begin{array}{cc}
0 & 0  \\
x_1  & 0 \\
\end{array}\right), \quad Y=0
\quad |x_1|^2=\zeta~.
\end{aligned}
\end{align}
The second solution corresponding to figure \ref{Tree:n3_U(2)_2} is given by
\begin{align} \label{eq:vacuumU(2)Nf1n3-2}
\begin{aligned}
&\sigma = (m,  m+m_Y)~,\\
&X=0,\quad
Y=\left(\begin{array}{cc}0 & 0  \\ y_1  & 0 \\ \end{array}\right),
\quad |y_1|^2=\zeta~.
\end{aligned}
\end{align}

\subsection*{$U(3)$ theory}
The $U(3)$ theory has two solutions and corresponding box diagrams are given in figure \ref{Tree:n3_U(3)}.
\begin{figure}[htbp]
\centering
\begin{subfigure}{0.3\textwidth}
\centering
\ytableausetup{mathmode, boxsize=2em}
 \begin{ytableau}
3  \\ 2 \\  1
  \end{ytableau}
\caption{ \label{Tree:n3_U(3)_1}}
\end{subfigure}
\begin{subfigure}{0.3\textwidth}
\centering
\ytableausetup{mathmode, boxsize=2em}
 \begin{ytableau}
\none \\ 2 \\  1 &3
  \end{ytableau}
\caption{ \label{Tree:n3_U(3)_2}}
\end{subfigure}
\caption{\label{Tree:n3_U(3)} The box diagrams for the $U(3)$ theory with $n=3$, $N_f=1$}
\end{figure}

The solution corresponding to figure \ref{Tree:n3_U(3)_1} is given by
\begin{align} \label{eq:vacuumU(3)Nf1n3-1}
\begin{aligned}
&\sigma=(m, m+m_X, m+2m_X)~,\\
&X=\left(\begin{array}{ccc}
0 & 0 &0\\
x_1 & 0 &0 \\
0  & x_2 & 0
\end{array}\right),\quad Y=0,\quad |x_i|^2=\left(2\zeta, \zeta\right)~,
\end{aligned}
\end{align}
and the solution corresponding to figure \ref{Tree:n3_U(3)_2} is given by
\begin{align} \label{eq:vacuumU(3)Nf1n3-2}
\begin{aligned}
&\sigma=(m,  m+m_X, m+m_Y)~,\\
&X=\left(\begin{array}{cccc}
0 & 0 & 0\\
x_1 & 0 & 0\\
0 & 0 &0
\end{array}\right),\quad
Y=\left(\begin{array}{ccc}
0 & 0 & 0 \\
0 & 0&0\\
y_1 & 0&0
\end{array}\right),\quad |x_1|^2=\zeta,\quad |y_1|^2=\zeta ~.
\end{aligned}
\end{align}

\subsection*{$U(4)$ theory}
The $U(4)$ theory has two solutions and corresponding box diagrams are given in figure \ref{Tree:n3_U(4)}.
\begin{figure}[htbp]
\centering
\begin{subfigure}{0.3\textwidth}
\centering
\ytableausetup{mathmode, boxsize=2em}
 \begin{ytableau}
4  \\  2 \\  1 & 3
  \end{ytableau}
\caption{ \label{Tree:n3_U(4)_1}}
\end{subfigure}
\begin{subfigure}{0.3\textwidth}
\centering
\ytableausetup{mathmode, boxsize=2em}
 \begin{ytableau}
\none  \\  2 & 4 \\  1 & 3
  \end{ytableau}
\caption{ \label{Tree:n3_U(4)_2}}
\end{subfigure}
\caption{\label{Tree:n3_U(4)} The box diagrams for the $U(4)$ theory with $n=3$, $N_f=1$}
\end{figure}

The solution corresponding to figure \ref{Tree:n3_U(4)_1} is given by
\begin{align} \label{eq:vacuumU(4)Nf1n3-1}
\begin{aligned}
&\sigma=(m, m+m_X, m+m_Y, m+2m_X)~,\\
&X=\left(\begin{array}{cccc}
0 & 0 & 0 & 0\\
x_1 & 0 & 0 &0\\
0 & 0 & 0 &0 \\
0 & x_2 & 0 &0
\end{array}\right),\quad Y=\left(\begin{array}{cccc}
0 & 0 & 0 & 0 \\
0 & 0 & 0&0\\
y_1 & 0 & 0 &0\\
0 & 0 & 0 & 0
\end{array}\right),\\
&|x_i|^2=\left(2\zeta, \zeta\right),\quad  |y_1|^2=\zeta ~,
\end{aligned}
\end{align}
and the solution corresponding to figure \ref{Tree:n3_U(4)_2} is given by
\begin{align} \label{eq:vacuumU(4)Nf1n3-2}
\begin{aligned}
& \sigma=(m, m+m_X,  m+m_Y, m+m_X+m_Y)~,\\
&X=\left(\begin{array}{cccc}
0 & 0 & 0 & 0\\
x_1 & 0 & 0 &0\\
0 & 0 & 0 &0 \\
0 & 0 & -\frac{x_1 y_2}{y_1} &0
\end{array}\right),\quad Y=\left(\begin{array}{cccc}
0 & 0 & 0 & 0 \\
0 & 0 & 0&0\\
y_1 & 0 & 0 &0\\
0 & y_2 & 0 & 0
\end{array}\right),\\
&|x_1|^2=\frac{3\zeta}{2},\quad |y_i|^2=\left( \frac{3\zeta}{2},  \frac{\zeta}{2}\right)~.
\end{aligned}
\end{align}

\subsection*{$U(5)$ theory}
The $U(5)$ theory has two solutions and corresponding box diagrams are given in figure \ref{Tree:n3_U(5)}.
\begin{figure}[htbp]
\centering
\begin{subfigure}{0.3\textwidth}
\centering
\ytableausetup{mathmode, boxsize=2em}
 \begin{ytableau}
4  \\  2 & 5 \\  1 & 3
  \end{ytableau}
\caption{ \label{Tree:n3_U(5)_1} }
\end{subfigure}
\begin{subfigure}{0.3\textwidth}
\centering
\ytableausetup{mathmode, boxsize=2em}
 \begin{ytableau}
4  \\  2 \\ 1 & 3 & 5
  \end{ytableau}
\caption{ \label{Tree:n3_U(5)_2}}
\end{subfigure}
\caption{\label{Tree:n3_U(5)} The box diagrams for the $U(5)$ theory with $n=3$, $N_f=1$}
\end{figure}

The solution corresponding to figure \ref{Tree:n3_U(5)_1} is given by
\begin{align} \label{eq:vacuumU(5)Nf1n3-1}
\begin{aligned}
& \sigma=(m, m+m_X, m+m_Y, m+2m_X, m+m_X+m_Y)~,\\
&X=\left(\begin{array}{ccccc}
0 & 0 & 0 & 0 &0\\
x_1 & 0 & 0 & 0&0\\
0 & 0 & 0 & 0&0 \\
0 & x_2 & 0 & 0&0\\
0 & 0 & -\frac{x_1 y_2}{y_1} & 0&0
\end{array}\right),\quad
Y=\left(\begin{array}{ccccc}
0 & 0 & 0 & 0 & 0 \\
0 & 0 & 0 &  0&0\\
y_1 & 0 & 0 &0&0\\
0 & 0 & 0 & 0&0\\
0 & y_2 & 0 & 0 & 0
\end{array}\right),\\
&|x_i|^2=\left(\frac{12\zeta}{5}, \zeta\right),\quad |y_i|^2=\left(\frac{8\zeta}{5}, \frac{2\zeta}{5} \right)~,
\end{aligned}
\end{align}
and the solution corresponding to figure \ref{Tree:n3_U(5)_2} is given by
\begin{align} \label{eq:vacuumU(5)Nf1n3-2}
\begin{aligned}
& \sigma=(m, m+m_X, m+m_Y, m+2m_X, m+2m_Y) ~,\\
&X=\left(\begin{array}{ccccc}
0 & 0 & 0 & 0 &0\\
x_1 & 0 & 0 & 0&0\\
0 & 0 & 0 & 0&0 \\
0 & x_2 & 0 & 0&0\\
0 & 0 & 0 & x_3 &0
\end{array}\right),\quad
Y=\left(\begin{array}{ccccc}
0 & 0 & 0 & 0 & 0 \\
0 & 0 & 0 &  0&0\\
y_1 & 0 & 0 &0&0\\
0 & 0 & 0 & 0&0\\
0 & 0 & y_2 & 0 & 0
\end{array}\right),\\
& |x_i|^2=\left(\alpha +2\zeta, \alpha +\zeta, \alpha \right),\quad |y_i|^2=\left( -\alpha+2\zeta,  -\alpha +\zeta \right)
\end{aligned}
\end{align}
where $\alpha$ is defined by $|x_3|^2=\alpha$ and it is fixed by the F-term equation $X^3+Y^2=0$, which is reduced to $x_1 x_2 x_3 + y_1 y_2=0$. $\alpha$ is the real positive solution of the F-term equation
\begin{align}
\alpha(\alpha+\zeta)(\alpha+2\zeta)=(\alpha-2\zeta)(\alpha-\zeta)
\end{align}
at $0 < \alpha < \zeta$.

\subsection*{$U(6)$ theory}
\begin{figure}[htbp]
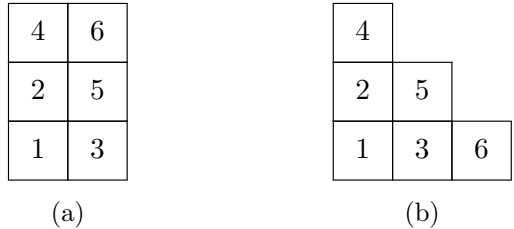

\centering
\begin{subfigure}{0.3\textwidth}
\centering
\ytableausetup{mathmode, boxsize=2em}
 \begin{ytableau}
4 & 6 \\ 2 & 5 \\ 1 & 3
  \end{ytableau}
\caption{ \label{Tree:n3_U(6)_1}}
\end{subfigure}
\begin{subfigure}{0.3\textwidth}
\centering
\ytableausetup{mathmode, boxsize=2em}
 \begin{ytableau}
4  \\ 2 & 5 \\ 1 & 3 & 6
  \end{ytableau}
\caption{ \label{Tree:n3_U(6)_2}}
\end{subfigure}
\caption{\label{Tree:n3_U(6)} The box diagrams for the $U(6)$ theory with $n=3$, $N_f=1$}
\end{figure}

The $U(6)$ theory has two solutions and corresponding box diagrams are given in figure \ref{Tree:n3_U(6)}.

The solution corresponding to figure \ref{Tree:n3_U(6)_1} is given by
\begin{align} \label{eq:vacuumU(6)Nf1n3-2}
\begin{aligned}
& \sigma=(m, m+m_X, m+m_Y, m+2m_X, m+m_X+m_Y, m+2m_X+m_Y)~,\\
&X=\left(\begin{array}{cccccc}
0 & 0 & 0 & 0 &0 &0\\
x_1 & 0 & 0 & 0&0 &0\\
0 & 0 & 0 & 0 &0&0\\
0 & x_2 & 0 & 0&0&0\\
0 & 0 & -\frac{x_1 y_2}{y_1} & 0&0&0\\
0 &0 & 0 & 0 & -\frac{x_2 y_3}{y_2} & 0
\end{array}\right),\quad
Y=\left(\begin{array}{cccccc}
0 & 0 & 0 & 0 & 0 &0\\
0 & 0 & 0 & 0 & 0&0\\
y_1 & 0 & 0 & 0 &0&0\\
0 & 0 & 0 & 0 & 0&0\\
0 & y_2 & 0 & 0 & 0 &0\\
0  &0 & 0 & y_3 & 0 &0
\end{array}\right),\\
&|x_i|^2=\left(3\zeta, \frac{4\zeta}{3}\right),\quad |y_i|^2=\left(2\zeta, \frac{2\zeta}{3}, \frac{\zeta}{3} \right)~,
\end{aligned}
\end{align}
and the solution corresponding to figure \ref{Tree:n3_U(6)_2} is given by
\begin{align} \label{eq:vacuumU(6)Nf1n3-1}
\begin{aligned}
& \sigma=(m, m+m_X, m+m_Y, m+2m_X, m+m_X+m_Y, m+2m_Y)~,\\
&X=\left(\begin{array}{ccccccc}
0 & 0 & 0 & 0 &0 &0\\
x_1 & 0 & 0 & 0&0&0\\
0 & 0 & 0 & 0 &0&0\\
0 & x_2 & 0 &0&0&0\\
0 &0 &  -\frac{x_1 y_2}{y_1} &0&0&0\\
0 &0 & 0 & x_3 & 0 & 0
\end{array}\right),\quad
Y=\left(\begin{array}{cccccc}
0 & 0 & 0 & 0 & 0 &0\\
0 & 0 & 0 & 0 & 0&0\\
y_1 & 0 & 0 & 0&0&0\\
0 & 0 & 0 & 0 & 0&0\\
0 & y_2 & 0 & 0 &0&0\\
0  &0 & y_3 & 0 &0&0
\end{array}\right),\\
&|x_i|^2=\left(\frac{5(\alpha +3\zeta)}{6}, \alpha +\zeta, \alpha \right),\quad |y_i|^2=\left(-\frac{5(\alpha-3\zeta)}{6},  -\frac{\alpha-3\zeta}{6} ,  -\alpha+\zeta\right)
\end{aligned}
\end{align}
where $\alpha$ is the real positive solution of the F-term equation
\begin{align}
\alpha(\alpha+3\zeta)(\alpha+\zeta)=(\alpha-3\zeta)(\alpha-\zeta)
\end{align}
at $0 < \alpha < \zeta$.

\subsection*{$U(7)$ theory}
\begin{figure}[htbp]
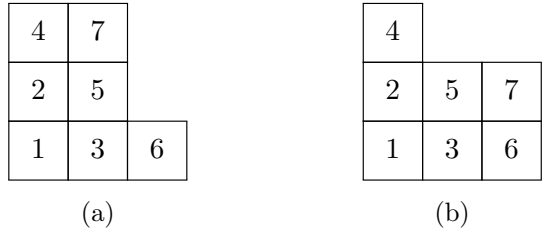

\centering
\begin{subfigure}{0.3\textwidth}
\centering
\ytableausetup{mathmode, boxsize=2em}
 \begin{ytableau}
4 & 7 \\ 2 & 5 \\ 1 & 3 & 6
  \end{ytableau}
\caption{ \label{Tree:n3_U(7)_1}}
\end{subfigure}
\begin{subfigure}{0.3\textwidth}
\centering
\ytableausetup{mathmode, boxsize=2em}
 \begin{ytableau}
4 \\ 2 & 5 & 7 \\ 1 & 3 & 6
  \end{ytableau}
\caption{ \label{Tree:n3_U(7)_2}}
\end{subfigure}
\caption{\label{Tree:n3_U(7)} The box diagrams for the $U(7)$ theory with $n=3$, $N_f=1$}
\end{figure}

The $U(7)$ theory has two solutions and corresponding box diagrams are given in figure \ref{Tree:n3_U(7)}.

The solution corresponding to figure \ref{Tree:n3_U(7)_1} is given by
\begin{align} \label{eq:vacuumU(7)Nf1n3-1}
\begin{aligned}
& \sigma=(m, m+m_X, m+m_Y, m+2m_X, m+m_X+m_Y, m+2m_Y, m+2m_X+m_Y)~,\\
&X=\left(\begin{array}{ccccccc}
0 & 0 & 0 & 0 &0 &0&0\\
x_1 & 0 & 0 & 0&0 &0&0\\
0 & 0 & 0 & 0 &0 &0&0\\
0 & x_2 & 0 & 0&0&0&0\\
0 & 0 & -\frac{x_1 y_2}{y_1} & 0&0&0&0\\
0 &0 & 0 & x_3 & 0 & 0&0\\
0 & 0 & 0 & 0 &- \frac{x_2 y_4}{y_2}&0&0
\end{array}\right),\quad
Y=\left(\begin{array}{ccccccc}
0 & 0 & 0 & 0 & 0 &0&0\\
0 & 0 & 0 & 0 & 0&0&0\\
y_1 & 0 & 0 & 0 & 0&0&0\\
0 & 0 & 0 & 0 & 0&0&0\\
0 & y_2 & 0 &0 & 0 &0&0\\
0  &0 &  y_3 & 0 & 0 &0&0\\
0  &0 & 0 & y_4 & 0 &0&0
\end{array}\right),\\
&|x_i|^2=\left(\frac{5\alpha +22\zeta}{7},  \frac{5(\alpha +2\zeta)(\alpha +3\zeta)}{5\alpha +22\zeta}, \alpha \right)\\
&|y_i|^2=\left(-\frac{5(\alpha-4\zeta)}{7},  -\frac{10(\alpha-4\zeta)(\alpha+3\zeta)}{7(5\alpha+22\zeta)},  -\alpha+\zeta,  -\frac{2\zeta(\alpha-4\zeta)}{5\alpha+22\zeta} \right)~,
\end{aligned}
\end{align}
where $\alpha$ is the real positive solution of the F-term equation
\begin{align}
\alpha(\alpha+2\zeta)(\alpha+3\zeta)=(\alpha-4\zeta)(\alpha-\zeta)
\end{align}
at $0 < \alpha < \zeta$.
The solution corresponding to figure \ref{Tree:n3_U(7)_2} is given by
\begin{align} \label{eq:vacuumU(7)Nf1n3-2}
\begin{aligned}
& \sigma=(m, m+m_X, m+m_Y, m+2m_X, m+m_X+m_Y, m+2m_Y, m+m_X+2m_Y)~,\\
&X=\left(\begin{array}{ccccccc}
0 & 0 & 0 & 0 &0 &0&0\\
x_1 & 0 & 0 & 0&0 &0&0\\
0 & 0 & 0 & 0 &0 &0&0\\
0 & x_2 & 0 & 0&0&0&0\\
0 & 0 & -\frac{x_1 y_2}{y_1} & 0&0&0&0\\
0 & 0 & 0 & x_3 & 0&0&0\\
0 & 0 & 0 & 0 & 0&\frac{x_1 y_2 y_4}{y_1 y_3}&0
\end{array}\right),\quad
Y=\left(\begin{array}{ccccccc}
0 & 0 & 0 & 0 & 0 &0&0\\
0 & 0 & 0 & 0 & 0&0&0\\
y_1 & 0 & 0 & 0 & 0&0&0\\
0 & 0 & 0 & 0 & 0&0&0\\
0 & y_2 & 0 & 0 & 0 &0&0\\
0  &0 & y_3 & 0 &0 &0&0\\
0  &0 & 0 & 0 &  y_4 &0&0
\end{array}\right),\\
&|x_i|^2=\left(\frac{5(\alpha +4\zeta)}{7},  \alpha +\zeta, \alpha \right)\\
&|y_i|^2=\left(-\frac{5\alpha-22\zeta}{7},  -\frac{2(\alpha-3\zeta)}{7},  -\frac{5(\alpha-3\zeta)(\alpha-2\zeta)}{5\alpha-22\zeta},  \frac{7\zeta(\alpha-2\zeta)}{5\alpha-22\zeta} \right)
\end{aligned}
\end{align}
where $\alpha$ is the real positive solution of the F-term equation
\begin{align}
\alpha(\alpha+4\zeta)(\alpha+\zeta)=(\alpha-3\zeta)(\alpha-2\zeta)
\end{align}
at $0 < \alpha < 2\zeta$.

\subsection*{$U(8)$ theory}
The $U(8)$ theory has a unique solution and corresponding box diagram is given in figure \ref{Tree:n3_U(8)}.
\begin{figure}[htbp]
\centering
\ytableausetup{mathmode, boxsize=2em}
 \begin{ytableau}
4 & 7 \\ 2 & 5 & 8 \\ 1 & 3 & 6
  \end{ytableau}
  \caption{\label{Tree:n3_U(8)} The box diagram for the $U(8)$ theory with $n=3$, $N_f=1$}
\end{figure}

The solution corresponding to figure \ref{Tree:n3_U(8)} is given by
\begin{align} \label{eq:vacuumU(8)Nf1n3-1}
\begin{aligned}
& \sigma=(m, m+m_X, m+m_Y, m+2m_X, m+m_X+m_Y, m+2m_Y, m+2m_X+m_Y, m+m_X+2m_Y)~,\\
&X=\left(\begin{array}{cccccccc}
0 & 0 & 0 & 0 &0 &0&0&0\\
x_1 & 0 & 0 & 0&0 &0&0&0\\
0 & 0 & 0 & 0 &0 &0&0&0\\
0 & x_2 & 0 & 0&0&0&0&0\\
0 & 0 & -\frac{x_1 y_2}{y_1} & 0&0&0&0&0\\
0 & 0 & 0 & x_3 & 0&0&0&0\\
0 & 0 & 0 & 0 & -\frac{x_2 y_4}{y_2} &0&0&0\\
0 & 0 & 0 & 0 & 0&\frac{x_1 y_2 y_4}{y_1 y_3}&0&0
\end{array}\right),\quad
Y=\left(\begin{array}{cccccccc}
0 & 0 & 0 & 0 & 0 &0&0&0\\
0 & 0 & 0 & 0 & 0&0&0&0\\
y_1 & 0 & 0 & 0 & 0&0&0&0\\
0 & 0 & 0 & 0 & 0&0&0&0\\
0 & y_2 & 0 & 0 & 0 &0&0&0\\
0  &0  & y_3& 0 & 0 &0&0&0\\
0  &0 & 0 &  y_4 &0 &0&0&0\\
0  &0 & 0 & 0 & y_5&0 &0&0
\end{array}\right),\\
&|x_i|^2=\left(\frac{5\alpha +28\zeta}{8},  \frac{5(\alpha+2\zeta)(\alpha+4\zeta)}{5\alpha +28\zeta}, \alpha \right)\\
&|y_i|^2=\left(-\frac{5\alpha-28\zeta}{8},  -\frac{15(\alpha-4\zeta)(\alpha+4\zeta)}{8(5\alpha+28\zeta)},  -\frac{5(\alpha-4\zeta)(\alpha-2\zeta)}{5\alpha-28\zeta},  -\frac{3\zeta(\alpha-4\zeta)}{5\alpha+28\zeta},  \frac{8\zeta(\alpha-2\zeta)}{5\alpha-28\zeta} \right)
\end{aligned}
\end{align}
where $\alpha$ is the real positive solution of the F-term equation
\begin{align}
\alpha(\alpha+2\zeta)(\alpha+4\zeta)=(\alpha-4\zeta)(\alpha-2\zeta)
\end{align}
at $0 < \alpha < 2\zeta$.

\subsection*{$U(9)$ theory}
The $U(9)$ theory has a unique solution and corresponding box diagram is given in figure \ref{Tree:n3_U(9)}.
\begin{figure}[htbp]
\centering
\ytableausetup{mathmode, boxsize=2em}
 \begin{ytableau}
4 & 7 & 9 \\ 2 & 5 & 8 \\ 1 & 3 & 6
  \end{ytableau}
  \caption{\label{Tree:n3_U(9)} The box diagram for the $U(9)$ theory with $n=3$, $N_f=1$}
\end{figure}

The solution corresponding to figure \ref{Tree:n3_U(9)} is given by
\begin{align} \label{eq:vacuumU(9)Nf1n3-1}
\begin{aligned}
& \sigma=(m, m+m_X, m+m_Y, m+2m_X, m+m_X+m_Y, m+2m_Y, m+2m_X+m_Y,\\
&\qquad \quad  m+m_X+2m_Y , m+2m_X+2m_Y)~,\\
&X=\left(\begin{array}{ccccccccc}
0 & 0 & 0 & 0 &0 &0&0&0&0\\
x_1 & 0 & 0 & 0&0 &0&0&0&0\\
0 & 0 & 0 & 0 &0 &0&0&0&0\\
0 & x_2 & 0 & 0&0&0&0&0&0\\
0 & 0 & -\frac{x_1 y_2}{y_1} & 0&0&0&0&0&0\\
0 & 0 & 0 & x_3 & 0&0&0&0&0\\
0 & 0 & 0 & 0 & -\frac{x_2 y_4}{y_2} &0&0&0&0\\
0 & 0 & 0 & 0 & 0 & \frac{x_1 y_2 y_4}{y_1 y_3} &0&0&0\\
0 & 0 & 0 & 0 & 0&0&0&\frac{x_2y_4y_6}{y_2 y_5}&0
\end{array}\right),\quad
Y=\left(\begin{array}{ccccccccc}
0 & 0 & 0 & 0 & 0 &0&0&0&0\\
0 & 0 & 0 & 0 & 0&0&0&0&0\\
y_1 & 0 & 0 & 0 & 0&0&0&0&0\\
0 & 0 & 0 & 0 & 0&0&0&0&0\\
0 & y_2 & 0 & 0 & 0 &0&0&0&0\\
0  &0 & y_3 & 0 & 0 &0&0&0&0\\
0  &0 & 0 & y_4 & 0 &0&0&0&0\\
0  &0 & 0 & 0 & y_5 &0&0&0&0\\
0  &0 & 0 & 0 & 0 &0&y_6&0&0
\end{array}\right)\\
&|x_i|^2=\left(\frac{5\alpha +36\zeta}{9},  \frac{5(\alpha+3\zeta)(\alpha+4\zeta)}{5\alpha +36\zeta}, \alpha \right)\\
&|y_i|^2=\left(-\frac{5\alpha-36\zeta}{9},  -\frac{4(5\alpha-108\zeta)}{9(5\alpha+36\zeta)},  -\frac{5(\alpha-4\zeta)(\alpha-3\zeta)}{5\alpha-36\zeta}, -\frac{6\zeta(\alpha-4\zeta)}{5\alpha+36\zeta}, \right.\\
&\qquad \qquad\left.   \frac{135\zeta(\alpha-4\zeta)(\alpha-3\zeta)(\alpha+4\zeta)}{(5\alpha-36\zeta)(5\alpha-108\zeta)},  \frac{\zeta(\alpha-3\zeta)(5\alpha+36\zeta)}{2(5\alpha-108\zeta)} \right)
\end{aligned}
\end{align}
where $\alpha$ is the real positive solution of the F-term equation
\begin{align}
\alpha(\alpha+3\zeta)(\alpha+4\zeta)=(\alpha-4\zeta)(\alpha-3\zeta)
\end{align}
at $0 < \alpha < 3\zeta$.

\section{Independent Monopole Opertors}
\label{sec:monopole}

In this appendix, we justify the claims made at eq. \eqref{eq:monop} about the independent monopole operators. To show that the independent monopole operators are given by \eqref{eq:monop}, we attempt to find the states in the Hilbert space on $S^2$ corresponding to those monopole operators, which are expected to be \eqref{eq:monopoles1}.

First we start from the theory without the superpotential. In general, for a theory with two adjoints and without superpotential, one can work out the global charges of bare monopole states as follows:
\begin{align}
\begin{aligned}
R &= N_f (1-r_Q) \sum_{i = 1}^{N_c} |m_i|+(1-r_X-r_Y) \sum_{i < j} |m_i-m_j|, \\
F_A &= -N_f \sum_{i = 1}^{N_c} |m_i|, \\
F_X &= F_Y = -\sum_{i,j = 1}^{N_c} |m_i-m_j|/2
\end{aligned}
\end{align}
where $R$, $F_A$, $F_X$ and $F_Y$ are the global charges under $U(1)_R \times U(1)_A \times U(1)_X \times U(1)_Y$. $r_{Q,X,Y}$ are the $R$-charges of $Q, X, Y$ respectively. For a scalar BPS state, the energy is the same as the $R$-charge, which will be mapped to the conformal dimension of a chiral operator in $\mathbb R^3$ by the radial quantization. We have found that the following chiral operators are independent and can span arbitrary bare monopole states:
\begin{align}
\begin{aligned}
& V_{0,0}^{([1]^p,[0]^{N_c-p})} \sim \left|[1]^p,[0]^{N_c-p}\right>, \,\, p=1
\cdots N_c
 \\
& V_{0,0}^{([0]^{N_c-p},[-1]^p)} \sim \left|[0]^{N_c-p},[-1]^p\right>.
\end{aligned}
\end{align}
$[m]^p$ indicates that $m$ is repeated $p$-times. An arbitrary bare monopole state is then spanned as follows:
\begin{align}
\label{eq:generic monopoles}
\left|m_1,m_2,\ldots,m_{N_c}\right> &\sim \left(\prod_{p = 1}^{n-1} \left(V_{0,0}^{([1]^p,[0]^{N_c-p})}\right)^{m_p-m_{p+1}}\right) \left(V_{0,0}^{([1]^n,[0]^{N_c-n})}\right)^{m_n} \nonumber \\
&\times \left(V_{0,0}^{([0]^{n},[-1]^{N_c-n})}\right)^{-m_{n+1}} \left(\prod_{p = 1}^{N_c-n-1} \left(V_{0,0}^{([0]^{N_c-p},[-1]^p)}\right)^{m_{N_c-p}-m_{N_c-p+1}}\right)
\end{align}
where $m_1 \geq m_2 \geq \ldots \geq m_n \geq 0 \geq m_{n+1} \geq \ldots \geq m_{N_c}$. Note that for the monopole states
\begin{align}
\begin{aligned}
&\left|[m]^p,[0]^{N_c-p}\right> \sim \left(V^{([1]^p,[0]^{N_c-p})}_{0,0}\right)^m, \\
&\left|[0]^{N_c-p},[-m]^p\right> \sim \left(V^{([0]^{N_c-p},[-1]^p)}_{0,0}\right)^m,
\end{aligned}
\end{align}
the gauge group is effectively broken to $U(p) \times U(N_c-p)$. One can dress those bare monopole states by the adjoints invariant under this unbroken gauge group, which yield additional independent chiral operators as we will see.
\\

From \eqref{eq:higher n} we expect two independent charge-2 monopole operators for $n = 5$, and more for the vanishing superpotential. Let us demonstrate them in the $U(4)$ theory with $N_f = 1$. We assign the trial $R$-charges $\frac{1}{6}$, $\frac{1}{3}$ and $\frac{5}{6}$ for the (anti-)fundamentals and the two adjoints respectively. Although we will focus on the positive charge sector, the negative charge sector can be explained in exactly the same way.

\subsubsection*{Topological Charge 1}

Firstly, for topological charge 1, we have the independent monopole operators $V^+_{s,t} \equiv V^{(1,0,0,0)}_{s,t}$ with $s,t = 0,1,2,3$ for the vanishing superpotential, which can be read off from $\mathrm{PL}[I]$ as follows. The topological charge-1 sector of $\mathrm{PL}[I]$ is given by
\begin{align}
\left.\mathrm{PL}[I]\right|_{w^1} &= x^\frac{1}{3} \tau^{-1} \upsilon_X^{-3} \upsilon_Y^{-3}+x^\frac{2}{3} \tau^{-1} \upsilon_X^{-2} \upsilon_Y^{-3}+x \tau^{-1} \upsilon_X^{-1} \upsilon_Y^{-3}+x^\frac{7}{6} \tau^{-1} \upsilon_X^{-3} \upsilon_Y^{-2}+x^\frac{4}{3} \tau^{-1} \upsilon_Y^{-3} \nonumber \\
& +x^\frac{3}{2} \left(-\tau^{-1} \upsilon_X^{-3} \upsilon_Y^{-4}+\tau^{-1} \upsilon_X^{-2} \upsilon_Y^{-2}\right)-x^\frac{5}{3} \tau \upsilon_Y^{-3}+x^\frac{11}{6} \left(-\tau^{-1} \upsilon_X^{-2} \upsilon_Y^{-4}+\tau^{-1} \upsilon_X^{-1} \upsilon_Y^{-2}\right) \nonumber \\
& +x^2 \left(-\tau^{-1} \upsilon_X^{-6} \upsilon_Y^{-7}-\tau^{-3} \upsilon_X^{-5} \upsilon_Y^{-7}-\tau^{-1} \upsilon_X^{-4} \upsilon_Y^{-3}-\tau \upsilon_X \upsilon_Y^{-3}+\tau^{-1} \upsilon_X^{-3} \upsilon_Y^{-1}\right)+\mathcal O(x^\frac{13}{6}).
\end{align}
The positive terms are from the following bosonic chiral operators:
\begin{align}
\label{eq:V}
V^+_{0,0}, \quad V^+_{1,0}, \quad V^+_{2,0}, \quad V^+_{0,1}, \quad V^+_{3,0}, \quad V^+_{1,1}, \quad V^+_{2,1}, \quad V^+_{0,2}, \quad \ldots
\end{align}
respectively. One should note that, for each $V^+_{s,1}$ with $s > 0$, there is an accompanying negative term $-x^{\frac{7}{6}+\frac{s}{3}} \tau^{-1} \upsilon_X^{-4+s} \upsilon_Y^{-4}$, which is from a fermionic state
\begin{align}
\label{eq:negative}
\mathrm{Tr}_2 X^{s-1} \psi_Y^\dagger \left|1,0,0,0\right>.
\end{align}
If we introduce the superpotential \eqref{eq:sup_D}, $\upsilon_i$'s should be turned off such that $V^+_{s,1}$ and \eqref{eq:negative} become indistinguishable by the global charges. Indeed, \eqref{eq:negative} implies a chiral ring relation
\begin{align}
V^+_{s,1} \sim \mathrm{Tr}_2 \left(X^s Y\right) \left|1,0,0,0\right> \quad = \quad 0
\end{align}
for $s > 0$, which is due to the F-term
\begin{align}
X Y+Y X = 0.
\end{align}
Also the operator $V^+_{0,2}$ accompanies negative terms $-\tau^{-1} \upsilon_X^{-6} \upsilon_Y^{-7}$ and $-\tau^{-1} \upsilon_X^{-4} \upsilon_Y^{-3}$. The first one is irrelevant here because it represents a relation among the operators having the same global charges as
\begin{align}
X M_{0,0} V^{(1,1,0,0)}_{0,0} V^{(0,0,0,-1)}_{0,0},
\end{align}
which includes a mesonic operator. The second one, on the other hand, comes from the fermionic state
\begin{align}
\label{eq:negative 2}
\mathrm{Tr_2} \psi_X^\dagger \left|1,0,0,0\right>.
\end{align}
If we introduce the superpotential \eqref{eq:sup_D}, \eqref{eq:negative 2} implies a chiral ring relation
\begin{align}
V^+_{0,2} \sim \mathrm{Tr}_2 Y^2 \left|1,0,0,0\right> \quad = \quad 0.
\end{align}
This relation originates from the F-term
\begin{align}
Y^2 = -X^5,
\end{align}
which vanishes due to the low gauge rank.

We also work out a few more higher order terms and find that, for topological charge 1, the remaining independent monopole operators are $V^+_{s,t}$ with $s = 0, 1, 2, 3$, $t = 0, 1$ and $s t = 0$ once we turn on the superpotential \eqref{eq:sup_D}. Note that $V^+_{4,0}$ and $V^+_{0,2}$, which are expected from \eqref{eq:higher n}, do not appear here because the gauge group rank is low. \eqref{eq:higher n} tells us that $V^+_{4,0}$ and $V^+_{0,2}$ still appear on the dual side as elementary chiral fields even though they do not exist on the original side as monopole operators. Thus these operators should be  truncated quantum mechanically as discussed in \cite{Kim:2013cma}.

\subsubsection*{Topological Charge 2}

Secondly, we examine operators of topological charge 2. The topological charge-2 sector of $\mathrm{PL}[I]$ is given by
\begin{align}
& \left.\mathrm{PL}[I]\right|_{w^2} \nonumber \\
&= x \tau^{-2} \upsilon_X^{-4} \upsilon_Y^{-4}+x^\frac{4}{3} \left(-\tau^{-2} \upsilon_X^{-4} \upsilon_Y^{-6}+\tau^{-2} \upsilon_X^{-3} \upsilon_Y^{-4}\right)+x^\frac{5}{3} \left(-\tau^{-2} \upsilon_X^{-3} \upsilon_Y^{-6}+2 \tau^{-2} \upsilon_X^{-2} \upsilon_Y^{-4}\right) \nonumber \\
& +x^\frac{11}{6} \left(-\tau^{-2} \upsilon_X^{-5} \upsilon_Y^{-5}+\tau^{-2} \upsilon_X^{-4} \upsilon_Y^{-3}\right)+x^2 \left(-2 \tau^{-2} \upsilon_X^{-2} \upsilon_Y^{-6}-\upsilon_X^{-2} \upsilon_Y^{-4}+\tau^{-2} \upsilon_X^{-1} \upsilon_Y^{-4}\right)+\mathcal O(x^\frac{13}{6}).
\end{align}
For $x^1$, we have the unique term coming from the bare monopole state
\begin{align}
\left|1,1,0,0\right>,
\end{align}
which is identified with the chiral operator $W^+_0 \equiv V^{(1,1,0,0)}_{0,0}$.
\\

For $x^\frac{4}{3}$, we have two terms. The first negative term, from which one can read off the global charges of the corresponding operators, is explained as follows. For those given global charges, the allowed monopole states are
\begin{align}
\left(\mathrm{Tr}_1 X\right)^2 \left|2,0,0,0\right>, \quad \left(\mathrm{Tr}_1 X\right) \left(\mathrm{Tr}_2 X\right) \left|2,0,0,0\right>, \quad \left(\mathrm{Tr}_2 X\right)^2 \left|2,0,0,0\right>, \quad \mathrm{Tr}_2 \left(X^2\right) \left|2,0,0,0\right>.
\end{align}
On the other hand, the allowed chiral operators composed of the lower dimensional monopole operators, such as \eqref{eq:V}, are
\begin{align}
(X_1)^2 \left(V^+_{0,0}\right)^2, \qquad X_2 \left(V^+_{0,0}\right)^2, \qquad X_1 V^+_{1,0} V^+_{0,0}, \qquad \left(V^+_{1,0}\right)^2, \qquad V^+_{2,0} V^+_{0,0}.
\end{align}
where $X_k = \mathrm{Tr} X^k$. Since the chiral operators are one more than the allowed monopole states, the five chiral operators should satisfy one nontrivial relation
\begin{align}
f(X_s,V^+_{s,0}) &= A \, (X_1)^2 \left(V^+_{0,0}\right)^2+B \, X_2 \left(V^+_{0,0}\right)^2+C \, X_1 V^+_{1,0} V^+_{0,0}+D \, \left(V^+_{1,0}\right)^2+E \, V^+_{2,0} V^+_{0,0} \nonumber \\
&= 0
\end{align}
with unfixed coefficients $A,B,C,D,E$. This relation indeed appears in $\mathrm{PL}[I]$ as the first negative term of order $x^\frac{4}{3}$.

The second term, which is positive, is also explained as follows. Again, for the corresponding global charges, the allowed monopole states are
\begin{align}
\mathrm{Tr}_1 X \left|1,1,0,0\right>, \qquad \mathrm{Tr}_2 X \left|1,1,0,0\right>
\end{align}
while the allowed chiral operator composed of the lower dimensional monopole operators is only $X_1 W^+_0$. Thus, we should introduce another independent operator $V^{(1,1,0,0)}_{1,0}$ and have the following state/operator identifications:
\begin{align}
\begin{aligned}
\mathrm{Tr}_1 X \left|1,1,0,0\right> \sim \gamma_1 X_1 W^+_0+\delta_1 V^{(1,1,0,0)}_{1,0}, \\
\mathrm{Tr}_2 X \left|1,1,0,0\right> \sim \gamma_2 X_1 W^+_0+\delta_2 V^{(1,1,0,0)}_{1,0}.
\end{aligned}
\end{align}
$V^{(1,1,0,0)}_{1,0}$ appears in $\mathrm{PL}[I]$ as the positive second term of order $x^\frac{4}{3}$. If we turn on the superpotential \eqref{eq:sup_D}, however, the relation $f(X_s,V^+_{s,0})$ and the operator $V^{(1,1,0,0)}_{1,0}$ can be mixed; i.e.,
\begin{align}
f(X_s,V^+_{s,0}) \sim V^{(1,1,0,0)}_{1,0}.
\end{align}
Thus, $V^{(1,1,0,0)}_{1,0}$ is not independent anymore and is spanned by the charge-1 monopole operators.
\\

For $x^\frac{5}{3}$, we have three contributions: one from a relation and two from operators. Let us examine the relation first. For the corresponding global charges, the allowed monopole states are
\begin{align}
& \left(\mathrm{Tr}_1 X\right)^3 \left|2,0,0,0\right>, \quad \left(\mathrm{Tr}_1 X\right)^2 \left(\mathrm{Tr}_2 X\right) \left|2,0,0,0\right>, \quad \left(\mathrm{Tr}_1 X\right) \left(\mathrm{Tr}_2 X\right)^2 \left|2,0,0,0\right>, \nonumber \\
& \left(\mathrm{Tr}_1 X\right) \left(\mathrm{Tr}_2 X^2\right) \left|2,0,0,0\right>, \quad \left(\mathrm{Tr}_2 X\right)^3 \left|2,0,0,0\right>, \quad \left(\mathrm{Tr}_2 X\right) \left(\mathrm{Tr}_2 X^2\right) \left|2,0,0,0\right>, \nonumber \\
& \left(\mathrm{Tr}_2 X^3\right) \left|2,0,0,0\right>
\end{align}
while the allowed chiral operators composed of lower dimensional monopole operators are
\begin{align}
& (X_1)^3 \left(V^+_{0,0}\right)^2, \quad X_2 X_1 \left(V^+_{0,0}\right)^2, \quad X_3 \left(V^+_{0,0}\right)^2, \quad (X_1)^2 V^+_{1,0} V^+_{0,0}, \quad X_2 V^+_{1,0} V^+_{0,0}, \nonumber \\
& X_1 \left(V^+_{1,0}\right)^2, \qquad X_1 V^+_{2,0} V^+_{0,0}, \qquad V^+_{2,0} V^+_{1,0}, \qquad V^+_{3,0} V^+_{0,0}.
\end{align}
We have two more chiral operators than the allowed monopole states. However, since we already have an imposed relation among those operators,
\begin{align}
X_1 f(X_s,V^+_{s,0}) = 0,
\end{align}
only one more new relation is required:
\begin{align}
g(X_s,V^+_{s,0}) = 0.
\end{align}
$g(X_s,V^+_{s,0})$ appears in $\mathrm{PL}[I]$ as the negative first term of order $x^\frac{5}{3}$.

Let us now examine the operator contributions, $2 \tau^{-2} \upsilon_X^{-2} \upsilon_Y^{-4}$. The allowed monopole states are
\begin{align}
\label{eq:monopole-2}
& \left(\mathrm{Tr}_1 X\right)^2 \left|1,1,0,0\right>, \quad \left(\mathrm{Tr}_1 X^2\right) \left|1,1,0,0\right>, \quad \left(\mathrm{Tr}_1 X\right) \left(\mathrm{Tr}_2 X\right) \left|1,1,0,0\right>, \nonumber \\
& \left(\mathrm{Tr}_2 X\right)^2 \left|1,1,0,0\right>, \quad \left(\mathrm{Tr}_2 X^2\right) \left|1,1,0,0\right>
\end{align}
while the allowed chiral operators composed of the lower dimensional monopole operators are
\begin{align}
(X_1)^2 W^+_0, \qquad X_2 W^+_0, \qquad X_1 V^{(1,1,0,0)}_{1,0}.
\end{align}
Thus, we should introduce two more independent chiral operators $V^{(1,1,0,0)}_{2,0}$ and ${V'}^{(1,1,0,0)}_{2,0}$, which span the monopole states \eqref{eq:monopole-2}  together with composites of the lower dimensional monopole operators.  We emphasize that while we don't specify the exact map between those two operators and the monopole states in \eqref{eq:monopole-2}, from the counting, we do know that those two, which carry global charges
\begin{align}
F_A = -2, \qquad F_X = -2, \qquad F_Y = -4,
\end{align}
are required. The super/subscripts indicate that the operators are mapped to the monopole states of type
$X^2 \left|1,1,0,0\right>$ although there are many possible ways of taking the trace on $X^2$ as shown in \eqref{eq:monopole-2}. Now if we turn on the superpotential \eqref{eq:sup_D}, the operators $V^{(1,1,0,0)}_{2,0}$, ${V'}^{(1,1,0,0)}_{2,0}$ and the relation $g(X_s,V^+_{s,0})$ can be mixed; i.e.,
\begin{align}
g(X_s,V^+_{s,0}) = \alpha V^{(1,1,0,0)}_{2,0}+\beta {V'}^{(1,1,0,0)}_{2,0}.
\end{align}
Therefore, only one combination of $V^{(1,1,0,0)}_{2,0}$ and ${V'}^{(1,1,0,0)}_{2,0}$ can survive and is identified as $W^+_1$.
\\

Higher order terms as well can be identified in a similar way. It turns out that, in the presence of the superpotential \eqref{eq:sup_D} with $n = 5$, the surviving charge-2 monopole operators are $W^+_0$ and $W^+_1$, which we have identified in terms of monopole states on $S^2$.

\subsubsection*{Topological Charge 3}

Lastly, let us examine operators of topological charge 3. The topological charge-3 sector of $\mathrm{PL}[I]$ is given by
\begin{align}
\left.\mathrm{PL}[I]\right|_{w^3} &= x^2 \left(-\tau^{-3} \upsilon_X^{-5} \upsilon_Y^{-7}+\tau^{-3} \upsilon_X^{-3} \upsilon_Y^{-3}\right)+\mathcal O(x^3).
\end{align}
The second term originates from the operator
\begin{align}
V^{(1,1,1,0)}_{0,0}
\end{align}
while the first term originates from the relation
\begin{align}
& h(X_s,V^{(1,0,0,0)}_{s,t},V^{(1,1,0,0)}_{s,t}) \nonumber \\
&\sim (X_1)^2 V^{(1,0,0,0)}_{0,0} V^{(1,1,0,0)}_{0,0}+X_2 V^{(1,0,0,0)}_{0,0} V^{(1,1,0,0)}_{0,0}+X_1 V^{(1,0,0,0)}_{1,0} V^{(1,1,0,0)}_{0,0}+X_1 V^{(1,0,0,0)}_{0,0} V^{(1,1,0,0)}_{1,0} \nonumber \\
& +V^{(1,0,0,0)}_{1,0} V^{(1,1,0,0)}_{1,0}+V^{(1,0,0,0)}_{2,0} V^{(1,1,0,0)}_{0,0}+V^{(1,0,0,0)}_{0,0} V^{(1,1,0,0)}_{2,0}+V^{(1,0,0,0)}_{0,0} V^{(1,1,0,0)}_{2',0},
\end{align}
which is expected from the fact that the number of the allowed chiral operators is eight while the number of the allowed monopole states:
\begin{align}
& \left(\mathrm{Tr}_1 X\right)^2 \left|2,1,0,0\right>, \quad \left(\mathrm{Tr}_1 X\right) \left(\mathrm{Tr}_2 X\right) \left|2,1,0,0\right>, \quad \left(\mathrm{Tr}_1 X\right) \left(\mathrm{Tr}_3 X\right) \left|2,1,0,0\right>, \nonumber \\
& \left(\mathrm{Tr}_2 X\right) \left(\mathrm{Tr}_3 X\right) \left|2,1,0,0\right>, \quad \left(\mathrm{Tr}_2 X\right)^2 \left|2,1,0,0\right>, \quad \left(\mathrm{Tr}_3 X\right)^2 \left|2,1,0,0\right>, \quad \left(\mathrm{Tr}_3 X^2\right) \left|2,1,0,0\right>
\end{align}
is seven. We have omitted the coefficients in the relation for simplicity. For the superpotential \eqref{eq:sup_D}, the operator $V^{(1,1,1,0)}_{0,0}$ and the relation $h(X_s,V^{(1,0,0,0)}_{s,t},V^{(1,1,0,0)}_{s,t})$ can be mixed:
\begin{align}
h(X_s,V^{(1,0,0,0)}_{s,t},V^{(1,1,0,0)}_{s,t}) \sim V^{(1,1,1,0)}_{0,0}
\end{align}
such that $V^{(1,1,1,0)}_{0,0}$ is written in terms of lower dimensional monopole operators. Therefore, there is no independent monopole operator of topological charge 3 once we turn on the superpotential.
\\

From the vortex partition function identity \eqref{eq:higher n} and the $N_c = 2,4$ examples, we expect that the independent monopole operators are give by
\begin{align}
\begin{array}{ll}
V^\pm_{s,t}, \qquad & s = 0, \ldots, n-1, \quad t = 0, 1, 2, \quad s t = 0, \\
W^\pm_u, \qquad & u = 0, \ldots, \frac{n-3}{2}
\end{array}
\end{align}
for high enough gauge ranks. Those operators are mapped to the following monopole states on $S^2$:
\begin{align}
\begin{aligned}
\label{eq:monopoles2}
V^\pm_{s,t} &\sim \mathrm{Tr} X^s Y^t \left|1,0,0,\ldots,0\right>, \\
W^\pm_u &\sim \mathrm{Tr} X^{2 u}\left|1,1,0\ldots,0\right>
\end{aligned}
\end{align}
by radial quantization. Note that other types of monopole operators, such as
\begin{align*}
\mathrm{Tr} X^{2 u+1} \left|1,1,0,\ldots,0\right> \qquad \text{and} \qquad \left|[1]^p,[0]^{N_c-p}\right> \quad \text{with} \quad p > 2,
\end{align*}
which appear in a generic two-adjoint theory without superpotential, do not appear in our theory because they are described by lower dimensional monopole operators by quantum relations. Also $V^\pm_{s,t}$ and $W^\pm_u$ in \eqref{eq:monopoles2} can be truncated if the gauge rank is low, as we have seen in the examples.

Indeed, the appearance of monopole operators of higher topological charge is in contrast to a theory with one or no adjoint matter. Without an adjoint matter, most of the monopole states are lifted due to the instanton effect. Only the following types of monopole states remain in the chiral ring:
\begin{align}
\left|m,0,\ldots,-n\right> \sim (V^+)^m (V^-)^n.
\end{align}
With one or more adjoint matters, the instanton superpotential is forbidden due to additional zero modes from the adjoints. All the monopole states therefore remain in the chiral ring unless there is a nontrivial superpotential. Nevertheless, for the one-adjoint case, they are still spanned by unit monopole operators
\begin{align}
V^+_s \sim \mathrm{Tr} X^s \left|1,0,\ldots,0\right>.
\end{align}
On the other hand, as we have seen, new independent monopole operators of higher topological charges are required to span arbitrary monopole flux for the two-adjoint case.
\\

\section{List of Growing Trees}
\label{sec:growing trees}

\begin{figure}[htbp]
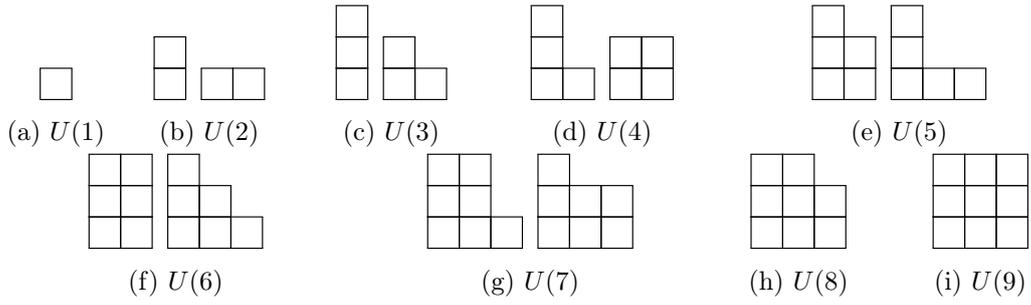

\centering
\begin{subfigure}{0.1\textwidth}
\centering
\ytableausetup{boxsize=0.4cm}
\ydiagram{0,0,1}
\caption{$U(1)$}
\end{subfigure}
\begin{subfigure}{0.15\textwidth}
\centering
\ydiagram{0,1,1} \, \ydiagram{0,0,2}
\caption{$U(2)$}
\end{subfigure}
\begin{subfigure}{0.15\textwidth}
\centering
\ydiagram{1,1,1} \, \ydiagram{0,1,2}
\caption{$U(3)$}
\end{subfigure}
\begin{subfigure}{0.2\textwidth}
\centering
\ydiagram{1,1,2} \, \ydiagram{0,2,2}
\caption{$U(4)$}
\end{subfigure}
\begin{subfigure}{0.3\textwidth}
\centering
\ydiagram{1,2,2} \, \ydiagram{1,1,3}
\caption{$U(5)$}
\end{subfigure}
\begin{subfigure}{0.3\textwidth}
\centering
\ydiagram{2,2,2} \, \ydiagram{1,2,3}
\caption{$U(6)$}
\end{subfigure}
\begin{subfigure}{0.3\textwidth}
\centering
\ydiagram{2,2,3} \, \ydiagram{1,3,3}
\caption{$U(7)$}
\end{subfigure}
\begin{subfigure}{0.15\textwidth}
\centering
\ydiagram{2,3,3}
\caption{$U(8)$}
\end{subfigure}
\begin{subfigure}{0.15\textwidth}
\centering
\ydiagram{3,3,3}
\caption{$U(9)$}
\end{subfigure}
\caption{\label{fig:n=3} The growing trees are listed for $n = 3$.}
\end{figure}

In this appendix, we give explicit lists of the growing trees for $n = 3, 5$ with a single flavor. For a single flavor, those are exactly the contributing forest graphs while for multiple flavors, the forest graphs are given by possible unions of growing trees that the total number of the nodes is the rank of the gauge group. Figure \ref{fig:n=3} is the list of the growing trees for $n = 3$. Figure \ref{fig:n=5} is the list of the growing trees for $n = 5$.
\begin{figure}[htbp]
\centering
\begin{subfigure}{0.1\textwidth}
\centering
\ydiagram{0,0,0,0,1}
\caption{$U(1)$}
\end{subfigure}
\begin{subfigure}{0.2\textwidth}
\centering
\ydiagram{0,0,0,1,1} \, \ydiagram{0,0,0,0,2}
\caption{$U(2)$}
\end{subfigure}
\begin{subfigure}{0.2\textwidth}
\centering
\ydiagram{0,0,1,1,1} \, \ydiagram{0,0,0,1,2}
\caption{$U(3)$}
\end{subfigure}
\begin{subfigure}{0.3\textwidth}
\centering
\ydiagram{0,1,1,1,1} \, \ydiagram{0,0,1,1,2} \, \ydiagram{0,0,0,2,2}
\caption{$U(4)$}
\end{subfigure}
\begin{subfigure}{0.3\textwidth}
\centering
\ydiagram{1,1,1,1,1} \, \ydiagram{0,1,1,1,2} \, \ydiagram{0,0,1,2,2}
\caption{$U(5)$}
\end{subfigure}
\begin{subfigure}{0.3\textwidth}
\centering
\ydiagram{1,1,1,1,2} \, \ydiagram{0,1,1,2,2} \, \ydiagram{0,0,2,2,2}
\caption{$U(6)$}
\end{subfigure}
\begin{subfigure}{0.3\textwidth}
\centering
\ydiagram{1,1,1,2,2} \, \ydiagram{1,1,1,1,3} \, \ydiagram{0,1,2,2,2}
\caption{$U(7)$}
\end{subfigure}
\begin{subfigure}{0.3\textwidth}
\centering
\ydiagram{1,1,2,2,2} \, \ydiagram{1,1,1,2,3} \, \ydiagram{0,2,2,2,2}
\caption{$U(8)$}
\end{subfigure}
\begin{subfigure}{0.4\textwidth}
\centering
\ydiagram{1,2,2,2,2} \, \ydiagram{1,1,2,2,3} \, \ydiagram{1,1,1,3,3}
\caption{$U(9)$}
\end{subfigure}
\begin{subfigure}{0.4\textwidth}
\centering
\ydiagram{2,2,2,2,2} \, \ydiagram{1,2,2,2,3} \, \ydiagram{1,1,2,3,3}
\caption{$U(10)$}
\end{subfigure}
\begin{subfigure}{0.4\textwidth}
\centering
\ydiagram{2,2,2,2,3} \, \ydiagram{1,2,2,3,3} \, \ydiagram{1,1,3,3,3}
\caption{$U(11)$}
\end{subfigure}
\begin{subfigure}{0.3\textwidth}
\centering
\ydiagram{2,2,2,3,3} \, \ydiagram{1,2,3,3,3}
\caption{$U(12)$}
\end{subfigure}
\begin{subfigure}{0.3\textwidth}
\centering
\ydiagram{2,2,3,3,3} \, \ydiagram{1,3,3,3,3}
\caption{$U(13)$}
\end{subfigure}
\begin{subfigure}{0.15\textwidth}
\centering
\ydiagram{2,3,3,3,3}
\caption{$U(14)$}
\end{subfigure}
\begin{subfigure}{0.15\textwidth}
\centering
\ydiagram{3,3,3,3,3}
\caption{$U(15)$}
\end{subfigure}
\caption{\label{fig:n=5} The growing trees are listed for $n = 5$.}
\end{figure}

\newpage

\bibliography{mybib}
\bibliographystyle{JHEP}
\end{document}